\documentclass[review]{elsarticle}

\usepackage{amssymb}
\usepackage{latexsym}

\usepackage{url}
\usepackage{xcolor}
\definecolor{newcolor}{rgb}{.8,.349,.1}

\usepackage{amsmath}
\usepackage{mathrsfs}
\usepackage{color}
\usepackage{lineno}
\usepackage[pdfencoding=auto]{hyperref}
\hypersetup{colorlinks=true}
\modulolinenumbers[5]
\usepackage{cite}
\usepackage[ruled,vlined]{algorithm2e}
\usepackage{setspace}
\usepackage{etoolbox}
\usepackage{graphicx}
\usepackage[labelformat=simple]{subcaption}

\usepackage{multirow}
\usepackage{times}

\AtBeginEnvironment{algorithm}{\setstretch{1.35}}

\newcounter{subassumption}[assumption]

\makeatletter
\renewcommand{\p@subassumption}{\theassumption}
\makeatother

\newcommand{\Ndim}{L}
\newcommand{\Psize}{K}
\newcommand{\Natom}{N_{\mathrm{d}}}
\newcommand{\Npatch}{N_{\mathrm{p}}}

\newcommand{\vf}[1]{#1}

\journal{ELSEVIER journal}

\begin{document}

\begin{frontmatter}

\title{Coupled dictionary learning for unsupervised change detection between multi-sensor \\remote sensing images\tnoteref{label1}}

\author[n7]{Vinicius {Ferraris}\corref{cor1}}
\ead{firstname.lastname@enseeiht.fr}
\author[n7]{Nicolas {Dobigeon}}
\author[n7]{Yanna {Cavalcanti}}
\author[n7]{Thomas {Oberlin}}
\author[n7]{Marie {Chabert}}
\address[n7]{University of Toulouse, IRIT/INP-ENSEEEIHT, 2 Rue Camichel, 31071 Toulouse, France}

\tnotetext[label1]{Part of this work has been supported by Coordenação de Aperfeiçoamento de Ensino Superior (CAPES), Brazil, the EU FP7 through the ERANETMED JC-WATER program [MapInvPlnt Project ANR-15-NMED-0002-02] and the ANR-3IA Artificial and Natural Intelligence Toulouse Institute (ANITI).}


\begin{abstract}
Archetypal scenarios for change detection generally consider two images acquired through sensors of the same modality. However, in some specific cases such as emergency situations, the only images available may be those acquired through sensors of different modalities. This paper addresses the problem of unsupervisedly detecting changes between two observed images acquired by sensors of different modalities with possibly different resolutions. These sensor dissimilarities introduce additional issues in the context of operational change detection that are not addressed by most of the classical methods. This paper introduces a novel framework to effectively exploit the available information by modelling the two observed images as a sparse linear combination of atoms belonging to a pair of coupled overcomplete dictionaries learnt from each observed image. As they cover the same geographical location, codes are expected to be globally similar, except for possible changes in sparse spatial locations. Thus, the change detection task is envisioned through a dual code estimation which enforces spatial sparsity in the difference between the estimated codes associated with each image. This problem is formulated as an inverse problem which is iteratively solved using an efficient proximal alternating minimization algorithm accounting for nonsmooth and nonconvex functions. The proposed method is applied to real images with simulated yet realistic and real changes. A comparison with state-of-the-art change detection methods evidences the accuracy of the proposed strategy.
\end{abstract}

\end{frontmatter}


\section{Introduction}

Ecosystems exhibit permanent variations at different temporal and spatial scales caused by natural, anthropogenic, or even both factors \citep{coppin_review_2004}. Monitoring spatial variations over a period of time is an important source of knowledge that helps understanding the possible transformations occurring on Earth's surface. Therefore, due to the importance of quantifying these transformations, change detection (CD) has been an ubiquitous issue addressed in the remote sensing and geoscience literature \citep{lu_change_2004}.

Remote sensing CD methods can be first classified with respect to (w.r.t.) their supervision \citep{bovolo_theoretical_2007}, depending on the availability of prior knowledge about the expected changes. More precisely, supervised CD methods require ground reference information about at least one of the observations. Conversely, unsupervised CD can be contextualized as automatic detection of changes without the need for any further external knowledge. Each class of CD methods present particular competitive advantages w.r.t. the others. For instance, supervised CD methods generally achieve better accuracy for predefined modalities whereas unsupervised methods are characterised by their flexibility and genericity. Nevertheless, implementing supervised methods require the acquisition of relevant ground information, which is a very challenging and expensive task, in terms of human and time resources \citep{bovolo_theoretical_2007}. Relaxing this constraints makes unsupervised methods more suitable for operational CD.

CD methods can also be categorized w.r.t. the imagery modalities the method is able to handle. As remote sensing encompasses many different types of imagery modalities (e.g., single- and multi-band optical images, radar, LiDAR), dedicated CD methods have been specifically developed for each one by exploiting its acquisition process and the intrinsic characteristics of the resulting data. Thus, due to differences in the physical meaning and statistical properties of images from different sensor modalities, a general CD method able to handle all modalities is particularly difficult to design and to implement. For this reason, most of the CD methods focus on a pair of images from one single target modality. In this case, the images are generally compared pixel-wisely using the underlying assumption of same spatial resolutions \citep{singh_review_1989,bovolo_time_2015}. Nevertheless, in some practical scenarios such as, e.g., emergency missions due to natural disasters, when the availability of data and the responsiveness are strong constraints, CD methods may have to handle observations of different modalities and/or resolutions. This highlights the need for robust and flexible CD techniques able to deal with this kind of observations.

\vf{The literature about multimodal CD is very limited, yet a few relevant references include the works by \citet{kawamura_automatic_1971}, \citet{bruzzone_neural-statistical_1999}, \citet{inglada_similarity_2002}, \citet{lu_change_2004}, \citet{alberga_comparison_2007}, \citet{mercier_conditional_2008} and \citet{prendes_new_2015}}. However, multimodal CD has always been an important topic since the initial development of CD methods.  Earlier work by \citet{kawamura_automatic_1971} described the potential of CD between a multimodal collection of datasets (e.g., photographic, infrared and radar), applied to weather prediction and land surveillance. Three features are extracted from the pair and the CD algorithm is trained on a learning set. According to \citet{lu_change_2004}, various methods dedicated to CD between images from different sources of data are grouped as geographical information system-based methods. For instance, \citet{solberg_markov_1996} proposed a supervised classification of multisource satellite images using Markov random fields. The work of \citet{bruzzone_neural-statistical_1999} uses compound classification to detect changes in multisource data. The method uses artificial neural networks to estimate the posterior probability of classes. Moreover,  \citet{inglada_similarity_2002} studies the relevance of several similarity measures between multisensor data. These measures are implemented in a CD context \citep{alberga_comparison_2007}. A preprocessing technique based on conditional copula that contributes to better statistically modeling multisensor images was proposed by \citet{mercier_conditional_2008}. Besides, \citet{brunner_earthquake_2010} presented a strategy to assess building damages using a pair of very high resolution (VHR) optical and radar images by geometrically modeling buildings in both modalities. \citet{chabert_logistic_2010} updated information databases by means of logistic regression. More recently, the work of \citet{prendes_new_2015} presented a supervised method to infer changes after learning a manifold defined by pairs of patches extracted from the two images. Although some of these methods present relatively high detection performance, they are often restrained to a specific modality or to a specific target application. For instance, \citet{Solano-Correa2018} proposed an approach to detect changes between multispectral images with different spatial and spectral resolutions by homogenization of radiometric and geometric image properties. However, this approach relies in particular on a taxonomy of possible radiometric changes observed in very high resolution images. Moreover, some methods are only suitable for building damage assessment taking benefit of their high-level modeling, but show a poor adaptability to other scenarios \citep{brunner_earthquake_2010,chabert_logistic_2010}. The other ones estimate some metrics from unchanged trained samples, which prevents their application within a fully unsupervised context \citep{bruzzone_neural-statistical_1999,prendes_new_2015,mercier_conditional_2008}.

Recently, an unsupervised multi-source CD method based on coupled dictionary learning was addressed by \citet{gong_coupled_2016}. In the proposed methodology, the CD is based on the reconstruction error of patches approximated thanks to estimated coupled dictionary and independent sparse codes. Atoms of the dictionary are learnt from pairs of patches jointly extracted from the observed images. Following the same principle, in \citet{lu_joint_2017}, a semi-supervised method was used to handle multispectral images based on joint dictionary learning. Both methods rely on the rationale that the coupled dictionary estimated from the observed images tends to produce stronger reconstruction errors in change regions rather than in unchanged ones. Because of the multi-modality, the problem has not been formulated in the image space, but rather in a latent space formed by the coupled dictionary atoms. However, both methods exhibit some crucial issues that may impair their relative performance. First, the underlying optimization problem is highly nonconvex and no convergence guarantees are ensured, even by using some traditional dictionary learning methods \citep{aharon_k-svd_2006}. Then, the considered CD problem has been split into two distinct steps: dictionary learning and code estimation. The errors in code estimation may produce false alarms in the final CD even with reliable dictionary estimates. Also, the statistical model of the noise -- inherent to each sensor modality -- has not been taken into consideration explicitly, which may dramatically impact the CD performance \citep{campbell_introduction_2011}. Finally, these methods do not consider  overlapping patches, which potentially would increase their robustness and thus do not explicitly handle the problem of possible differences in spatial resolutions \citep{ferraris_robust_2017,ferraris_detecting_2017}. The adequacy between the size of patches and the image scale is not discussed, although it may have a negative impact on the dictionary coupling and thus on the detection performance.

Overcoming these limitations, this paper proposes a similar methodology to learn coupled dictionaries able to conveniently model multimodal remote sensing images. Specifically, contrary to the aforementioned methods, the problem is fully formulated without splitting the learning and coding steps. Also, an appropriate statistical model is derived to describe the image from each specific remote sensing modality. Besides, the proposed method explicitly allows patch overlapping within the overall estimation process. To couple images with different resolutions, additional scaling matrices inspired by the work by \citet{seichepine_soft_2014} are jointly estimated within the whole process. Finally, as the problem is highly nonconvex, it is iteratively solved based on the proximal alternating linearized minimization (PALM) algorithm \citep{bolte_proximal_2014}, which ensures convergence towards a  critical point for some nonconvex nonsmooth problems. Note that the proposed patch-based method departs from segmentation-based methods which generally extract change information at an object-level, whose resolution is implicitly defined by the chosen segmentation procedure \citep{Feng2018}. Instead, the proposed method linearly decomposes overlapping square patches onto an appropriate common latent space, which allows CD to be operated at a pixel level.

This manuscript is organized as follows. Generic and well-admitted image models are introduced in Section \ref{sec:models}. Capitalizing on these image models, Section \ref{sec:DL} formulated the CD problem as a coupled dictionary learning. Section \ref{sec:min_alg} proposes an algorithmic solution to minimize the resulting CD-based objective function. Section \ref{sec:experiments} reports experimental results obtained on synthetic images, considering three distinct simulation scenarios. Experiments conducted on real images are presented in Section \ref{subsec:reference_images}. Finally, Section \ref{sec:conclusion} concludes the manuscript.

\section{Image models}
\label{sec:models}

\subsection{Forward model}
\label{subsec:GFM}

Let us consider that the image formation process inherent to all digital remote sensing imagery modalities is modeled as a sequence of transformations, denoted $\mathit{T}[\cdot]$. This sequence applies to the original scene to produce the sensor output image. This output image is referred to as the observed image and is denoted by $\mathbf{Y} \in \mathbb{R}^{\Ndim \times N}$
  consisting of $N$ voxels $\mathbf{y}_{i} \in \mathbb{R}^{\Ndim}$ stacked lexicographically that is from left to right, row by row. The voxel dimension $\Ndim$ may represent different quantities depending on the modality of the data. For instance, it stands for the number of spectral bands in the case of multiband optical images \citep{ferraris_robust_2017} or for the number of polarization modes in the case of polarimetric synthetic aperture radar (POLSAR) images. The observed image provides a limited representation of the original scene with properties imposed by the image signal processor characterizing the sensor. The original scene cannot be exactly represented because of its continuous nature, but it can be conveniently approximated by a latent (i.e., unobserved) image $\mathbf{X} \in \mathbb{R}^{\Ndim \times N}$ related to the observed image as  follows
\begin{equation}
\label{eq:sensortransf}
	\mathbf{Y} = \mathit{T}[\mathbf{X}].
\end{equation}
The sequence of transformations $\mathit{T}[\cdot]$ operated by the sensor over the latent image is often referred to as the \emph{degradation process}. It may represent resolution degradations accounting for the spatial and/or spectral characteristics of the sensor \citep{ferraris_detecting_2017,ferraris_robust_2017}. In this paper, it specifically models the intrinsic noise corruption associated to the sensor modality \citep{sun_alternating_2014}. The latent image $\mathbf{X}$ can be understood, in this context, as a noise-free version of the observed image $\mathbf{Y}$ with the same resolution.

More precisely, the transformation $\mathit{T}[\cdot]$ underlies the likelihood function $p(\mathbf{Y}|\mathbf{X})$  which statistically models the observed image $\mathbf{Y}$ conditionally to the latent image $\mathbf{X}$ by taking into account the noise statistics. The noise statistical model mainly depends on the modality and rely on some classical distributions, e.g., the Gaussian distribution for optical images or the Gamma distribution for multi-look SAR images. Moreover, as already pointed out by \citet{fevotte_nonnegative_2009} in a different application context, for a wide family of distributions, this likelihood function relies on a divergence measure $\mathcal{D}(\cdot|\cdot)$ between the observed and latent images, which finally defines an explicit data-fitting term through a negative-log transformation
\begin{equation}
			- \log p(\mathbf{Y}|\mathbf{X}) = \phi^{-1} \mathcal{D}(\mathbf{Y}|\mathbf{X}) + \theta
\end{equation}
where $\phi$ and $\theta$ are parameters characterizing the distributions. In \ref{ap:dft}, the divergence measures $\mathcal{D}(\cdot|\cdot)$ are derived for two of the most common remote sensing image modalities, namely optical multiband and SAR images, considered in this work.

\subsection{Latent image sparse model}\label{subsec:sparse_model}

Sparse representations have been an ubiquitous and well-admitted tool to model images in various applications and task-driven contexts \citep{mairal_sparse_2014}. Indeed, natural images are known to be compressible in a transformed domain, i.e., they can be efficiently represented by a few expansion coefficients acting on basis functions  \citep{mallat_wavelet_2009}. This finding has motivated numerous works on image understanding, compression and denoising \citep{olshausen_sparse_1997,chen_atomic_2001}. In earlier works, this transformed domain, equivalently defined by the associated basis functions, was generally fixed in advance and chosen in agreement with the expected spatial content of the images \citep{mallat_wavelet_2009}. Thus, the basis functions belonged to pre-determined families with specific representation abilities, such as cosines, wavelets, contourlets, shearlets, among others. More recently, the seminal contribution by Aharon \emph{et al.} proposed a new paradigm by learning an overcomplete dictionary jointly with a sparse code \citep{aharon_k-svd_2006}. This dictionary learning-based approach exploits the key property of self-similarity characterizing the images to provide an adaptive representation. Indeed, it aims at identifying elementary patches that can be linearly and sparsely combined to approximate the observed image patches. In this paper, following the approach by \citet{aharon_k-svd_2006}, we propose to resort to this dictionary-based representation to model the latent image $\mathbf{X}$. \vf{More precisely, the image is first decomposed into a set of $N_{\mathrm{p}}$ 3D-patches with $1 \leq N_{\mathrm{p}}\leq N$. %
Let $\mathcal{R}_{i} : \mathbb{R}^{\Ndim \times N} \rightarrow \mathbb{R}^{\Psize^{2}\Ndim}$ denote a binary operation modeling the extraction, from the image, of the $i$th patch ($i \in \{1,\ldots,N_{\mathrm{p}}\}$) such that
\begin{equation}
\label{eq:patch_extraction}
\mathbf{p}_{i} = \mathcal{R}_{i}\mathbf{X}
\end{equation}
where $\mathbf{p}_{i} \in \mathbb{R}^{\Psize^{2}\Ndim}$ stands for the $i$th $\Psize \times \Psize \times \Ndim$-pixel patch in its vectorized form. The integer $K>1$ defines the spatial size of the patches, i.e., its number of rows and columns before being vectorized. Note that the number of patches $\Npatch$ is such that $1 \leq \Npatch \leq N$ and patches may overlap. The choice of the number $N_{\mathrm{p}}$ of patches will be more deeply discussed in Section \ref{sec:DL} in the specific context of CD.} The conjugate of the patch-extraction operator\footnote{Note that, despite a slight abuse of notation, the operator $\mathcal{R}$ (resp., $\mathcal{R}^T$) does not stand for a matrix, but rather for a linear operator acting on the image $\mathbf{X}$ (resp., the patch $\mathbf{p}_{i}$) directly.}, denoted $\mathcal{R}^{T}_{i}$, acts on $\mathbf{p}_{i}$ to produce a zero-padded image composed by the unique patch $\mathbf{p}_{i}$ located at the $i$th spatial position.

In accordance with dictionary-based representation principles, these patches are assumed to be approximately and independently modeled as sparse combinations of atoms belonging to an overcomplete dictionary $\mathbf{D}  = \left[\mathbf{d}_{1}, \cdots, \mathbf{d}_{\Natom}\right] \in \mathbb{R}^{\Psize^{2}\Ndim \times \Natom}$
\begin{equation}
\label{eq:patchRepresentation}
	\mathbf{p}_{i}|\mathbf{D},\mathbf{a}_{i} \sim \mathcal{N}\left(\mathbf{D}\mathbf{a}_{i},\sigma^2 \mathbf{I}_{\Natom}\right)
\end{equation}
where \vf{$\Natom >0 $ stands for the user-defined number of atoms composing the dictionary, commonly referred to as dictionary size and} $\mathbf{a}_{i} \in \mathbb{R}^{\Natom}$ represents the code (coefficients) of the current patch over the dictionary, $\boldsymbol{\Sigma}=\sigma^2 \mathbf{I}_{\Natom}$ is the error covariance matrix and . \vf{Let $\mathbf{P} \in \mathbb{R}^{\Psize^{2}\Ndim \times \Npatch} = \left[\mathbf{p}_{1}, \cdots, \mathbf{p}_{\Npatch}\right]$ denote the matrix that stacks the set of all, possibly overlapping, patches extracted from the latent image $\mathbf{X}$ at $\Npatch$ spatial positions arranged on a generally regular spatial grid and enumerated in a lexicographical order (i.e., from left to right and top to bottom of the image). The matrix $\mathbf{A} \in \mathbb{R}^{\Natom \times \Npatch} = \left[\mathbf{a}_{1}, \cdots, \mathbf{a}_{\Npatch}\right]$ is the code matrix in which each column represents the code for each corresponding column of $\mathbf{P}$.} The overcompletness property of the dictionary, occurring when the number of atoms is greater than the effective dimensionality of the input space, $\Natom \gg \Psize^{2} \Ndim $, allows for the sparsity of the representation \citep{olshausen_sparse_1997}. The overcompletness implies redundancy and non-orthogonality between atoms. This property is not necessary for the decomposition, but has been proved to be very useful in some applications like denoising and compression \citep{aharon_k-svd_2006}.
Given the image patch matrix $\mathbf{P}$, dictionary learning aims at recovering the set of atoms $\mathbf{D}$ and the associated code matrix $\mathbf{A}$ and it is generally tackled through a 2-step procedure. First, inferring the code matrix $\mathbf{A}$ associated with the patch matrix $\mathbf{P}$ and the dictionary $\mathbf{D}$ can be formulated as a set of $\Npatch$ sparsity-penalized optimization problems. Sparsity of the code vectors $\mathbf{a}_i=\left[a_{1i},\ldots,a_{\Natom i}\right]^T$ ($i=1,\ldots,\Npatch$) can be promoted by minimizing its $\ell_{0}$-norm. However, since this leads to a non-convex problem \citep{chen_atomic_2001}, it is generally substituted by the corresponding convex relaxation, i.e., an $\ell_1$-norm. Within a probabilistic framework, taking into account the expected non-negativeness of the code, this choice can be formulated by assigning a single-side exponential (i.e., Laplacian) prior distribution to the code components, assumed to be a priori independent
 \begin{equation}
	\label{eq:code_model}
	\mathbf{a}_{i} \sim \prod_{j=1}^{\Natom}\mathcal{L}(a_{ji};\lambda)
\end{equation}
where $\lambda$ is the hyperparameter adjusting the sparsity level over the code.
Conversely, learning the dictionary $\mathbf{D}$ given the code $\mathbf{A}$ can also be formulated as an optimization problem. As the number of solutions for the dictionary learning problem can be extremely large, one common assumption is to constrain the energy of each atom, thereby preventing $\mathbf{D}$ to become arbitrarily large \citep{mairal_online_2009}. Moreover, in the particular context considered in this work, to promote the positivity of the reconstructed patches, the atoms are also constrained to positive values. Thus, each atom will be constrained to the set
\begin{equation}\label{eq:dict_dist}
  \mathcal{S} \triangleq \left\{\mathbf{D} \in \mathbb{R}^{\Psize^{2}\Ndim \times \Natom}_{+} \; \mid \forall j \in\left\{ 1,\ldots,\Natom,\right\} \; \left\|\mathbf{d}_j\right\|^2_{2} = 1\right\}.
\end{equation}
\subsection{Optimization problem}
\label{subsec:sdl}
Adopting a Bayesian probabilistic formulation of the image model introduced in Sections \ref{subsec:GFM} and \ref{subsec:sparse_model}, the posterior probability of the unknown variables $\mathbf{X}$, $\mathbf{D}$ and $\mathbf{A}$ can be derived using the probability chain rule \citep{gelman_bayesian_2004}
\begin{equation}
\label{eq:single_bayesian_model}
p(\mathbf{X},\mathbf{D},\mathbf{A}|\mathbf{Y}) \propto p(\mathbf{Y}|\mathbf{X}) p(\mathbf{X}|\mathbf{D},\mathbf{A}) p(\mathbf{D})p(\mathbf{A})
\end{equation}
where $p(\mathbf{Y}|\mathbf{X})$ is the likelihood function relating the observation data to the latent image through the direct model \eqref{eq:sensortransf}, $p(\mathbf{X}|\mathbf{D},\mathbf{A})$ is the dictionary-based prior model of the latent image, $p(\mathbf{D})$ and $p(\mathbf{A})$ are the (hyper-)prior distributions associated with the dictionary and the sparse code. Under a maximum a posteriori (MAP) paradigm, the joint MAP estimator $\left\{\hat{\mathbf{X}}_{\text{MAP}},\hat{\mathbf{D}}_{\text{MAP}},\hat{\mathbf{A}}_{\text{MAP}}\right\}$ can be derived by minimizing the negative log-posterior, leading to the following minimization problem
\begin{equation}
	\label{eq:MAP}		
    \left\{\hat{\mathbf{X}}_{\text{MAP}},\hat{\mathbf{D}}_{\text{MAP}},\hat{\mathbf{A}}_{\text{MAP}} \right\} \in \mathop{\rm argmin}\limits_{\mathbf{X},\mathbf{D},\mathbf{A}}   \mathcal{J}\left(\mathbf{X},\mathbf{D},\mathbf{A}\right)
\end{equation}
with
\begin{equation}
\label{eq:objective_single}
\begin{aligned}
  \mathcal{J}\left(\mathbf{X},\mathbf{D},\mathbf{A}\right) &= \mathcal{D}(\mathbf{Y}|\mathbf{X}) \\
  &+\frac{\sigma^{2}}{2}\sum_{i=1}^{\Npatch}\left\|\mathcal{R}_{i}\mathbf{X} - \mathbf{D}\mathbf{a}_{i}\right\|_{\mathrm{F}}^{2} + \\
  & + \lambda \left\|\mathbf{A}\right\|_1 + \iota_{\mathcal{S}}(\mathbf{D})
\end{aligned}
\end{equation}
where $\iota_{\mathcal{S}}$ represents the indicator function on the set $\mathcal{S}$,
 \begin{equation}
	\iota_{\mathcal{S}}(z) = \left\{\begin{matrix}
0 \quad \text{if} \ z \in \mathcal{S}\\
+\infty \quad \text{elsewhere} \\
\end{matrix}\right.
\end{equation}
and $\mathcal{D}(\cdot|\cdot)$ is the data-fitting term associated with the image modality.

This model has been widely advocated in the literature, e.g., for denoising images of various modalities \citep{elad_image_2006,ma_dictionary_2013}. Particularly, in \citet{ma_dictionary_2013}, an additional regularization $\Psi\left(\mathbf{X}\right)$ of the latent image was introduced as the target modalities may present strong fluctuations due to their inherent image formation process, i.e. Poissonian or multiplicative gamma processes. The final objective function \eqref{eq:objective_single} can thus be rewritten as
\begin{equation}
	\label{eq:objective_single_den}
	\begin{aligned}
  \mathcal{J}\left(\mathbf{X},\mathbf{D},\mathbf{A}\right) &=
  \mathcal{D}(\mathbf{Y}|\mathbf{X}) \\
  & + \frac{\sigma^{2}}{2}\sum_{i=1}^{\Npatch}\left\|\mathcal{R}_{{i}}\mathbf{X} - \mathbf{D}\mathbf{a}_{i}\right\|_{\mathrm{F}}^{2} + \Psi\left(\mathbf{X}\right) \\
  &+ \lambda \left\|\mathbf{A}\right\|_1 + \iota_{\mathcal{S}}(\mathbf{D})
	\end{aligned}
\end{equation}
where, for instance, $\Psi\left(\mathbf{X}\right)$ can stand for a total-variation (TV) regularization \citep{ma_dictionary_2013}.

The next section expands the proposed image models to handle a pair of observed images in the specific context of CD.

\section{From change detection to coupled dictionary learning}
\label{sec:DL}

\subsection{Problem statement}
\label{subsec:ps}

Let us consider two geographically aligned observed images $\mathbf{Y}_{1} \in \mathbb{R}^{\Ndim_{1} \times N_{1}}$ and $\mathbf{Y}_{2} \in \mathbb{R}^{\Ndim_{2} \times N_{2}}$ acquired by two sensors $\mathsf{S}_{1}$ and $\mathsf{S}_{2}$ at times $t_1$ and $t_2$, respectively. The ordering of acquisition times is indifferent, i.e., either $t_2 < t_1$ or $t_2 > t_1$ are possible cases and the order does not impact the applicability the proposed method. The problem addressed
in this paper consists in detecting significant changes between these two observed images. This is a challenging task mainly due to the possible dissimilarities in terms of spatial and/or spectral resolutions and of modality. Indeed, resolution dissimilarity prevents any use of classical CD algorithms without homogenization of the resolutions as a preprocessing step \citep{singh_review_1989,bovolo_time_2015}. Moreover modality dissimilarity, which makes most of the CD algorithms inoperative because their inability of handling images of different nature \citep{ferraris_detecting_2017,ferraris_robust_2017}. To alleviate this issue, this work proposes to improve and generalize the CD methods introduced by \citet{seichepine_soft_2014,gong_coupled_2016,lu_joint_2017}. Following the widely admitted forward model described in Section \ref{subsec:GFM} and adopting consistent notations, the observed images $\mathbf{Y}_{1}$ and $\mathbf{Y}_{2}$ can be related to two latent images $\mathbf{X}_{1} \in \mathbb{R}^{\Ndim_{1} \times N_{1}}$ and $\mathbf{X}_{2} \in \mathbb{R}^{\Ndim_{2} \times N_{2}}$
\begin{subequations}
\label{eq:jointobsmodel}
		\begin{align}
			&\mathbf{Y}_{1} = \mathit{T}_{1}[\mathbf{X}_{1}]  \label{eq:jointobsmodel1}\\
			&\mathbf{Y}_{2} = \mathit{T}_{2}[\mathbf{X}_{2}] \label{eq:jointobsmodel2}
		\end{align}
\end{subequations}
where $\mathit{T}_{1}$ and $\mathit{T}_{2}$ denote two degradation operators imposed by the sensors $\mathsf{S}_{1}$ and $\mathsf{S}_{2}$. Note that \eqref{eq:jointobsmodel} is a double instance of the model \eqref{eq:sensortransf}. In particular, in the CD context considered in this work, the two latent images $\mathbf{X}_{1}$ and $\mathbf{X}_{2}$ are supposed to represent the same geographical region provided the observed images have been beforehand co-registered.

Both latent images can be represented thanks to a dedicated dictionary-based decomposition, as stated in Section \ref{subsec:sparse_model}. More precisely, a pair of homologous
patches extracted from each image represents the same geographical spot. Each patch can be reconstructed from a sparse linear combination of atoms of an image-dependent dictionary. In the absence of changes between the two observed images, the sparse codes associated with the corresponding latent image are expected to be approximately the same and the two learned dictionaries are coupled \citep{yang_image_2010,yang_coupled_2012,zeyde_single_2010}. This coupling can be understood as the ability of deriving a joint representation for homologous multiple observations in a latent coupled space \citep{gong_coupled_2016}. Akin to the manifold proposed by \citet{prendes_new_2015}, this representation offers the opportunity to analyze images of different modalities in a common dual space. In the case where a pair of homologous patches has been extracted from two images representing the same scene, given perfect estimated coupled dictionaries, each patch should be exactly reconstructed thanks to the same sparse code. In other words, the pair of patches is an element of the latent coupled space.  Nevertheless, in the case where the pair of homologous patches does not represent exactly the same scene, owing to a change that occurs between acquisitions, perfect reconstruction cannot be achieved using the same code. This means that the pair of patches does not belong to the coupled spaces. Using the same code for reconstruction amounts to estimate the point in the coupled spaces that best approximates the patch pair. Thereby, relaxing this constraint in some possible change locations may provide an accurate reconstruction of both images while
spatially mapping change locations. In the specific context of CD addressed in this work, this finding suggests to evaluate any change between the two observed, or equivalently latent, images by comparing the corresponding codes
\begin{equation}
\label{eq:code_relaxation}
	\Delta\mathbf{A} = \mathbf{A}_{2} - \mathbf{A}_{1}
\end{equation}
where $\Delta\mathbf{A}=\left[\Delta\mathbf{a}_1,\ldots,\Delta\mathbf{a}_{\Npatch} \right] $ and $\Delta\mathbf{a}_{i} \in \mathbb{R}^{\Natom}$ denotes the code change vector associated with the $i$th patch, $\quad i=1,\ldots,\Npatch$. Then, to spatially locate the changes, a natural approach consists in monitoring the magnitude of $\Delta\mathbf{A}$, summarized by the code change energy image \citep{bovolo_theoretical_2007}
\begin{equation}
\label{eq:change_energy_matrix}
  \mathbf{e} =\left[e_1,\ldots,e_{\Npatch}\right]\in \mathbb{R}^{\Npatch}
\end{equation}
with
\begin{equation*}
  e_i = \left\|\Delta\mathbf{a}_i\right\|_2, \quad i=1,\ldots,\Npatch.
\end{equation*}
\vf{Note that, in the case of analyzing a pair of optical images, \citet{zanetti_rayleigh-rice_2015} proposed to describe the components $e_i$ of the energy vector $\mathbf{e}$ thanks to a Rayleigh-Rice mixture model whose parameters can be estimated to locate the changes. Conversely, in this work we propose to derive the CD rule directly from this magnitude. When the CD problem in the $i$th patch is formulated as the binary hypothesis testing
\begin{equation*}
\label{eq:test}
 \left\{
		\begin{array}{rcl}
			\mathcal{H}_{0,i} &:& \text{no change occurs in the $i$th patch}  \\
			\mathcal{H}_{1,i} &:& \text{a change occurs in the $i$th patch}
		\end{array}
        \right.
\end{equation*}
a patch-wise statistical test can be written by thresholding the code change energy
\begin{equation*}
    \label{eq:decision_rule}
  e_i \overset{\mathcal{H}_{1,i}}{\underset{\mathcal{H}_{0,i}}{\gtrless}} \tau
\end{equation*}
where the threshold $\tau \in [0,\infty]$ implicitly adjusts the target probability of false alarm or, reciprocally, the probability of detection.
The final binary CD map denoted ${\mathbf{m}} = \left[m_1,\ldots,m_{\Npatch}\right] \in \{0,1\}^{\Npatch}$ can be derived as
\begin{equation*}
	\label{eq:CVArule}
 {m}_i = \left\{\begin{array}{lll}
             1 & \mbox{if } e_i \geq \tau & (\mathcal{H}_{1,i})\\
			 0 & \mbox{otherwise}          & (\mathcal{H}_{0,i}).
				\end{array}\right.
\end{equation*}
The spatial resolution of this CD map is defined by the number $\Npatch$ of homologous patches extracted from the latent images $\mathbf{X}_1$ and $\mathbf{X}_2$. This number can be tailored by the user according to the adopted strategy of patch extraction. In practice, to reach the highest resolution, overlapping patches should be extracted according to the regular grid defined by the observed image of highest resolution, i.e., $\Npatch = \max\left\{N_1, N_2\right\}$.}

Finally, to solve the multimodal image CD problem, the key issue lies in the joint estimation of the pair of representation codes $\left\{\mathbf{A}_{1},\mathbf{A}_{2}\right\}$ or, equivalently, to the joint estimation of one code matrix and of the change code matrix, i.e. of $\left\{\mathbf{A}_{1},\Delta\mathbf{A}\right\}$, as well as of the pair of coupled dictionary $\left\{\mathbf{D}_{1},\mathbf{D}_{2}\right\}$ and consequently of the pair of latent images $\left\{\mathbf{X}_{1},\mathbf{X}_{2}\right\}$ from the joint forward model \eqref{eq:jointobsmodel}. The next paragraph introduces the CD-driven optimization problem to be solved.

\subsection{Coupled dictionary learning for CD}
\label{subsec:cdl}

The single dictionary estimation problem presented on Section \ref{subsec:sdl} can be generalized to take into account the modeling presented in Section \ref{subsec:ps}. Nevertheless, some previous considerations must be carefully handled in order to provide good coupling of the two dictionaries.

As the prior information about the dictionaries constrains each atom into the set $\mathcal{S}$ of unitary energy defined by \eqref{eq:dict_dist}, an unbiased estimation of the code change vector would allow a pair of unchanged homologous patches to be reconstructed  with exactly the same code, while changed patches would exhibit differences in their code. Obviously, this can only be achieved if the coupled dictionaries represent data with the same dynamics and resolutions. However, when analyzing images of different modalities and/or resolutions, this assumption can be not fulfilled. To alleviate this issue, we propose to resort to the strategy proposed by \citet{seichepine_soft_2014}, by introducing an additional diagonal scaling matrix constrained to the set $\mathcal{C} \triangleq \left\{\mathbf{S} \in \mathbb{R}^{N_{\mathrm{d}1} \times N_{\mathrm{d}1}}_{+} \; \mid \mathbf{S} = \mathrm{diag}(\mathbf{s}),\ \mathbf{s}\succeq 0 \right\}$ where $N_{\mathrm{d}1}$ is the size of the dictionary $\mathbf{D}_1$. This scaling matrix gathers the code energy differences originated from different modalities for each pair of coupled atoms. This is essential to ensure that the sparse codes of the two observed images are directly comparable, following \eqref{eq:code_relaxation}, and then properly estimated. Therefore, considering a pair of homologous patches, their joint representation model derived from \eqref{eq:patchRepresentation} can be written as
\begin{equation}
	\label{eq:coupled_patchRepresentation}
    \begin{aligned}
	&\mathbf{p}_{1{i}} = \mathcal{R}_{1{i}}\mathbf{X}_{1}  \approx \mathbf{D}_{1}\mathbf{S}\mathbf{a}_{1{i}}\\
	&\mathbf{p}_{2{i}} = \mathcal{R}_{2{i}}\mathbf{X}_{2} \approx \mathbf{D}_{2}\mathbf{a}_{2{i}} = \mathbf{D}_{2}\left(\mathbf{a}_{1{i}} + \Delta\mathbf{a}_{i}\right)
	\end{aligned}
\end{equation}
where $\left\{\mathbf{p}_{1{i}},\mathbf{p}_{2{i}}\right\}$
represents the pair of homologous patches and $\mathbf{S}$ is the diagonal scaling matrix.

Since the codes $\mathbf{A}_{1}$ and $\mathbf{A}_{2}$ are now element-wise comparable, a natural choice to enforce coupling between them should be the equality $\mathbf{A}_{1} = \mathbf{A}_{2} = \mathbf{A}$. This has been a classical assumption in various coupled dictionary learning applications \citep{yang_image_2010,zeyde_single_2010,yang_coupled_2012}. Nevertheless, in a CD context, some spatial positions may not contain the same objects. To account for possible changes in some specific locations while most of the patches remain unchanged, as in \citet{ferraris_robust_2017},  the code change energy matrix $\mathbf{e}$ defined by \eqref{eq:change_energy_matrix} is expected to be sparse. As a consequence, the corresponding regularizing function is chosen as the sparsity-inducing
$\ell_1$-norm of the code change energy matrix $\mathbf{e}$ or, equivalently, as the $\ell_{2,1}$-norm of the code change matrix
\begin{equation}
	\label{eq:phi_2}
	\phi_{2}\left(\Delta\mathbf{A}\right) = \left\|\Delta\mathbf{A}\right\|_{2,1} = \sum_{i=1}^{\Npatch} \left\|\Delta \mathbf{a}_i\right\|_2.
\end{equation}
This regularization is a specific instance of the non-overlapping group-lasso penalization \citep{bach_optimization_2011} which has been considered in various applications to promote structured sparsity \citep{wright_sparse_2009,fevotte_nonlinear_2015,ferraris_robust_2017}.

Then, a Bayesian model extending the one derived for a single image \eqref{eq:single_bayesian_model} leads to the posterior distribution of the parameters of interest
\begin{equation}
\begin{aligned}
\label{eq:coupled_bayesian_model}
p&\left(\mathbf{X}_{1},\mathbf{X}_{2},\mathbf{D}_{1},\mathbf{D}_{2},\mathbf{S},\mathbf{A}_{1},\Delta\mathbf{A}|\mathbf{Y}_{1},\mathbf{Y}_{2}\right)  \\ &\propto p(\mathbf{Y}_{1}|\mathbf{X}_{1}) p(\mathbf{Y}_{2}|\mathbf{X}_{2}) \\
&\times p(\mathbf{X}_{1}|\mathbf{D}_{1},\mathbf{S},\mathbf{A}_{1}) p(\mathbf{X}_{2}|\mathbf{D}_{2},\mathbf{A}_{1},\Delta\mathbf{A})\\
&\times p(\mathbf{D}_{1})p(\mathbf{D}_{2})p(\mathbf{S})p(\mathbf{A}_{1})p(\Delta\mathbf{A}).
\end{aligned}
\end{equation}
By incorporating all previously defined prior distributions (or, equivalently, regularizations), the joint MAP estimator $\hat{\boldsymbol{\Theta}}_{\text{MAP}} = \left\{\hat{\mathbf{X}}_{1,{\text{MAP}}},\hat{\mathbf{X}}_{2,{\text{MAP}}},\hat{\mathbf{D}}_{1,{\text{MAP}}},\hat{\mathbf{D}}_{2,{\text{MAP}}},\hat{\mathbf{S}}_{\text{MAP}},\hat{\mathbf{A}}_{1,{\text{MAP}}},\Delta\hat{\mathbf{A}}_{{\text{MAP}}}\right\}$ of the quantities of interest can be obtained by minimizing the negative log-posterior, leading to the following minimization problem
\begin{equation}
	\label{eq:map_coupled}
    \begin{aligned}
    &\hat{\boldsymbol{\Theta}}_{\text{MAP}} \in
    &\mathop{\rm argmin}\limits_{\boldsymbol{\Theta}}   \mathcal{J}\left(\boldsymbol{\Theta}\right)
    \end{aligned}
\end{equation}
with
\begin{equation}
\begin{aligned}
	\label{eq:objective_coupled_den}
  \mathcal{J}\left(\boldsymbol{\Theta}\right) &\triangleq  \mathcal{J}\left(\mathbf{X}_{1},\mathbf{X}_{2},\mathbf{D}_{1},\mathbf{D}_{2},\mathbf{S},\mathbf{A}_{1},\Delta\mathbf{A}\right) \\
  &=\mathcal{D}(\mathbf{Y}_{1}|\mathbf{X}_{1}) + \mathcal{D}(\mathbf{Y}_{2}|\mathbf{X}_{2}) \\
  &+\frac{\sigma_{1}^2}{2}\sum_{i=1}^{\Npatch}\left\|\mathcal{R}_{1{i}}\mathbf{X}_{1} - \mathbf{D}_{1}\mathbf{S}\mathbf{a}_{1{i}}\right\|_{\mathrm{F}}^{2} + \Psi\left(\mathbf{X}_{1}\right) \\
  &+  \frac{\sigma_{2}^2}{2}\sum_{i=1}^{\Npatch}\left\|\mathcal{R}_{2{i}}\mathbf{X}_{2} - \mathbf{D}_{2}\left(\mathbf{a}_{1{i}} + \Delta\mathbf{a}_{i}\right)\right\|_{\mathrm{F}}^{2}  + \Psi\left(\mathbf{X}_{2}\right)\\
  &+ \lambda \left\|\mathbf{A}_{1}\right\|_1 + \lambda \left\|\mathbf{A}_{1}+\Delta\mathbf{A}\right\|_1  + \gamma \left\|\Delta\mathbf{A}\right\|_{2,1}\\
 &+\iota_{\mathcal{S}}(\mathbf{D}_{1}) + \iota_{\mathcal{S}}(\mathbf{D}_{2}) + \iota_{\mathcal{C}}(\mathbf{S}).
\end{aligned}
\end{equation}
The next section describes an iterative algorithm which solves the minimization problem in \eqref{eq:map_coupled}.

\section{Minimization Algorithm}
\label{sec:min_alg}

Given the nature of the optimization problem \eqref{eq:map_coupled}, which is genuinely nonconvex and nonsmooth, the adopted minimization strategy relies on the proximal alternating linearized minimization (PALM) scheme \citep{bolte_proximal_2014}. PALM is an iterative, gradient-based algorithm which generalizes the Gauss-Seidel method. It performs iterative proximal gradient steps w.r.t. each block of variables from $\boldsymbol{\Theta}$ and ensures convergence to a local critical point $\boldsymbol{\Theta}^{*}$. It has been successfully applied in many matrix factorization cases \citep{bolte_proximal_2014,cavalcanti_unmixing_2017,thouvenin_online_2016}. Now, the goal is to generalize the single factorization to coupled factorization. The resulting CD-driven coupled dictionary learning (CDL) algorithm, whose main steps are described in the following paragraphs, is summarized in Algorithm \ref{algo:PALM_SCDL_Diff}.
\begin{algorithm}
    \DontPrintSemicolon
    \KwData{$\mathbf{Y}$}
    \KwIn{$\mathbf{A}_{1}^{(0)}$, $\Delta\mathbf{A}^{(0)}$, $\mathbf{D}_{1}^{(0)}$, $\mathbf{D}_{2}^{(0)}$, $\mathbf{S}^{(0)}$, $\mathbf{X}_{1}^{(0)}$, $\mathbf{X}_{2}^{(0)}$}
    $k \leftarrow 0$\;
    \Begin{
		\While{stopping criterion not satisfied}{
		\tcp{Code update}
    	\label{algostep:SCDLC_A} $\mathbf{A}^{(k+1)} \leftarrow \text{Update}\left(\mathbf{A}^{(k)}\right)$ \tcp*{cf. (\ref{eq:code_Update})}
      	\label{algostep:SCDLC_dA} $\Delta\mathbf{A}^{(k+1)} \leftarrow \text{Update}\left(\Delta\mathbf{A}^{(k)}\right)$ \tcp*{cf. (\ref{eq:deltacode_Update})}
        \tcp{Dictionary update}
		\label{algostep:SCDLC_Da} $\mathbf{D}_{1}^{(k+1)} \leftarrow \text{Update}\left(\mathbf{D}_{1}^{(k)}\right)$ \tcp*{cf. (\ref{eq:dictionary_Update})}
		\label{algostep:SCDLC_Db} $\mathbf{D}_{2}^{(k+1)} \leftarrow \text{Update}\left(\mathbf{D}_{2}^{(k)}\right)$ \tcp*{cf. (\ref{eq:dictionary_Update})}
    	 \tcp{Scale update}
		\label{algostep:SCDLC_Sa} $\mathbf{S}^{(k+1)} \leftarrow \text{Update}\left(\mathbf{S}^{(k)}\right)$ \tcp*{cf. (\ref{eq:scale_Update})}
        \tcp{Latent image update}
    	\label{algostep:SCDLC_Xa} $\mathbf{X}_{1}^{(k+1)} \leftarrow \text{Update}\left(\mathbf{X}_{1}^{(k)}\right)$\tcp*{cf. (\ref{eq:latent_Update})}
		\label{algostep:SCDLC_Xb} $\mathbf{X}_{2}^{(k+1)} \leftarrow \text{Update}\left(\mathbf{X}_{2}^{(k)}\right)$\tcp*{cf. (\ref{eq:latent_Update})}
		$k \leftarrow k+1$\;
    }
    $\hat{\mathbf{A}}_{1} \leftarrow \mathbf{A}_{1}^{(k+1)}$, $\Delta\hat{\mathbf{A}} \leftarrow \Delta\mathbf{A}^{(k+1)}$,\\
    $\hat{\mathbf{D}}_{1} \leftarrow \mathbf{D}_{1}^{(k+1)}$, $\hat{\mathbf{D}}_{2} \leftarrow \mathbf{D}_{2}^{(k+1)}$,
    \\ $\hat{\mathbf{S}} \leftarrow \mathbf{S}^{(k+1)}$, \\
    $\hat{\mathbf{X}}_{1}  \leftarrow \mathbf{X}_{1} ^{(k+1)}$, $\hat{\mathbf{X}}_{2}  \leftarrow \mathbf{X}_{2} ^{(k+1)}$
		}
    \KwResult{$\hat{\mathbf{A}}_{1}$, $\Delta\hat{\mathbf{A}}$, $\hat{\mathbf{D}}_{1}$, $\hat{\mathbf{D}}_{2}$, $\hat{\mathbf{S}}$, $\hat{\mathbf{X}}_{1}$, $\hat{\mathbf{X}}_{2}$}
    \caption{PALM-CDL \label{algo:PALM_SCDL_Diff}}
    \end{algorithm}

\subsection{PALM implementation}
The PALM algorithm was proposed by \citet{bolte_proximal_2014} for solving a broad class of problems involving the minimization of the sum of finite collections of possibly nonconvex and nonsmooth functions. Particularly, the target optimization function is composed by a coupling function gathering the block of variables, denoted $H(\cdot)$, and regularization functions for each block. Nonconvexity constraint is assumed for either coupling or regularization functions. One of the main advantages of the PALM algorithm over classical optimization algorithms is that each bounded sequence generated by PALM converges to a critical point. The rationale of the method can be seen as an alternating minimization approach for the proximal forward-backward algorithm \citep{combettes_signal_2005}. Some assumptions are required in order to solve this problem with all guarantees of convergence (c.f \citep[Assumption~1, Assumption~2]{bolte_proximal_2014}). The most restrictive one \citep[Assumption~2(ii)]{bolte_proximal_2014} requires that the partial gradient of the coupling function $H(\cdot)$ is globally Lipschitz continuous for each block of variable keeping the remaining ones fixed. Indeed, it is a classical assumption for proximal gradient methods which guarantees a sufficient descent property.

Therefore, given the objective function to be minimized \eqref{eq:objective_coupled_den} and considering the same structure proposed by \citet{bolte_proximal_2014} and the Lipschitz property for linear combinations of functions \citep{eriksson_lipschitz_2004}, let us define the coupling function $H(\Theta)$ as
\begin{multline}
	\label{eq:objective_coupled_term}
  H\left(\boldsymbol{\Theta}\right) \triangleq  H\left(\mathbf{X}_{1},\mathbf{X}_{2},\mathbf{D}_{1},\mathbf{D}_{2},\mathbf{S},\mathbf{A}_{1},\Delta\mathbf{A}\right)  \\
 =\Psi\left(\mathbf{X}_{1}\right)   + \Psi\left(\mathbf{X}_{2}\right) + \frac{\sigma_{1}^2}{2}\sum_{i=1}^{\Npatch}\left\|\mathcal{R}_{1{i}}\mathbf{X}_{1} - \mathbf{D}_{1}\mathbf{S}\mathbf{a}_{1{i}}\right\|_{\mathrm{F}}^{2}\\
  +\frac{\sigma_{2}^2}{2}\sum_{i=1}^{\Npatch}\left\|\mathcal{R}_{2{i}}\mathbf{X}_{2} - \mathbf{D}_{2}\left(\mathbf{a}_{1{i}} + \Delta\mathbf{a}_{i}\right)\right\|_{\mathrm{F}}^{2} +  \lambda \left\|\mathbf{A}_{1}+\Delta\mathbf{A}\right\|_1.
\end{multline}
This coupling function defined accordingly does not fulfill
\citep[Assumption~2(ii)]{bolte_proximal_2014} because some of its terms are nonsmooth, specifically the TV regularizations $\Psi(\cdot)$ and the $\ell_{1}$-norm sparsity promoting regularizations applied to $\mathbf{A}_{2}$. Thus, to ensure such a coupling function is in agreement with the required assumptions, smooth relaxations of $\Psi(\cdot)$ and $\left\|\cdot\right\|_1$ are applied by using the pseudo-Huber function  \citep{fountoulakis_second_2016,jensen_implementation_2012}.

The remaining terms of \eqref{eq:objective_coupled_den} are composed of the regularization functions associated with each variable block. Within the PALM structure, a gradient step applied to the coupling function w.r.t. a given variable block is followed by proximal step associated with the corresponding regularization functions. As a consequence, those regularization functions must be proximal-like where their proximal mappings or projections must exist and have closed-form solutions. It is important to keep in mind that, even if the convergence is guaranteed for all optimization orderings, it should not vary during iterations. Thus, the updating rules for each optimization variable in Algorithm \ref{algo:PALM_SCDL_Diff} are defined. More details about the proximal operators and projections involved in this section are given in \ref{ap:proj}.

\subsection{Optimization with respect to \texorpdfstring{$\mathbf{A}_{1}$}{A}}

Considering the single block optimization variable $\mathbf{A}_{1}$, and assuming that the remaining variables are fixed, the PALM updating step can be written
\begin{equation}
		\mathbf{A}_{1}^{(k+1)} = \mathrm{prox}^{L_{\mathbf{A}_{1}}}_{\lambda\left\|\cdot\right\|_1 + \geq0}\left(\mathbf{A}_{1}^{(k)} - \frac{1}{L_{\mathbf{A}_{1}}^{(k)} }\nabla_{\mathbf{A}_{1}} H(\boldsymbol{\Theta}) \right)
        \label{eq:code_Update}
\end{equation}
with
\begin{equation}
\begin{aligned}
		\nabla_{\mathbf{A}_{1}} H(\boldsymbol{\Theta}) &= \sigma_{1}^2\mathbf{S}^{T}\mathbf{D}_{1}^{T}\left( \mathbf{D}_{1}\mathbf{S}\mathbf{A}_{1} - \mathbf{P}_{1}\right) \\
&+\sigma_{2}^2\mathbf{D}_{2}^{T}\left( \mathbf{D}_{2}\left(\mathbf{A}_{1} + \Delta\mathbf{A}\right)-\mathbf{P}_{2}\right)  \\
&+ \lambda \frac{\left[\mathbf{A}_{1} + \Delta\mathbf{A}\right]_{i}}{\sqrt{\left[\mathbf{A}_{1} + \Delta\mathbf{A}\right]_{i}^{2} + \epsilon_{\mathbf{A}_{1}}^{2}}}
        \label{eq:nabla_code}
\end{aligned}
\end{equation}
where $[\cdot]_i/[\cdot]_i$ should be understood as an element-wise operation and $L_{\mathbf{A}_{1}}^{(k)}$ is the associated Lipschitz constant
\begin{equation}
		L_{\mathbf{A}_{1}}^{(k)} = \sigma_{1}^2\left\|\mathbf{S}^{T}\mathbf{D}_{1}^{T}\mathbf{D}_{1}\mathbf{S}\right\| +  \sigma_{2}^2\left\|\mathbf{D}_{2}^{T}\mathbf{D}_{2}\right\| + \frac{\lambda}{\epsilon_{\mathbf{A}_{1}}}.
        \label{eq:lip_code}
\end{equation}
Note that $\mathrm{prox}^{L_{\mathbf{A}_{1}}}_{\lambda\left\|\cdot\right\|_1 + \geq0}(\cdot)$ can be simply computed by considering the positive part of the soft-thresholding operator \citep{parikh_proximal_2014}.

\subsection{Optimization with respect to \texorpdfstring{$\Delta\mathbf{A}$}{DA}}

Similarly, considering the single block optimization variable $\Delta\mathbf{A}$ and consistent notations, the PALM update can be derived as
\begin{equation}
		\Delta\mathbf{A}^{(k+1)} = \mathrm{prox}^{L_{\Delta\mathbf{A}}^{(k)} }_{\left\|\cdot\right\|_{2,1}}\left(\Delta\mathbf{A}^{(k)} - \frac{1}{L_{\Delta\mathbf{A}}^{(k)} }\nabla_{\Delta\mathbf{A}} H(\boldsymbol{\Theta}) \right)
        \label{eq:deltacode_Update}
\end{equation}
where
\begin{equation}
\begin{aligned}
		\nabla_{\Delta\mathbf{A}} H(\boldsymbol{\Theta}) &= \sigma_{2}^2\mathbf{D}_{2}^{T}\left( \mathbf{D}_{2}\left(\mathbf{A}_{1} + \Delta\mathbf{A}\right)-\mathbf{P}_{2}\right) \\
&+ \lambda \frac{\left[\mathbf{A}_{1} + \Delta\mathbf{A}\right]_{i}}{\sqrt{\left[\mathbf{A}_{1} + \Delta\mathbf{A}\right]_{i}^{2} + \epsilon_{\mathbf{A}_{1}}^{2}}}
        \label{eq:nabla_deltacode}
\end{aligned}
\end{equation}
and
\begin{equation}
		L_{\Delta\mathbf{A}}^{(k)} = \sigma_{2}^2\left\|\mathbf{D}_{2}^{T}\mathbf{D}_{2}\right\| + \frac{\lambda}{\epsilon_{\mathbf{A}_{1}}}.
        \label{eq:lip_deltacode}
\end{equation}
The proximal operator $\mathrm{prox}^{L_{\Delta\mathbf{A}}^{(k)} }_{\left\|\cdot\right\|_{2,1}}(\cdot)$ can be simply computed as a group soft-thresholding operator \citep{ferraris_robust_2017}, where each group is composed by each column of $\Delta\mathbf{A}$.

\subsection{Optimization with respect to \texorpdfstring{$\mathbf{D}_{\alpha}$}{Da}}
\label{subsec:optim_dictionary}
As before, considering the single block optimization variable $\mathbf{D}_{\alpha}$ with $\alpha = \left\{1,2\right\}$, the PALM updating steps can be written as
\begin{equation}
	\label{eq:dictionary_Update}
		\mathbf{D}_{\alpha}^{(k+1)} = \mathcal{P}_{\mathcal{S}}\left(\mathbf{D}_{\alpha}^{(k)} - \frac{1}{L_{\mathbf{D}_{\alpha}}^{(k)} }\nabla_{\mathbf{D}_{\alpha}} H(\boldsymbol{\Theta}) \right)
\end{equation}
where
\begin{equation}
		\nabla_{\mathbf{D}_{\alpha}} H(\boldsymbol{\Theta}) = \sigma^2_{\alpha}\left( \mathbf{D}_{\alpha}\bar{\mathbf{A}}_{\alpha} -\mathbf{P}_{\alpha}\right)\bar{\mathbf{A}}_{\alpha}^{T}
        \label{eq:nabla_dict}
\end{equation}
and
$L_{\mathbf{D}_{\alpha}}^{(k)} $ is the Lipschitz constant
\begin{equation}
		L_{\mathbf{D}_{\alpha}}^{(k)} = \sigma_{\alpha}^{2}\left\|\bar{\mathbf{A}}_{\alpha}\bar{\mathbf{A}}_{\alpha}^{T}\right\|
        \label{eq:lip_dict}
\end{equation}
with $\bar{\mathbf{A}}_{1} = \mathbf{S}\mathbf{A}_{1}$ and $\bar{\mathbf{A}}_{2} = \mathbf{A}_{1} + \Delta\mathbf{A}$. Note that the projection $\mathcal{P}_{\mathcal{S}}(\cdot)$ can be computed as in \citet{mairal_online_2009}, keeping only the values greater than zero.

\subsection{Optimization with respect to \texorpdfstring{$\mathbf{S}$}{S}}
\label{subsec:optim_scaling}
The updating rule of the scaling matrix $\mathbf{S}$ can be written as
\begin{equation}
		\mathbf{S}^{(k+1)} = \mathcal{P}_{\mathcal{C}}\left(\mathbf{S}^{(k)} - \frac{1}{L_{\mathbf{S}^{(k)}}}\nabla_{\mathbf{S}} H(\boldsymbol{\Theta}) \right)
        \label{eq:scale_Update}
\end{equation}
where
\begin{equation}
		\nabla_{\mathbf{S}}H(\boldsymbol{\Theta}) = \sigma_{1}^2\mathbf{D}_{1}^{T}\left( \mathbf{D}_{1}\mathbf{S}\mathbf{A}_{1} - \mathbf{P}_{1}\right)\mathbf{A}_{1}^{T}
        \label{eq:nabla_scale}
\end{equation}
and $L_{\mathbf{S}}^{(k)}$ is the Lipschitz constant related to $\nabla_{\mathbf{S}}f(\boldsymbol{\Theta})$
\begin{equation}
		L_{\mathbf{S}}^{(k)} = \sigma_{1}^2\left\|\mathbf{D}_{1}^{T}\mathbf{D}_{1}\mathbf{A}_{1}\mathbf{A}_{1}^{T}\right\|.
        \label{eq:lip_scale}
\end{equation}
The projection $\mathcal{P}_{\mathcal{C}}(\cdot)$ constrains all diagonal elements of $\mathbf{S}$ to be nonzero.

\subsection{Optimization with respect to \texorpdfstring{$\mathbf{X}_{\alpha}$}{X}}
\label{subsec:optim_latent}

Finally, the updates of the latent images $\mathbf{X}_{\alpha}$ ($\alpha \in \left\{1,2\right\}$) are achieved as follows
\begin{equation}
 		\mathbf{X}_{\alpha}^{(k+1)} = \mathrm{prox}^{L_{\mathbf{X}_{\mathrm{\alpha}}}^{(k)}}_{\mathcal{D}_{\mathrm{\alpha}}(\mathbf{Y}_{\mathrm{\alpha}}|\cdot)}\left(\mathbf{X}_{\alpha}^{(k)} - \frac{1}{L_{\mathbf{X}_{\alpha}}^{(k)} }\nabla_{\mathbf{X}_{\alpha}} H(\boldsymbol{\Theta}) \right)
       	\label{eq:latent_Update}\\
\end{equation}
with
\begin{equation}
\begin{aligned}
\nabla_{\mathbf{X}_{\alpha}}H(\boldsymbol{\Theta}) &= \sigma_{\alpha}^2\sum_{i=1}^{\Npatch}\mathcal{R}_{\alpha{i}}^{T}\left(\mathcal{R}_{\alpha{i}}\mathbf{X}_{\alpha} - \mathbf{D}_{\alpha}\bar{\mathbf{a}}_{\alpha{i}}\right) \\
       &- \tau_{\alpha}\mathrm{div}\left( \frac{\left[\nabla\mathbf{X}_{1}\right]_{i}}{\sqrt{\left[\nabla\mathbf{X}_{\alpha}\right]_{i}^{2} + \epsilon_{\mathbf{X}_{\alpha}}^{2}}}\right)
        \label{eq:nabla_latent}
        \end{aligned}
\end{equation}
and
\begin{equation}
		L_{\mathbf{X}_{\mathrm{\alpha}}}^{(k)} = \sigma_{\mathrm{\alpha}}^{2}\left\|\sum_{i=1}^{\Npatch}\mathcal{R}_{\alpha i}^{T}\mathcal{R}_{\alpha i}\right\|+ \frac{8\tau_{\mathrm{\alpha}}}{\epsilon_{\mathbf{X}_{\mathrm{\alpha}}}}
        \label{eq:lip_latent}
\end{equation}
and where $\mathrm{div}(\cdot)$  stands for the discrete divergence \citep{chambolle_algorithm_2004}. Note that, $\mathrm{prox}^{L_{\mathbf{X}_{\mathrm{\alpha}}}^{(k)}}_{\mathcal{D}_{\mathrm{\alpha}}(\mathbf{Y}_{\mathrm{\alpha}}|\cdot)}$ represents the proximal mapping for the divergence measure associated with the likelihood function characterizing the modality of the observed image $\mathbf{Y}_{\alpha}$. For the most common remote sensing modalities, e.g., optical and radar, these divergences are well documented and \ref{ap:dft} presents the corresponding proximal operators.

\section{Performance analysis}\label{sec:experiments}

Real datasets with corresponding ground truth are too scarce to statistically assess the performance of CD algorithms.  Indeed, this assessment would require a huge number of pairs of images acquired at two different dates, geometrically co-registered and presenting changes. These pairs should also be accompanied by a ground truth (i.e., a binary CD mask locating the actual changes) to allow quantitative figures-of-merit to be computed. As a consequence, in a first step, we illustrate the algorithm and state-of-the-art method outcomes over such rare examples corresponding to real images, real changes and associated ground truth (section \ref{subsec:real_images_wgt}). This first set of experiments is conducted on images of same spatial resolutions. Thus we also exhibit another set of examples that involves images with different resolutions, but alas without ground truth. For this second example, the accuracy of the proposed method cannot be quantified, but can be evaluated through visual inspection (section \ref{subsec:real_images_wogt}). In a second step, to conduct a complete performance analysis, the algorithm and comparable methods will be tested on image pairs that are representative of possible scenarios considered in this paper, i.e., coming from multimodal images. This test set is composed of real images, however simulated changes and associated ground truth (section \ref{subsec:synthetic_images}). 

\subsection{Compared methods} \label{subsec:compared}
	As the number of unsupervised multimodal CD methods is rather reduced, the proposed technique has been compared to the unsupervised fuzzy-based (F) method proposed by \citet{gong_coupled_2016}, that is able to deal with multimodal images, to the robust fusion (RF) method proposed by \citet{ferraris_robust_2017} which deals exclusively with multi-band optical images and with unsupervised segmentation-based (S) technique proposed by \citet{Huang2015}. The fuzzy-based method by \citet{gong_coupled_2016} relies on a coupled dictionary learning methodology using a modified K-SVD \citep{aharon_k-svd_2006} with an iterative patch selection procedure to provide only unchanged patches for the coupled dictionary training phase. Then, the sparse code for each observed image is estimated separately from each other allowing to compute the cross-image reconstruction errors. Finally, a local fuzzy C-Means is applied to the mean of the cross-image reconstruction errors in order to separate change and unchanged classes. Equivalently to the proposed one, this method makes no assumption about the joint observation model. On the other hand, the robust fusion method by \citet{ferraris_robust_2017} is based on a more constrained joint observation model, considering that the two latent images share the same resolutions and differ only in changed pixels. Finally, the method proposed by \citet{Huang2015} replaces the pixel-based approach used on all previous methods to a feature-based approach. In this approach, features are derived from the segmentation of each image. Then, a difference map is generated based on metrics computed for the matching features extracted on previous steps at different scales. This strategy is used in order to provide finer details. At the end, the change map is generated using the histogram of difference map. The final change maps estimated by these algorithms are denoted as $\hat{\mathbf{m}}_{\mathrm{F}}$, $\hat{\mathbf{m}}_{\mathrm{RF}}$ and $\hat{\mathbf{m}}_{\mathrm{S}}$, respectively, while the proposed PALM-CDL method provides a change map denoted $\hat{\mathbf{m}}_{\mathrm{CDL}}$.
	
	\subsection{Figures-of-merit}	
\label{subsec:figures_of_merit}
The CD performance of the different methods has been assessed through the empirical receiver operating characteristics (ROC) curves, representing the estimated pixel-wise probability of detection ($\mathrm{PD}$) as a function of the probability of false alarm ($\mathrm{PFA}$). Moreover, two quantitative criteria derived from these ROC curves have been computed, namely, i) the area under the curve (AUC), corresponding to the integral of the ROC curve and ii) the distance between the no detection point $(\mathrm{PFA} = 1, \mathrm{PD} = 0)$ and the point at the interception of the ROC curve with the diagonal line defined by $\mathrm{PFA} = 1 - \mathrm{PD}$. For both metrics, the greater the criterion, the better the detection.

\newcommand{\subfwidth}{0.49\columnwidth}
\newcommand{\figsize}{0.85\columnwidth}
\newcommand{\subfigwidthROC}{0.32\columnwidth}
\newcommand{\figwidthROC}{0.85\textwidth}
\newcommand{\one}[1]{\bf{\textcolor[rgb]{0.00,0.00,1.00}{#1}}}
\newcommand{\two}[1]{\textcolor[rgb]{0.00,0.00,1.00}{#1}}
\setlength{\tabcolsep}{5pt}
\renewcommand{\arraystretch}{1.3}

\subsection{Illustration through real images with real changes}\label{subsec:real_images}
\label{subsec:reference_images}
In a first step, experiments are conducted on real images with real changes to emphasize the reliability of the proposed CD method and to illustrate the performance of the proposed algorithmic framework. Three distinct scenarios involving $3$ pairs of images of different modalities and resolutions, are considered namely,

\begin{itemize}
  \item Scenario 1 considers two optical images: the acquisition process is very similar for the two images and the image formation processes is characterized by an additive Gaussian noise corruption for both sensors.
  \item Scenario 2 considers two SAR images: the image formation process is not the same as for optical images, in particular differing on the noise model, i.e., multiplicative Gamma noise instead of additive Gaussian model. 
  \item Scenario 3 considers a SAR image and an optical image: for this more challenging situation, there is no similarity between the noise corruption models for the two sensors. 
\end{itemize}

To summarize, Scenarios 1 and 2 are dedicated to a pair of images with the same modality, but with a variation on the properties of images between scenarios, e.g., the noise statistics. Note that the proposed CDL algorithm has not been designed to specifically handle these conventional scenarios. However, they are still considered to evaluate the performance of the proposed method, in particular w.r.t. the methods specifically designed to address scenario 1 or 2. Conversely, Scenario 3 handles images of different modalities. All considered images have been manually geographically aligned to fulfil the requirements imposed by the considered CD setup. 

\subsubsection{Case of same resolutions and ground truth}\label{subsec:real_images_wgt}

For this first set of experiments, images of same spatial resolutions are considered. They are accompanied by a ground truth in the form of an actual change map $\mathbf{m}$ to be estimated.\\

\begin{figure}
\centering
			\begin{subfigure}{\subfwidth}
					\centering	
					\includegraphics[width=\figsize]{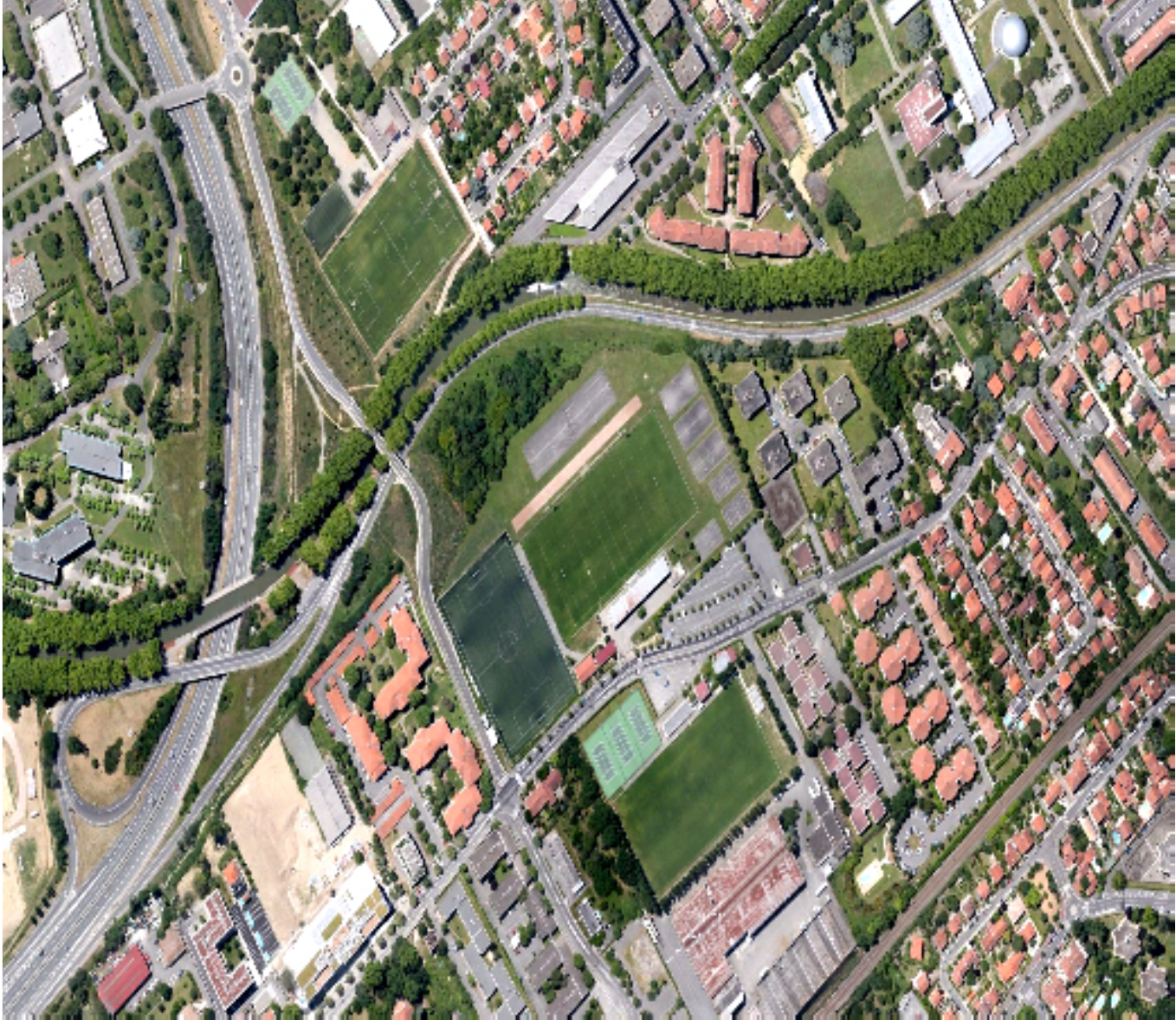}
					\caption{$\mathbf{Y}_{t_1}$}
					\label{fig:Yt1_1}
			\end{subfigure}
			\begin{subfigure}{\subfwidth}
					\centering	
					\includegraphics[width=\figsize]{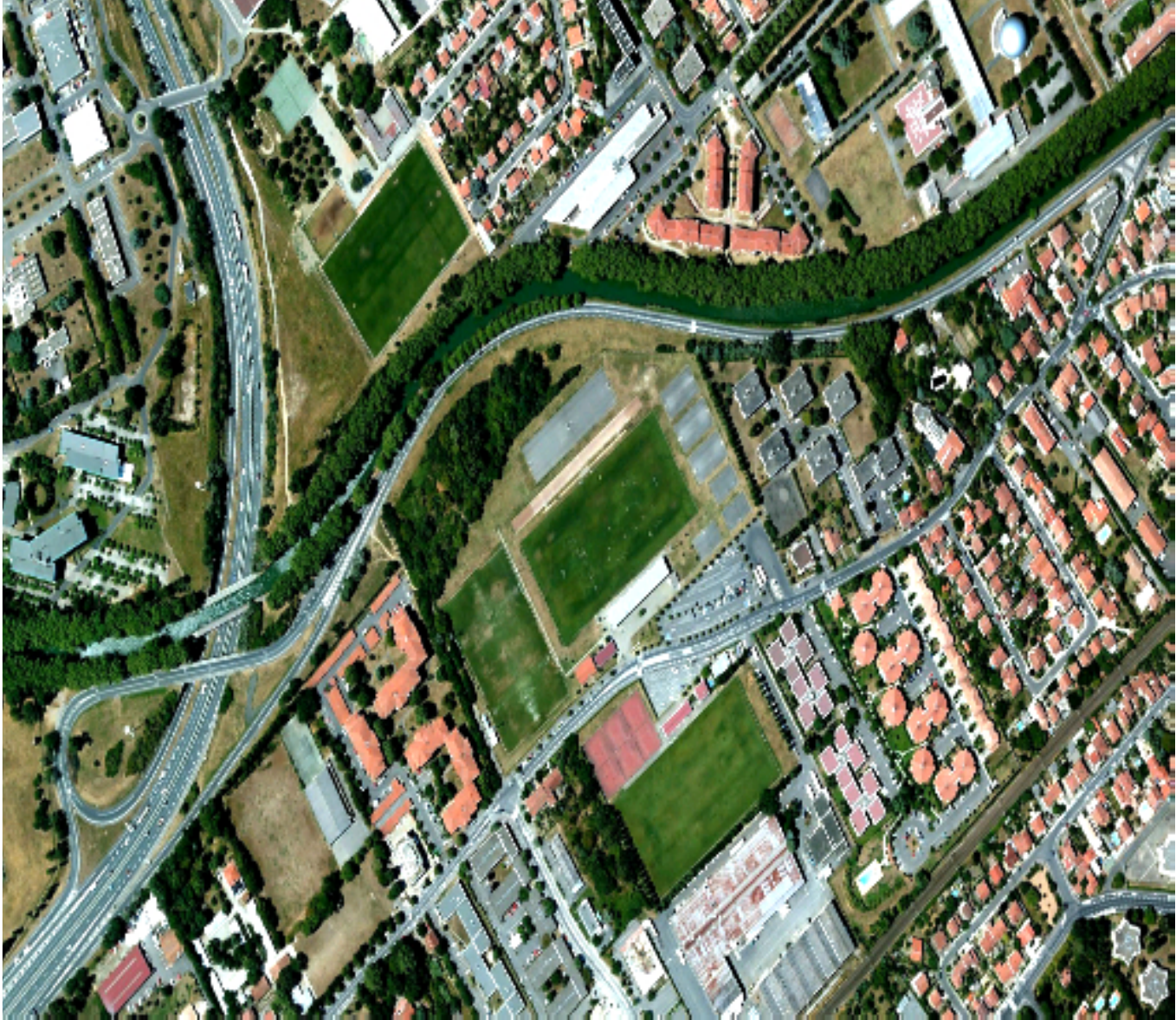}
					\caption{$\mathbf{Y}_{t_2}$}
					\label{fig:Yt2_1}
			\end{subfigure}
			\begin{subfigure}{\subfwidth}
					\centering
					\includegraphics[width=\figsize]{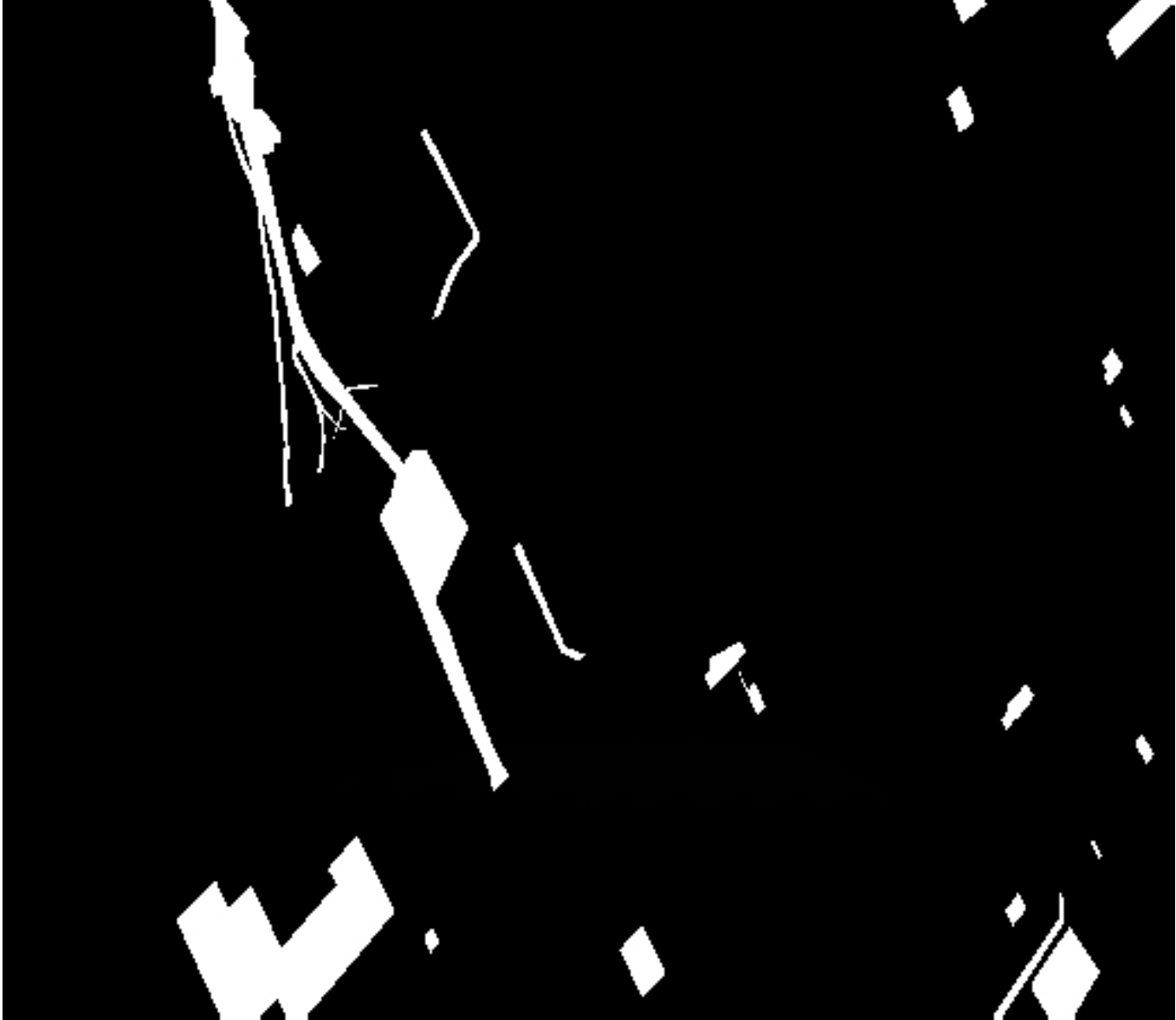}
					\caption{$\mathbf{m}$}
					\label{fig:mask_1}
			\end{subfigure}
			\begin{subfigure}{\subfwidth}
					\centering
					\includegraphics[width=\figsize]{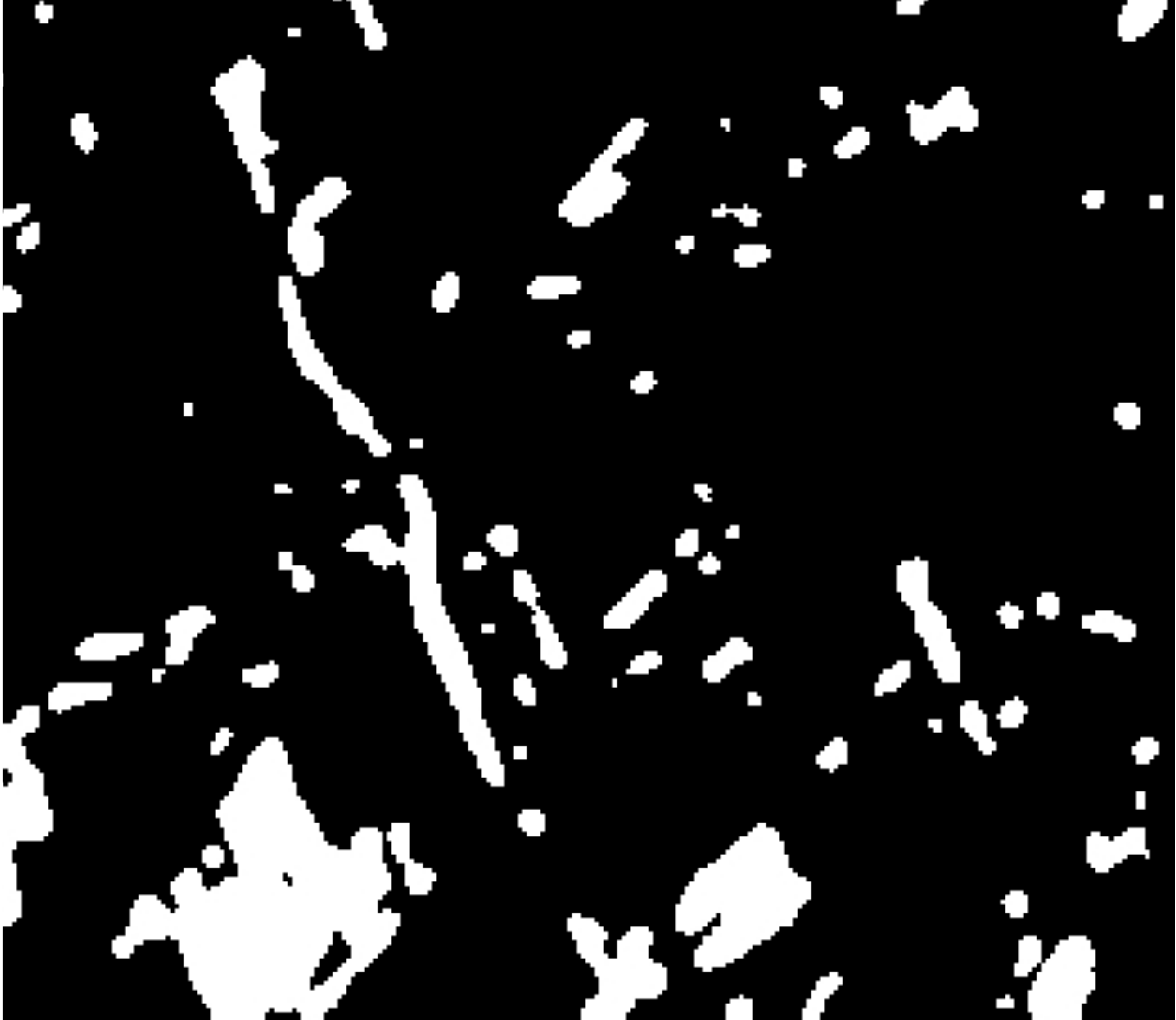}
					\caption{$\hat{\mathbf{m}}_{\mathrm{F}}$}
					\label{fig:FMAP_1}
			\end{subfigure}
			\begin{subfigure}{\subfwidth}
					\centering
					\includegraphics[width=\figsize]{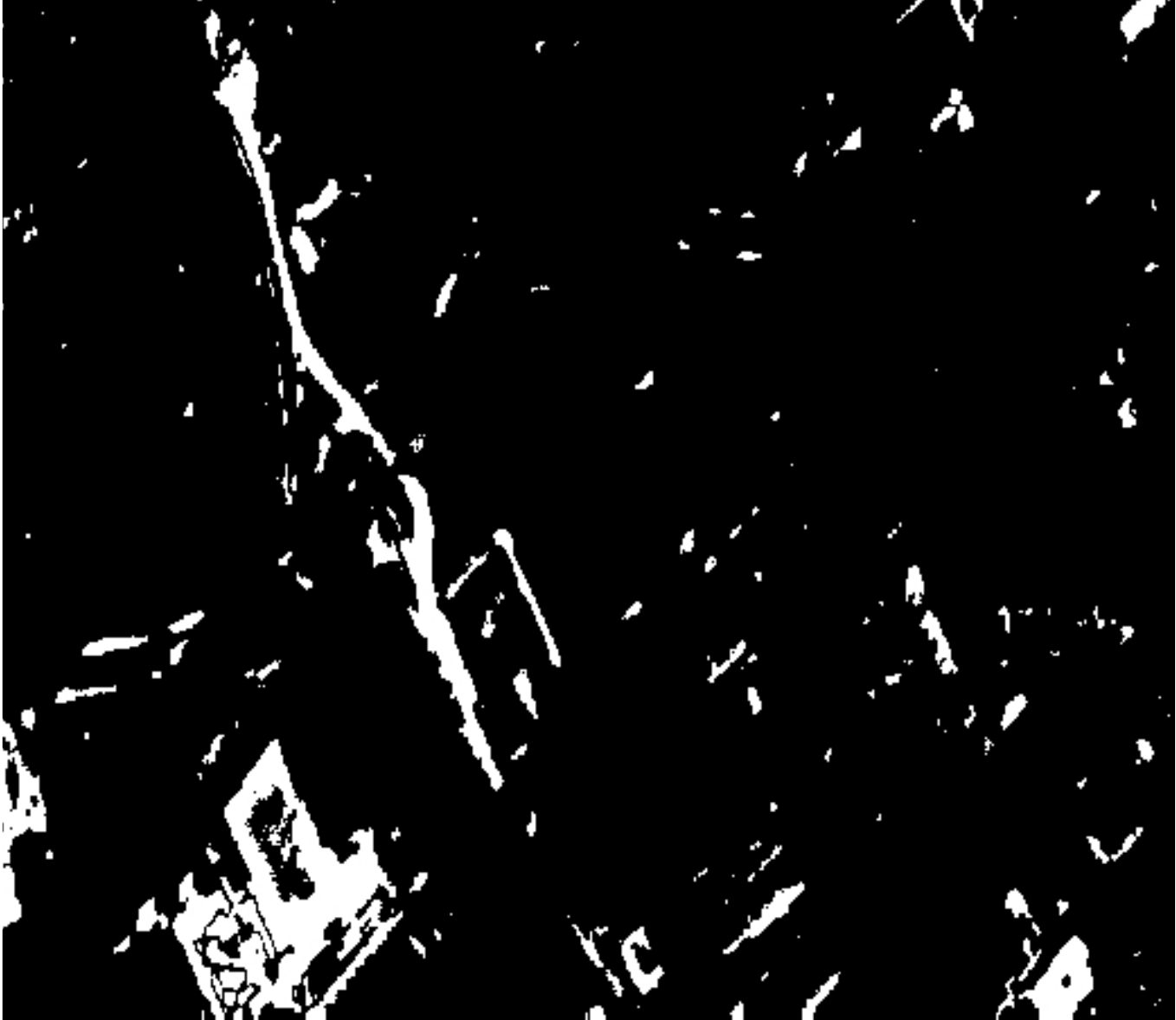}
					\caption{$\hat{\mathbf{m}}_{\mathrm{RF}}$}
					\label{fig:RFMAP_1}
			\end{subfigure}
						\begin{subfigure}{\subfwidth}
					\centering
					\includegraphics[width=\figsize]{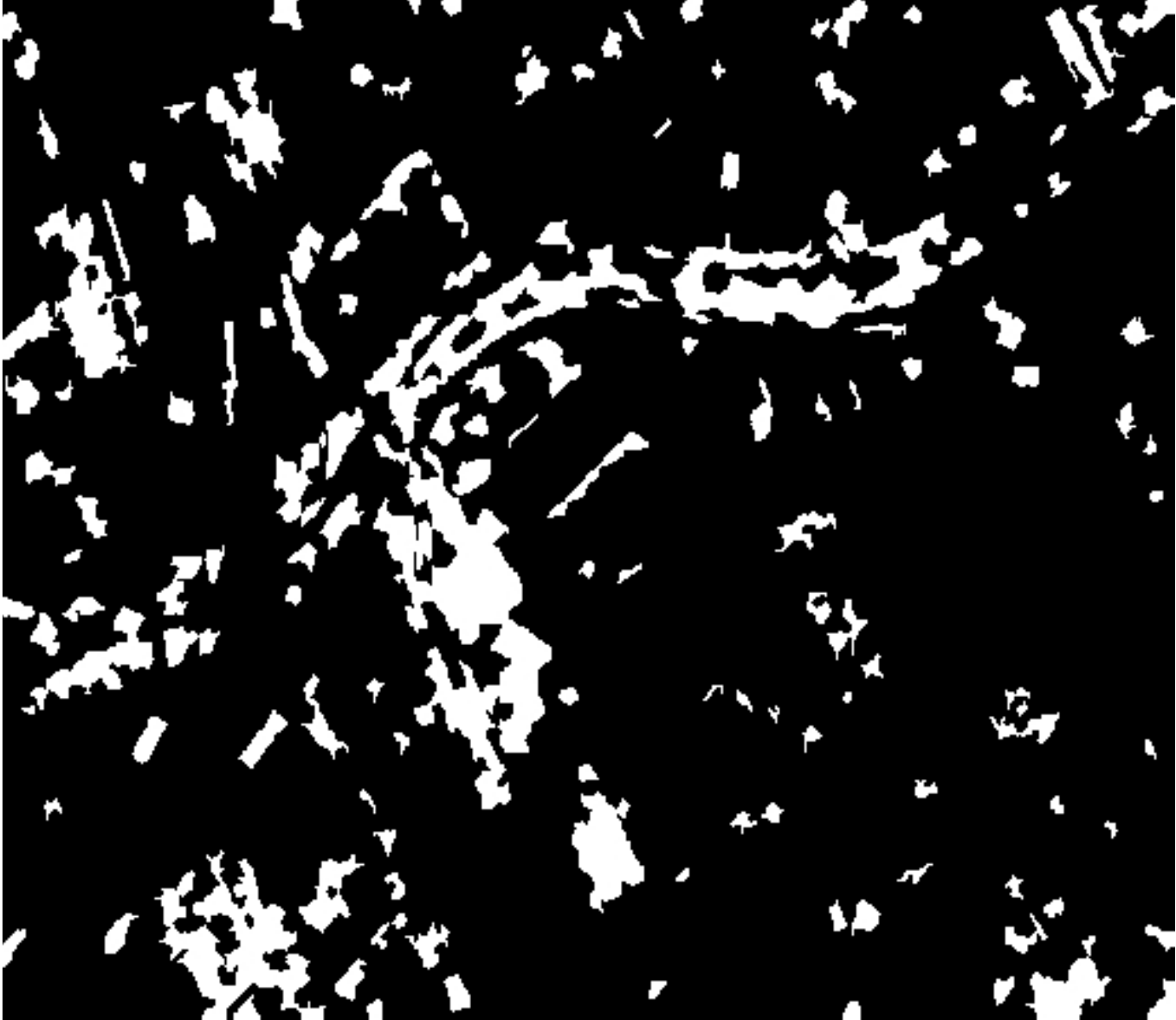}
					\caption{$\hat{\mathbf{m}}_{\mathrm{S}}$}
					\label{fig:SMAP_1}
			\end{subfigure}
			\begin{subfigure}{\subfwidth}
					\centering
					\includegraphics[width=\figsize]{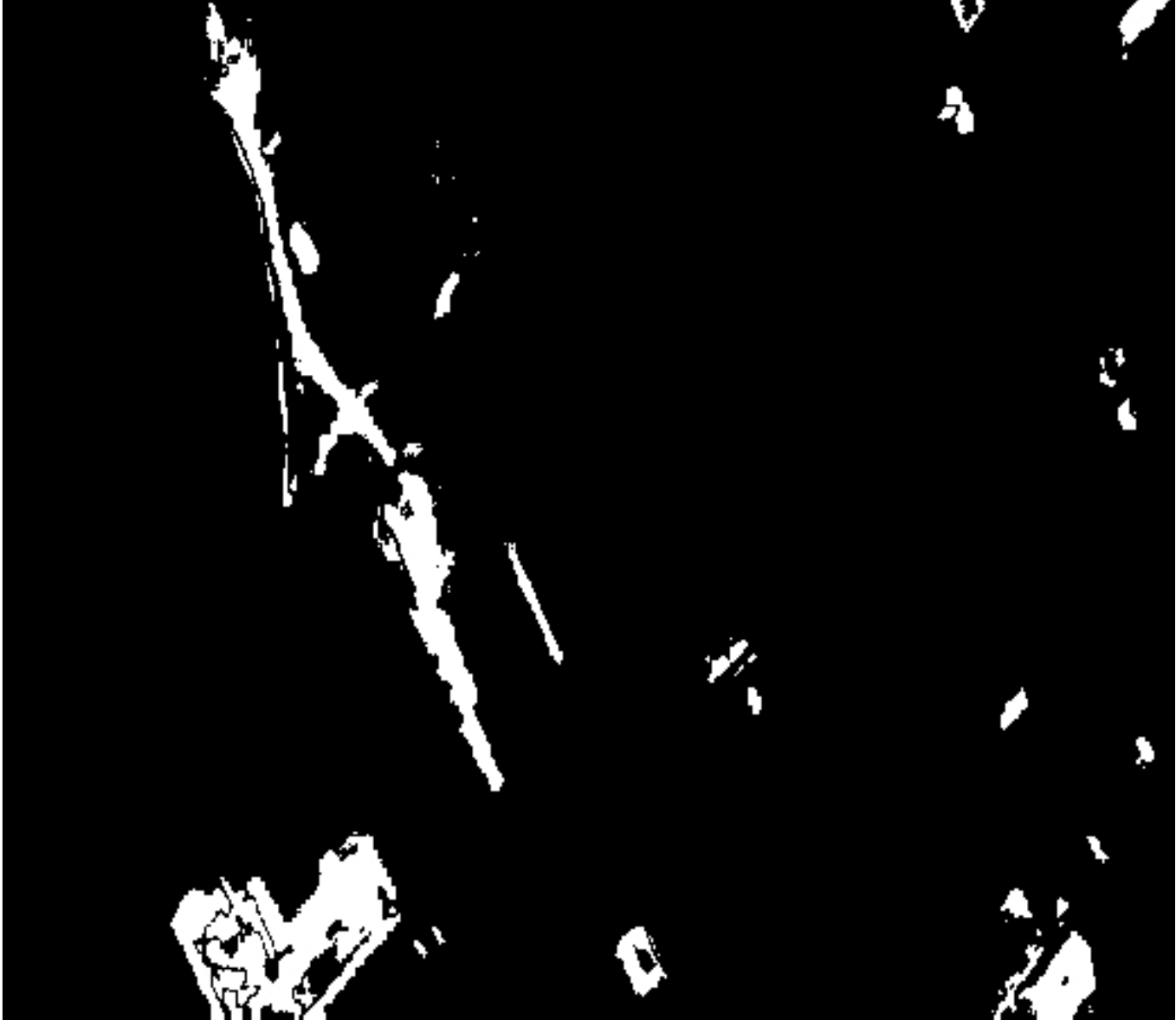}
					\caption{$\hat{\mathbf{m}}_{\mathrm{CDL}}$}
					\label{fig:CDLMAP_1}
			\end{subfigure}

\caption{Real images affected by real changes  with ground truth, Scenario 1: \protect\subref{fig:Yt1_1} observed MS optical image $\mathbf{Y}_{t_1}$ from the south of Toulouse acquired before the construction of a new road , \protect\subref{fig:Yt2_1}  observed MS optical image $\mathbf{Y}_{t_2}$  acquired after its construction, \protect\subref{fig:mask_1} groud-truth mask $\mathbf{m}$ indicating changed areas constructed by photointerpretation,\protect\subref{fig:FMAP_1} change map $\hat{\mathbf{m}}_{\mathrm{F}}$ of the fuzzy method, \protect\subref{fig:RFMAP_1} change map $\hat{\mathbf{m}}_{\mathrm{RF}}$ of the robust fusion method, \protect\subref{fig:SMAP_1} change map $\hat{\mathbf{m}}_{\mathrm{S}}$ of the segmentation-based method, and \protect\subref{fig:CDLMAP_1} change map $\hat{\mathbf{m}}_{\mathrm{CDL}}$ of proposed method.}%
	\label{fig:real_1}%
\end{figure}

\noindent \textbf{Scenario 1: optical vs. optical --} The observed images are two multispectral (MS) optical images with 3 channels representing an urban region in the south of Toulouse, France, before (Figure \ref{fig:Yt1_1}) and after (Figure \ref{fig:Yt2_1}) the construction of a road. These $960 \times 1560$-pixel images are both characterized by a $50$cm spatial resolution. The ground-truth change mask $\mathbf{m}$ is represented in Figure \ref{fig:mask_1}. Figure \ref{fig:real_1} depicts the observed images at each date, the ground-truth change mask and the change maps estimated by the four compared methods. 

\begin{figure*}[t]
    	\centering
        	\begin{subfigure}{\subfigwidthROC}
					\centering	
					\includegraphics[width=\figwidthROC]{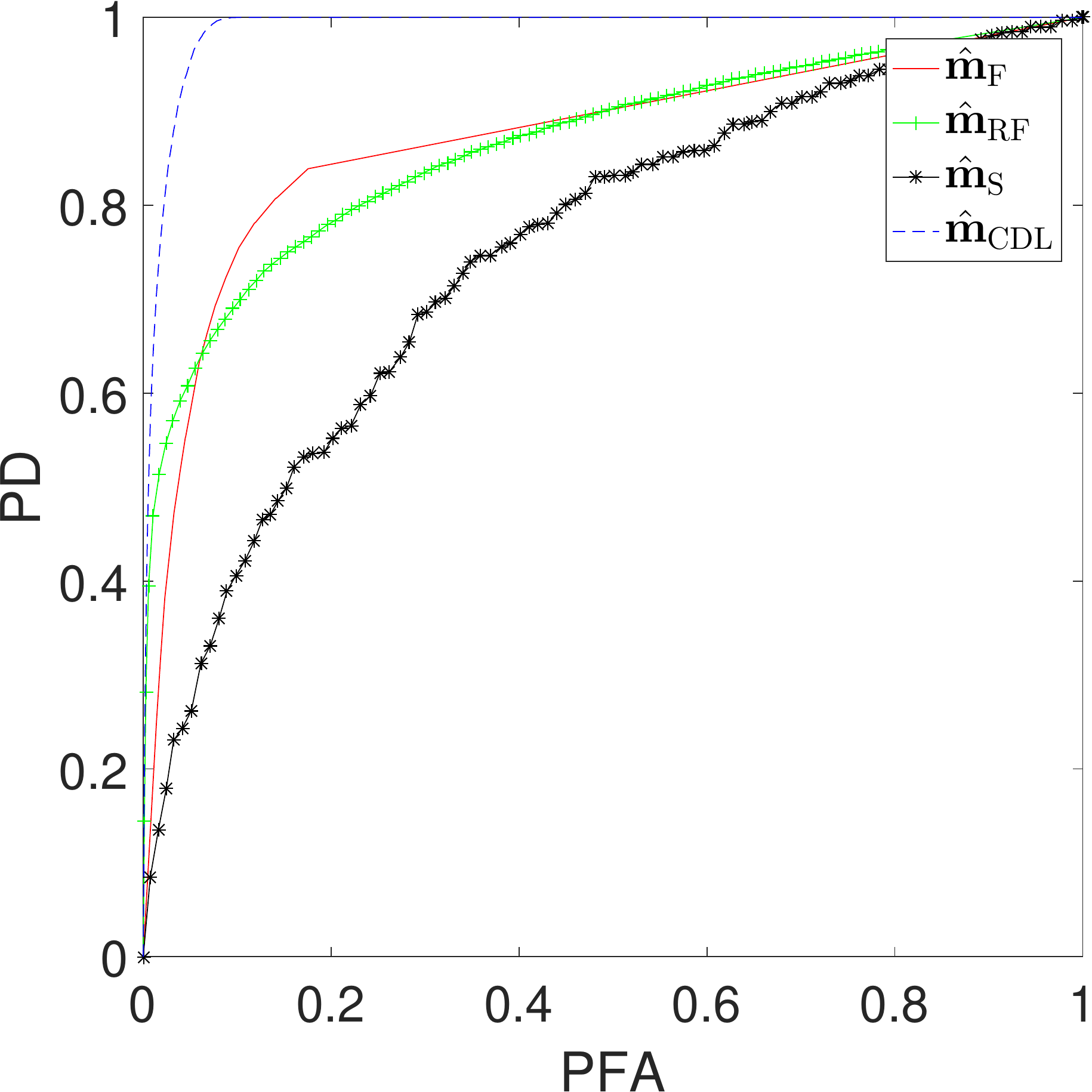}
					\caption{}
					\label{fig:rocOPTOPT}
			\end{subfigure}
        	\begin{subfigure}{\subfigwidthROC}
					\centering	
					\includegraphics[width=\figwidthROC]{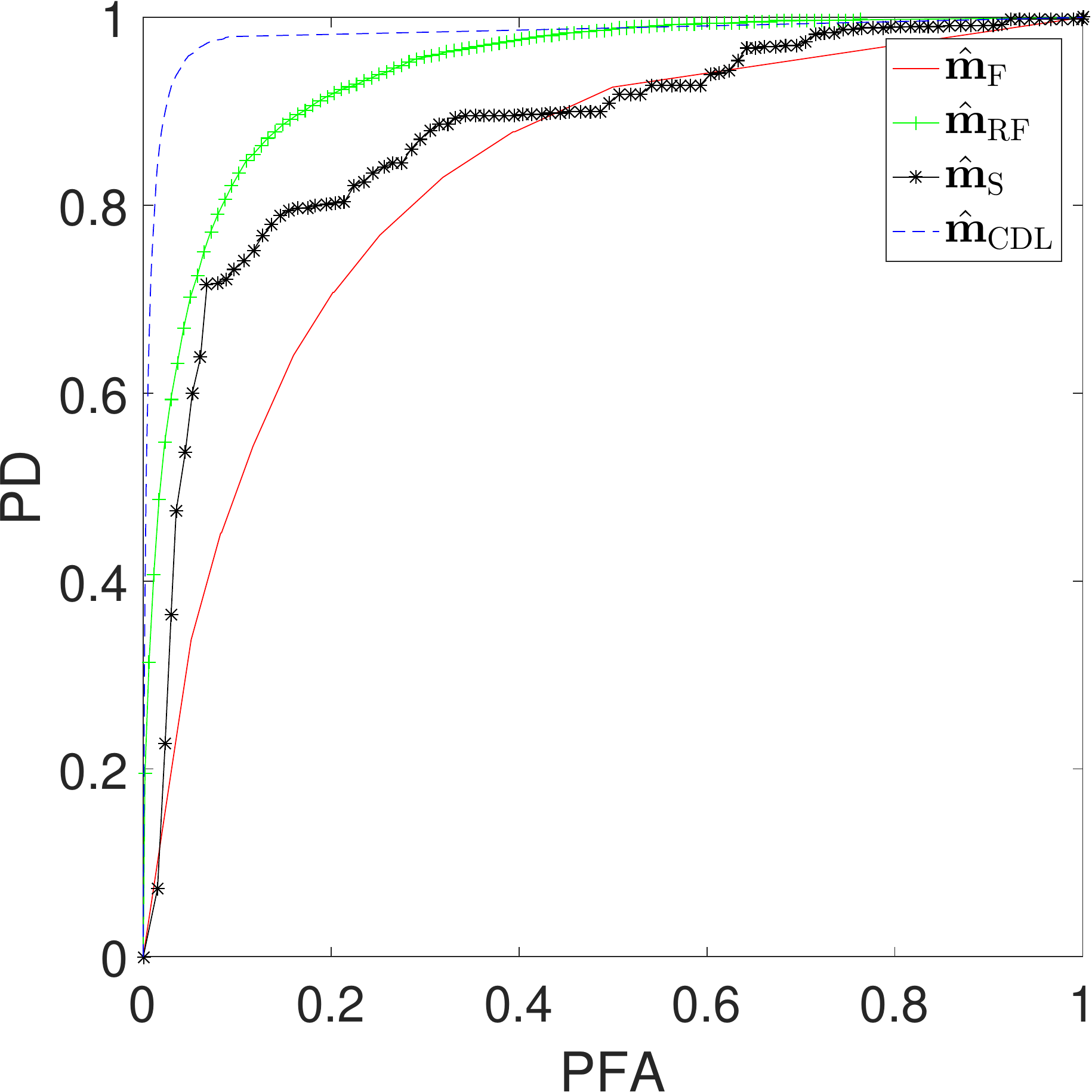}
					\caption{}
					\label{fig:rocSARSAR}
			\end{subfigure}
        	\begin{subfigure}{\subfigwidthROC}
					\centering	
					\includegraphics[width=\figwidthROC]{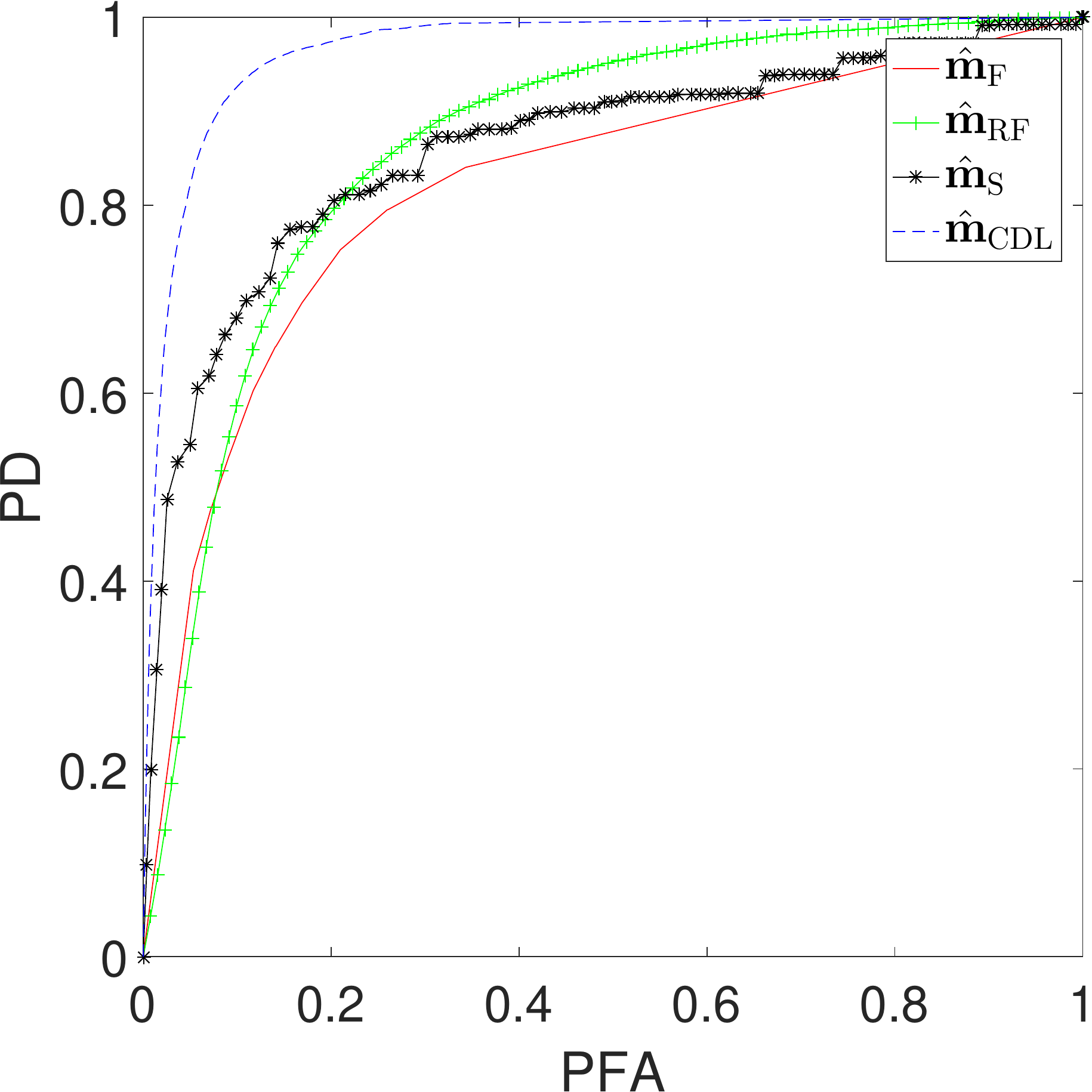}
					\caption{}
					\label{fig:rocSAROPT}
			\end{subfigure}
			\caption{Real images affected by real changes  with ground truth: ROC curves for  \protect\subref{fig:rocOPTOPT} Scenario 1,  \protect\subref{fig:rocSARSAR} Scenario 2, \protect\subref{fig:rocSAROPT} Scenario 3.}%
            \label{fig:ROC_real}%
\end{figure*}

The quantitative results for Scenario 1 are reported in Table \ref{table:ROCOPTOPT} (lines 1 and 2) and the corresponding ROC curves in Figure \ref{fig:rocOPTOPT}. The analysis of these results shows that the proposed method outperforms state-of-the-art methods for this scenario which involves common changes in urban areas and in MS optical images. Note that, besides the changes, this kind of situation involves a lot of small differences between the two observed images due to the variations in sun the illumination, in the vegetation cover, etc. These effects sometimes are classified as changes, increasing the false alarm rate, especially for state-of-the-art methods. Besides, the proposed method still provides the best detection for this dataset.\\ 

\begin{table}[h]
    \caption{Real images affected by real changes  with ground truth  for Scenarios 1--3: quantitative detection performance (AUC  and distance).}
    \centering
    \begin{tabular}{|c|c|c|c|c|c|}
    \cline{3-6}
    \multicolumn{2}{c|}{} & $\hat{\mathbf{m}}_{\mathrm{F}}$ & $\hat{\mathbf{m}}_{\mathrm{RF}}$ & $\hat{\mathbf{m}}_{\mathrm{S}} $ & $\hat{\mathbf{m}}_{\mathrm{CDL}}$  \\
    \hline
	\hline
	\multirow{2}{*}{\rotatebox{00}{Sc. 1}}     &AUC   & $\two{0.870426}$ & $0.866061$ & $0.750601$ & $\one{0.987379}$\\
                                               &Dist. & $\two{0.831983}$ & $0.788579$ & $0.692469$ & $\one{0.950695}$\\
\hline
\multirow{2}{*}{\rotatebox{00}{Sc. 2}}            &AUC   & $0.823414$ & $\two{0.93982}$ & $0.874743$ & $\one{0.981355}$\\
                                               &Dist. & $0.757076$ & $\two{0.869387}$ & $0.80188$ & $\one{0.954995}$\\
\hline
\multirow{2}{*}{\rotatebox{00}{Sc. 3}}         &AUC   & $0.818246$ & $\two{0.862729}$ & $0.862025$ & $\one{0.966283}$\\
                                               &Dist. & $0.769877$ & $0.79658$ & $\two{0.80078}$ & $\one{0.912191}$\\
\hline  
    \end{tabular}
  \label{table:ROCOPTOPT}
\end{table}	
	
\begin{figure}
\centering
			\begin{subfigure}{\subfwidth}
					\centering	
					\includegraphics[width=\figsize]{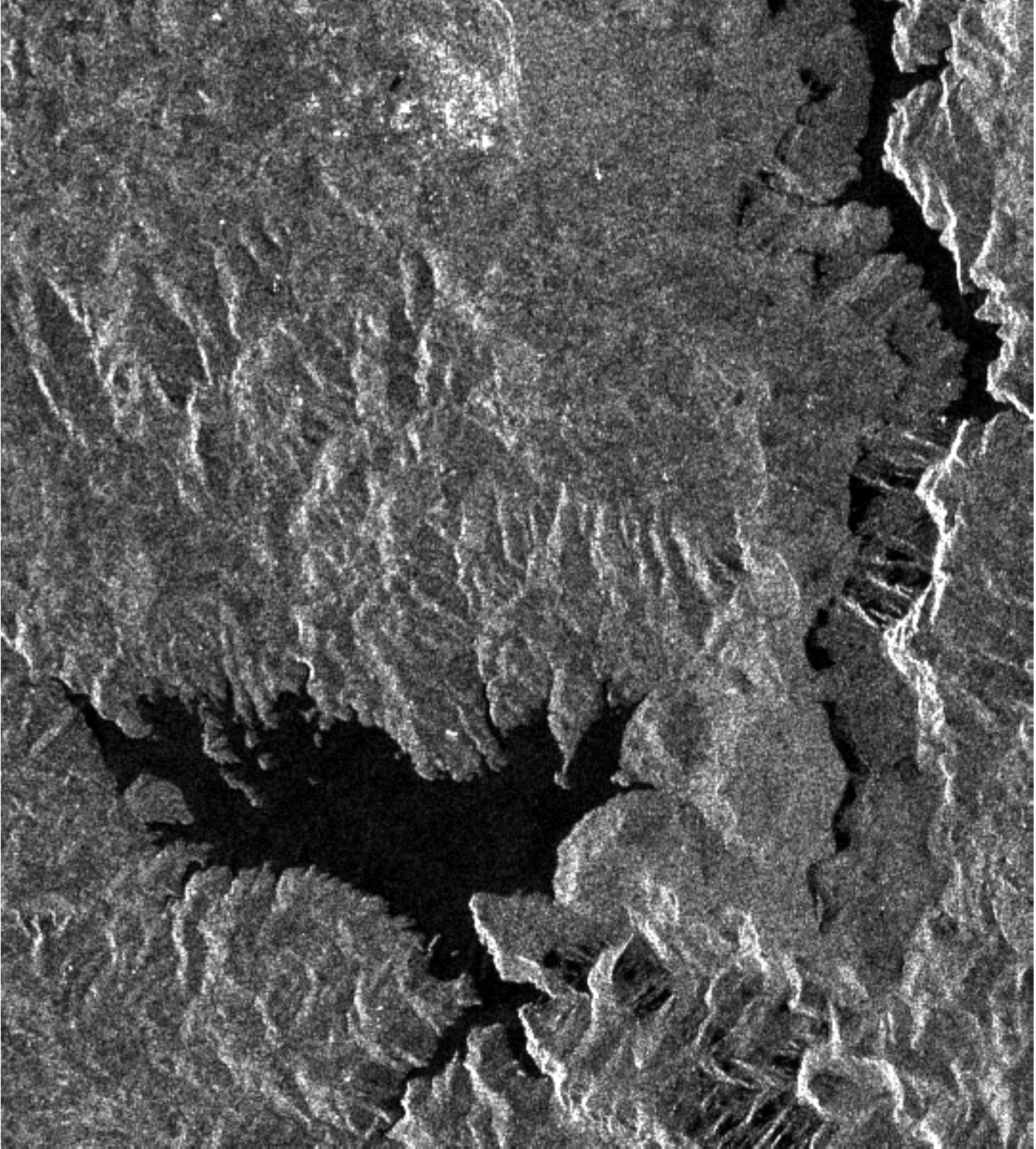}
					\caption{$\mathbf{Y}_{t_1}$}
					\label{fig:Yt1_2}
			\end{subfigure}
			\begin{subfigure}{\subfwidth}
					\centering	
					\includegraphics[width=\figsize]{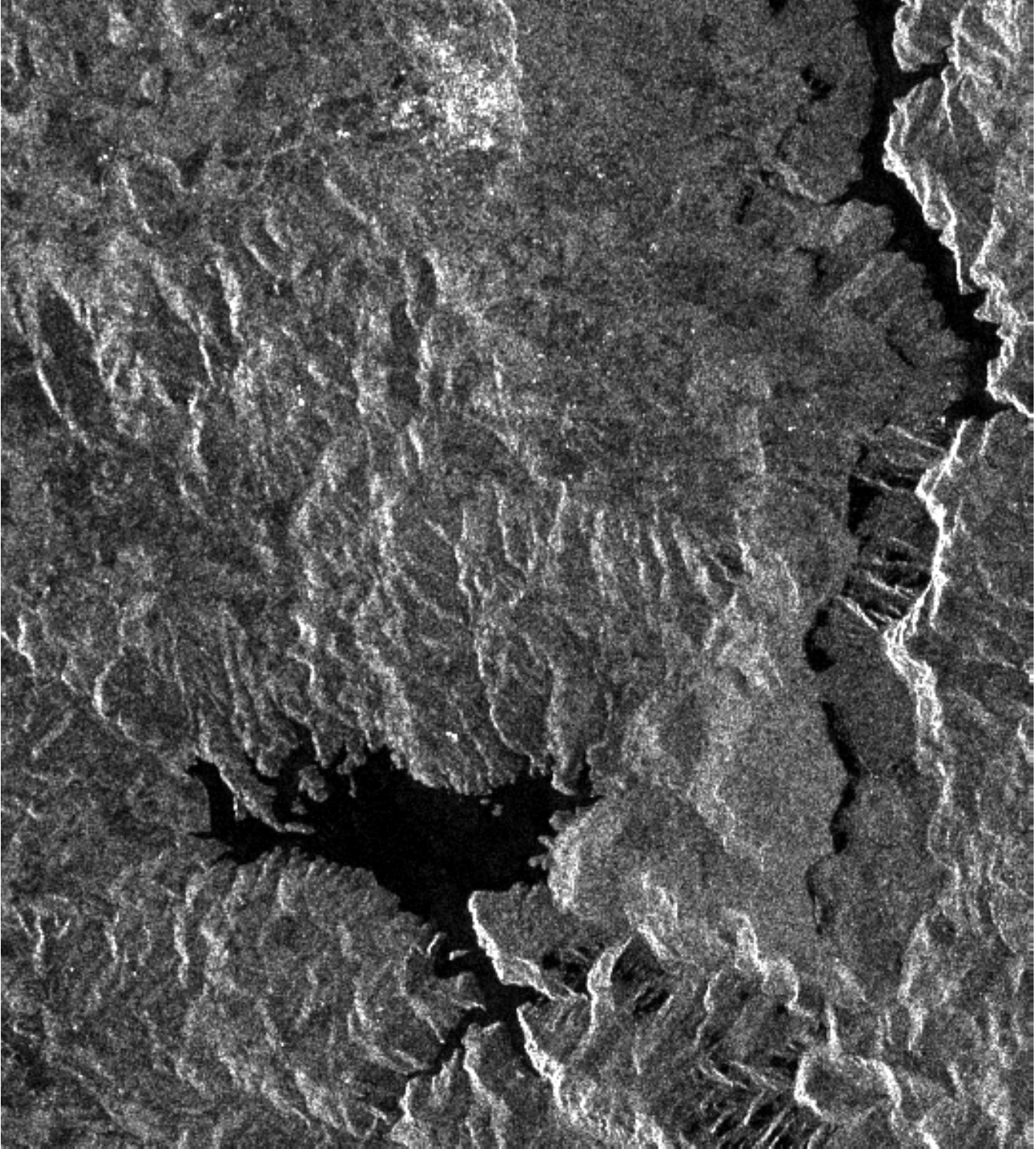}
					\caption{$\mathbf{Y}_{t_2}$}
					\label{fig:Yt2_2}
			\end{subfigure}
			\begin{subfigure}{\subfwidth}
					\centering
					\includegraphics[width=\figsize]{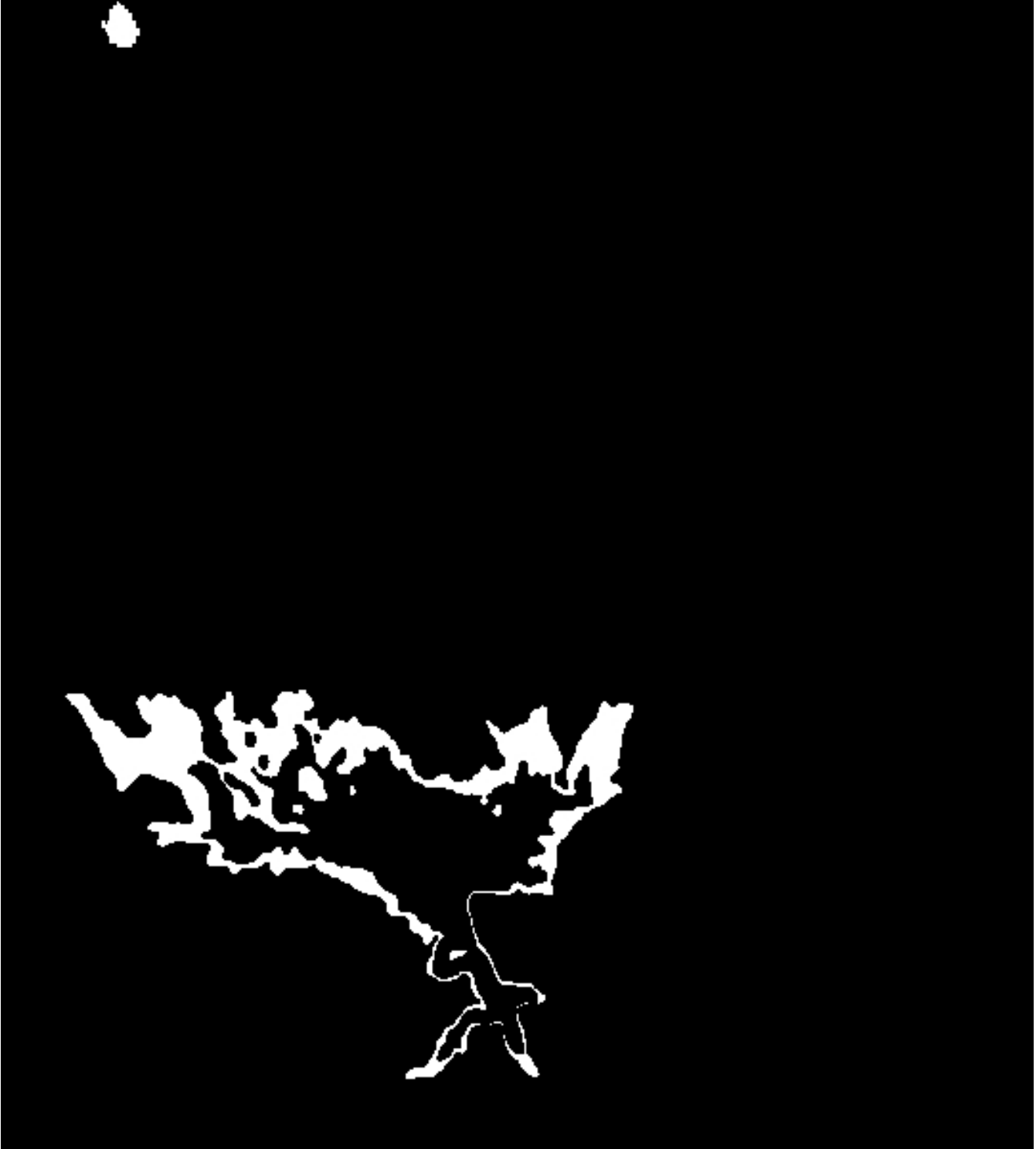}
					\caption{$\mathbf{m}$}
					\label{fig:mask_2}
			\end{subfigure}
			\begin{subfigure}{\subfwidth}
					\centering
					\includegraphics[width=\figsize]{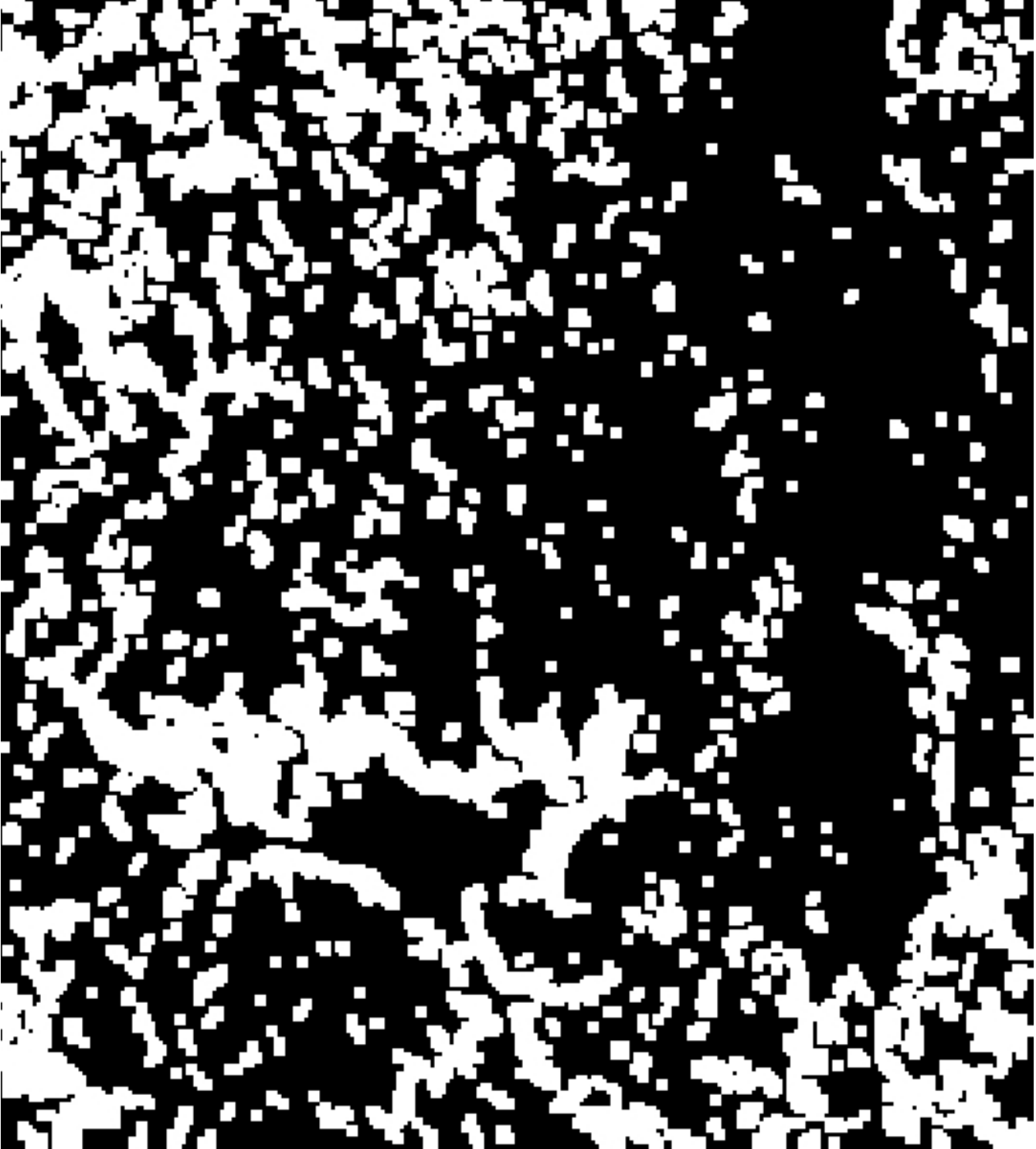}
					\caption{$\hat{\mathbf{m}}_{\mathrm{F}}$}
					\label{fig:FMAP_2}
			\end{subfigure}
			\begin{subfigure}{\subfwidth}
					\centering
					\includegraphics[width=\figsize]{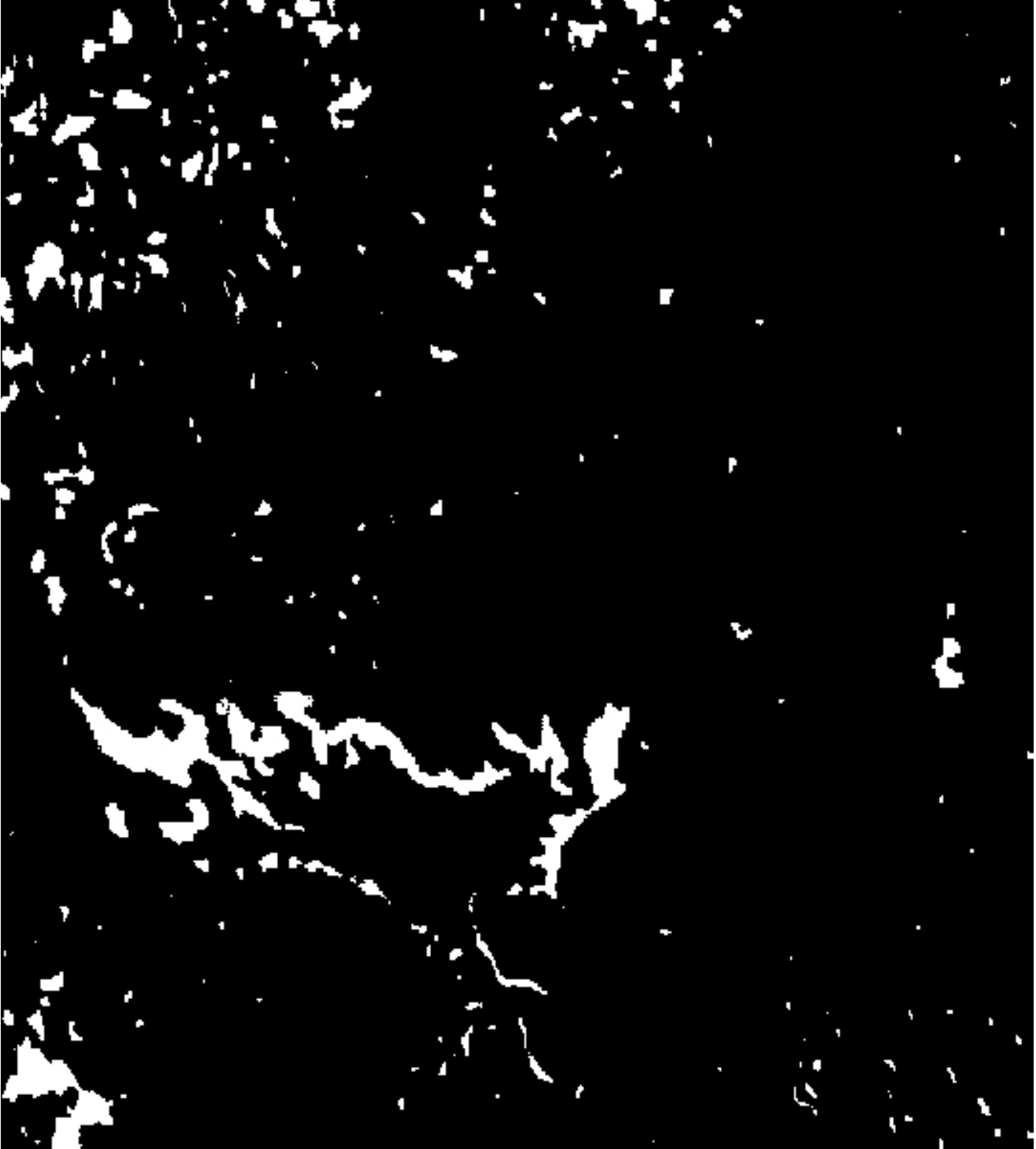}
					\caption{$\hat{\mathbf{m}}_{\mathrm{RF}}$}
					\label{fig:RFMAP_2}
			\end{subfigure}
						\begin{subfigure}{\subfwidth}
					\centering
					\includegraphics[width=\figsize]{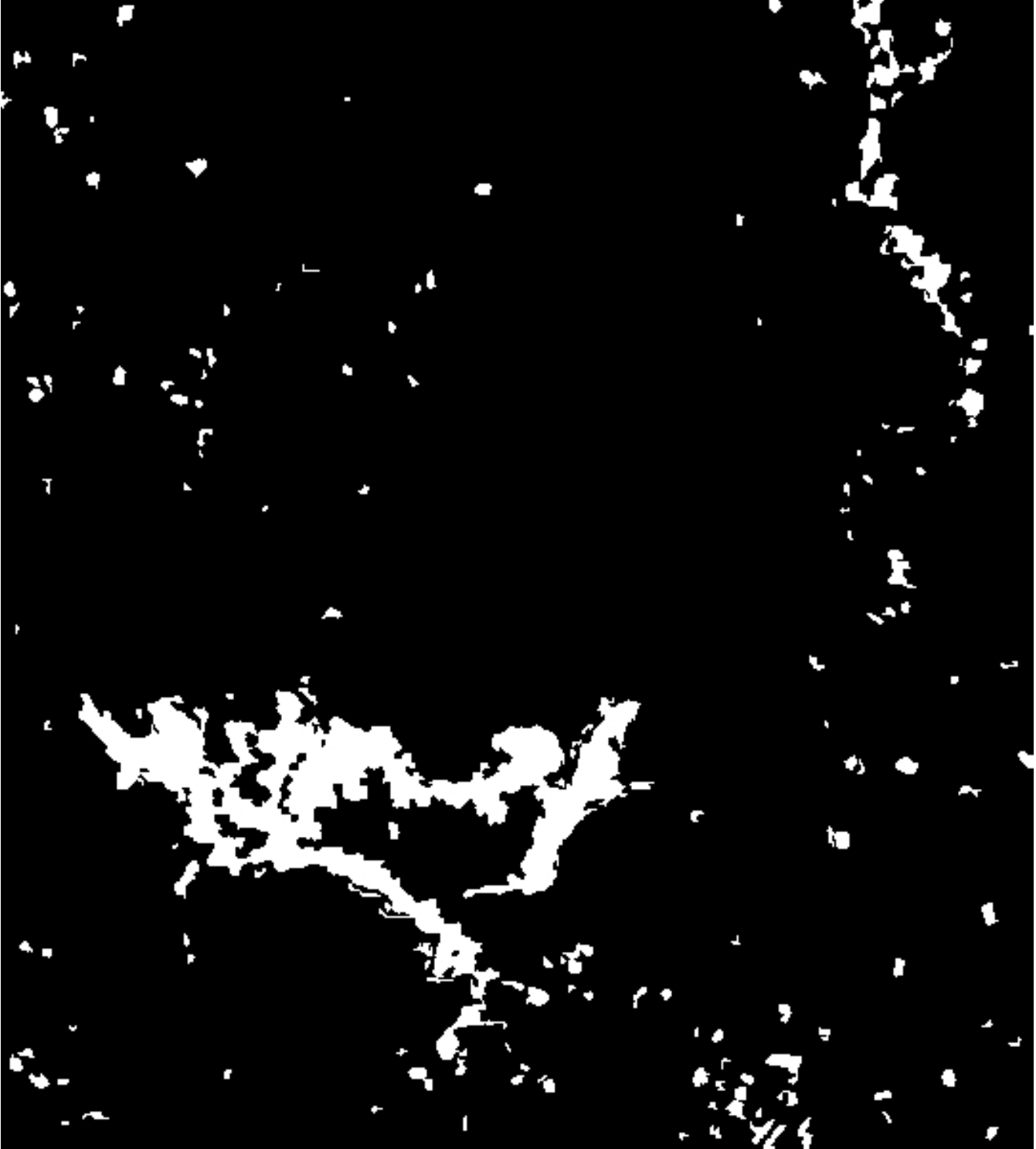}
					\caption{$\hat{\mathbf{m}}_{\mathrm{S}}$}
					\label{fig:SMAP_2}
			\end{subfigure}
			\begin{subfigure}{\subfwidth}
					\centering
					\includegraphics[width=\figsize]{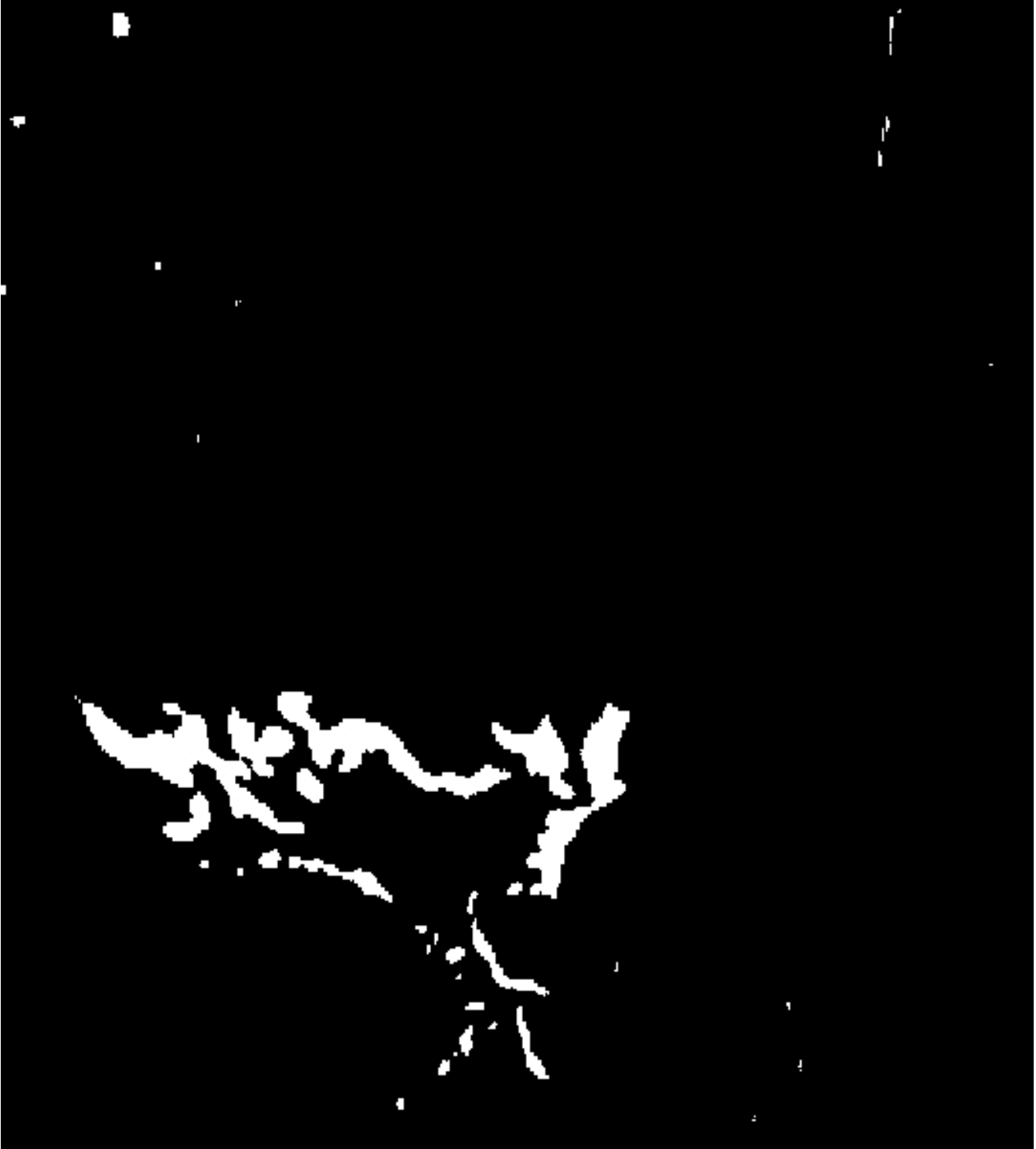}
					\caption{$\hat{\mathbf{m}}_{\mathrm{CDL}}$}
					\label{fig:CDLMAP_2}
			\end{subfigure}
\caption{Real images affected by real changes  with ground truth, Scenario 2: \protect\subref{fig:Yt1_2} observed radar image $\mathbf{Y}_{t_1}$ from the Lake Mulargia acquired in 05/21/2016 by Sentinel 1 , \protect\subref{fig:Yt2_2} observed radar image $\mathbf{Y}_{t_2}$  from the Lake Mulargia acquired in 10/30/2016 by Sentinel 1, \protect\subref{fig:mask_2} groud-truth mask $\mathbf{m}$ indicating changed areas constructed by photointerpretation,\protect\subref{fig:FMAP_2} change map $\hat{\mathbf{m}}_{\mathrm{F}}$ of the fuzzy method, \protect\subref{fig:RFMAP_2} change map $\hat{\mathbf{m}}_{\mathrm{RF}}$ of the robust fusion method, \protect\subref{fig:SMAP_2} change map $\hat{\mathbf{m}}_{\mathrm{S}}$ of the segmentation-based method, and \protect\subref{fig:CDLMAP_2} change map $\hat{\mathbf{m}}_{\mathrm{CDL}}$ of proposed method.}%
	\label{fig:real_2}%
\end{figure}

\noindent\textbf{Scenario 2: SAR vs. SAR --} For this scenario, two intensity radar images acquired over the Lake Mulargia region in Sardegna, by the Sentinel-1 satellite in 05/21/2016 (Figure \ref{fig:Yt1_2}) and 10/30/2016 (Figure \ref{fig:Yt2_2}) are considered. These $1200 \times 1800$-pixel images are both characterized by a $10$m spatial resolution. This dataset mostly presents seasonal changes, in particular the variation of flooding areas around the Lake Mulargia. The ground-truth change mask  is represented in Figure \ref{fig:mask_2}. Figure \ref{fig:real_2} depicts the two observed images, the ground-truth change mask and the change maps estimated by the four compared methods. 

The quantitative results for Scenario 2 are reported in Table \ref{table:ROCOPTOPT} (lines 3 and 4) and the corresponding ROC curves in Figure \ref{fig:rocSARSAR}. The analysis of these results shows that the proposed method also outperforms the state-of-the-art methods for this scenario. It is a good indication of its flexibility w.r.t. image modalities. Note that, due to the multiplicative noise and the consequent strong fluctuations, the state-of-the-art methods present a lot of false alarms. This effect seems to be attenuated by the proposed method, probably thanks the TV regularization.\\

\begin{figure}
	\centering
			\begin{subfigure}{\subfwidth}
					\centering	
					\includegraphics[width=\figsize]{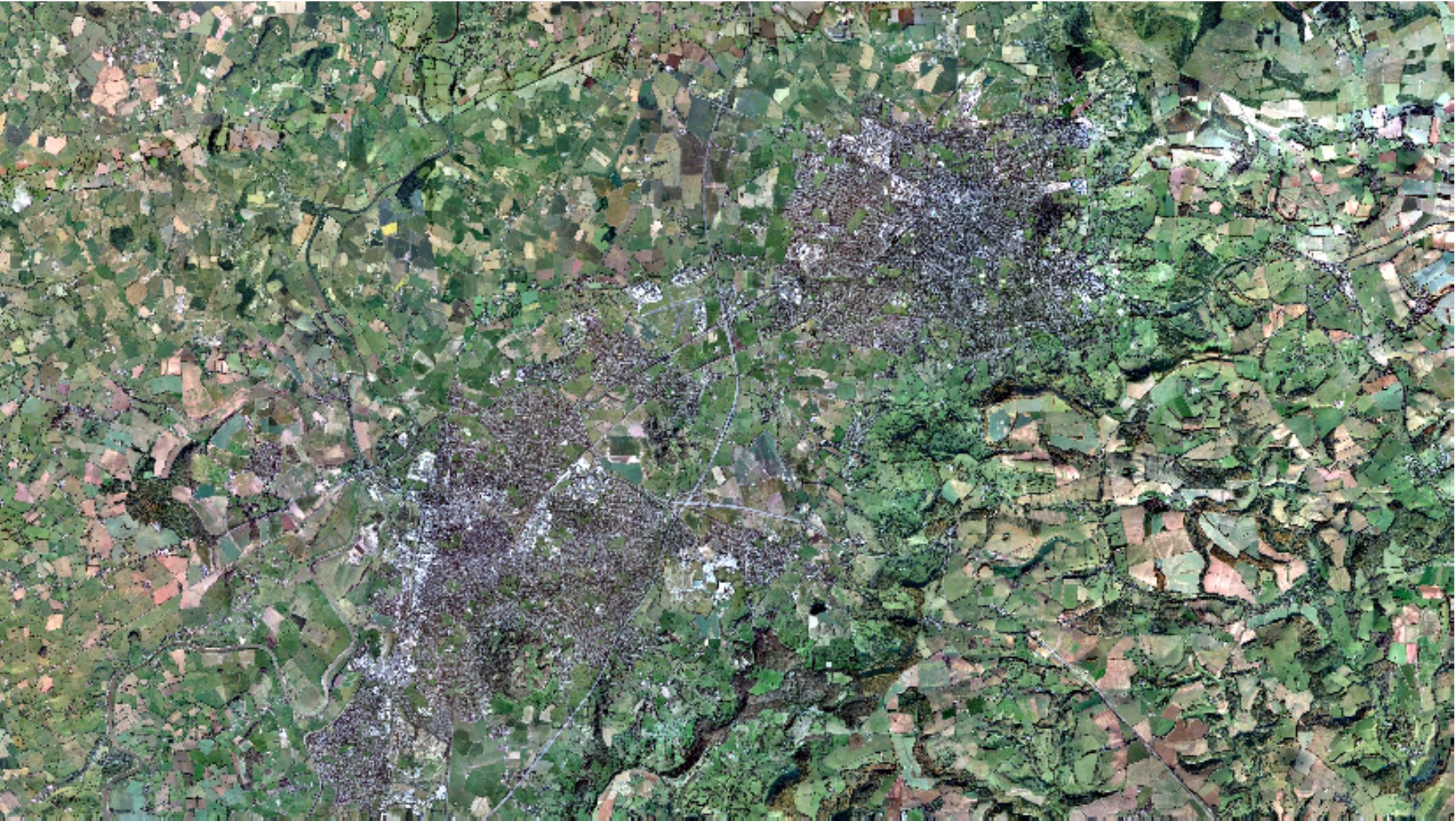}
					\caption{$\mathbf{Y}_{t_1}$}
					\label{fig:Yt1_3}
			\end{subfigure}
			\begin{subfigure}{\subfwidth}
					\centering	
					\includegraphics[width=\figsize]{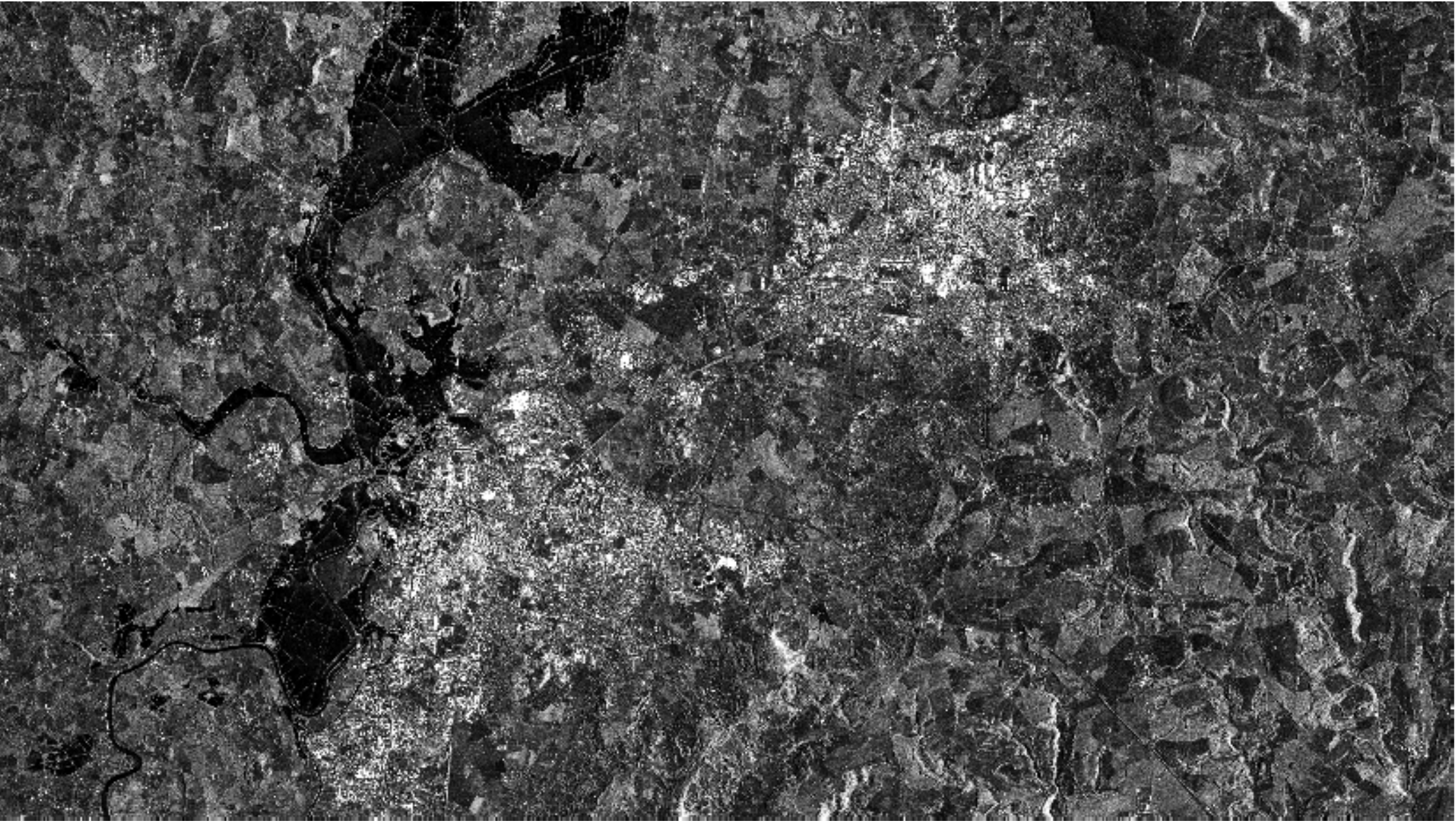}
					\caption{$\mathbf{Y}_{t_2}$}
					\label{fig:Yt2_3}
			\end{subfigure}
			\begin{subfigure}{\subfwidth}
					\centering
					\includegraphics[width=\figsize]{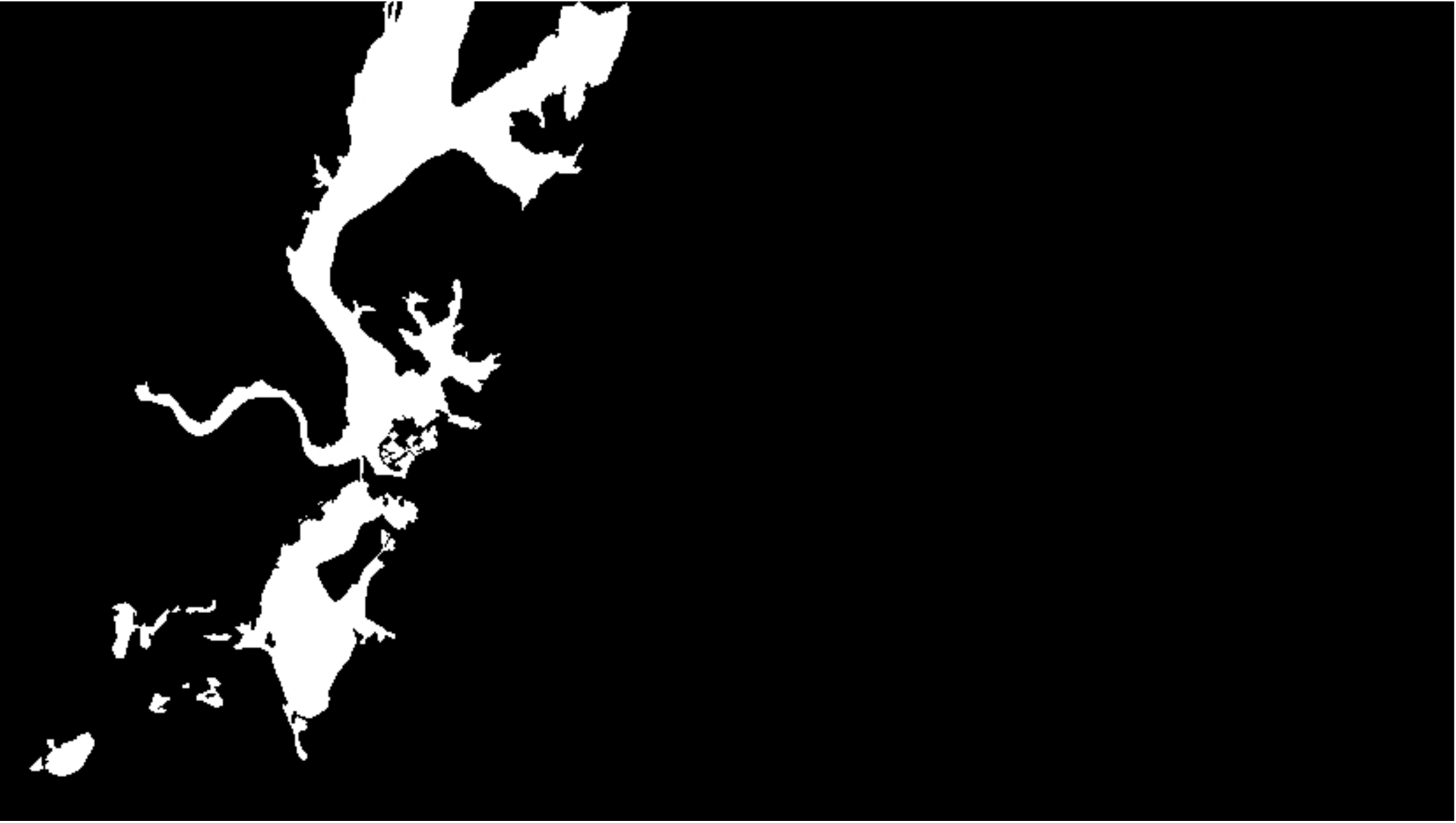}
					\caption{$\mathbf{m}$}
					\label{fig:mask_3}
			\end{subfigure}
			\begin{subfigure}{\subfwidth}
					\centering
					\includegraphics[width=\figsize]{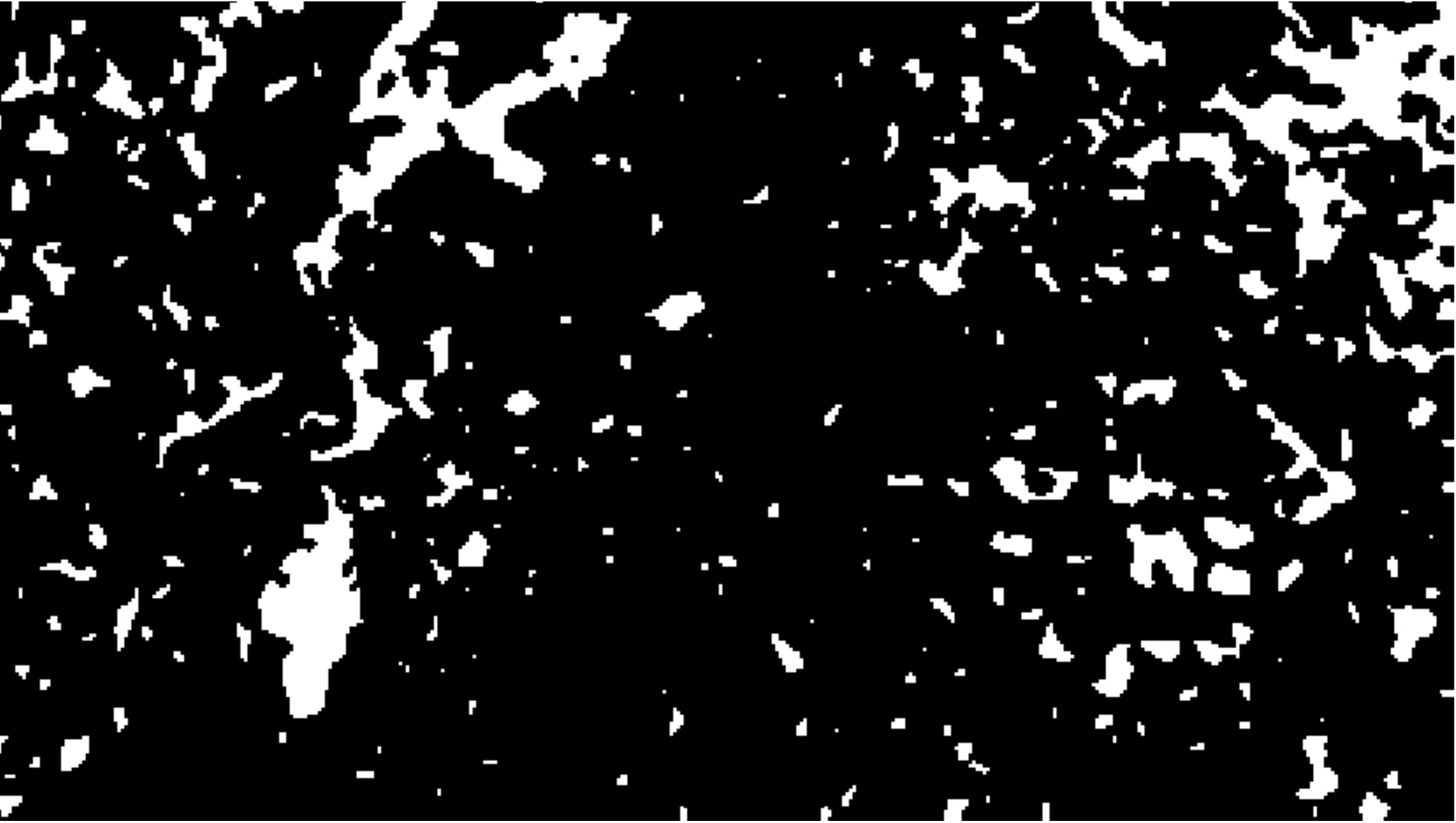}
					\caption{$\hat{\mathbf{m}}_{\mathrm{F}}$}
					\label{fig:FMAP_3}
			\end{subfigure}
			\begin{subfigure}{\subfwidth}
					\centering
					\includegraphics[width=\figsize]{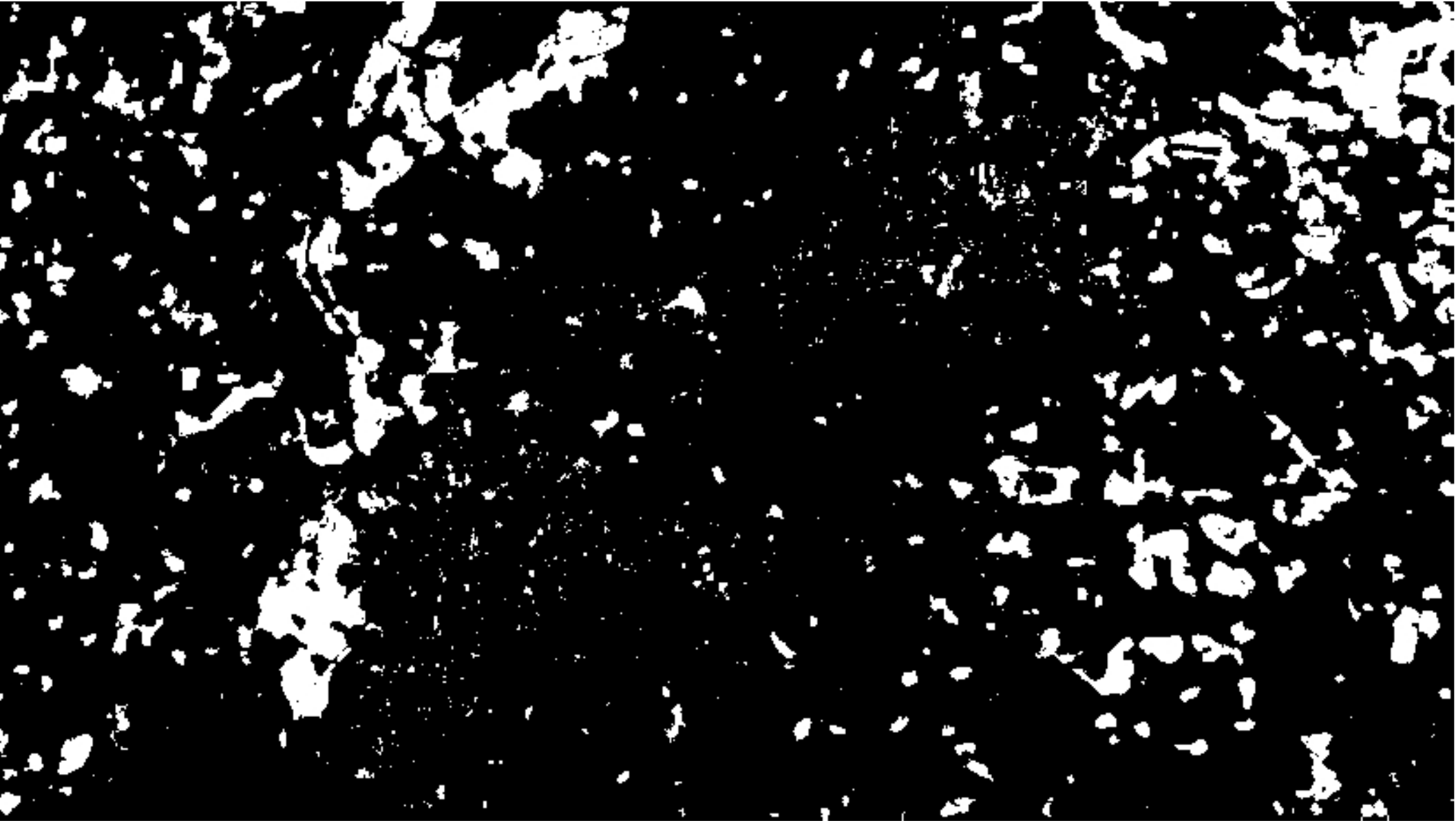}
					\caption{$\hat{\mathbf{m}}_{\mathrm{RF}}$}
					\label{fig:RFMAP_3}
			\end{subfigure}
						\begin{subfigure}{\subfwidth}
					\centering
					\includegraphics[width=\figsize]{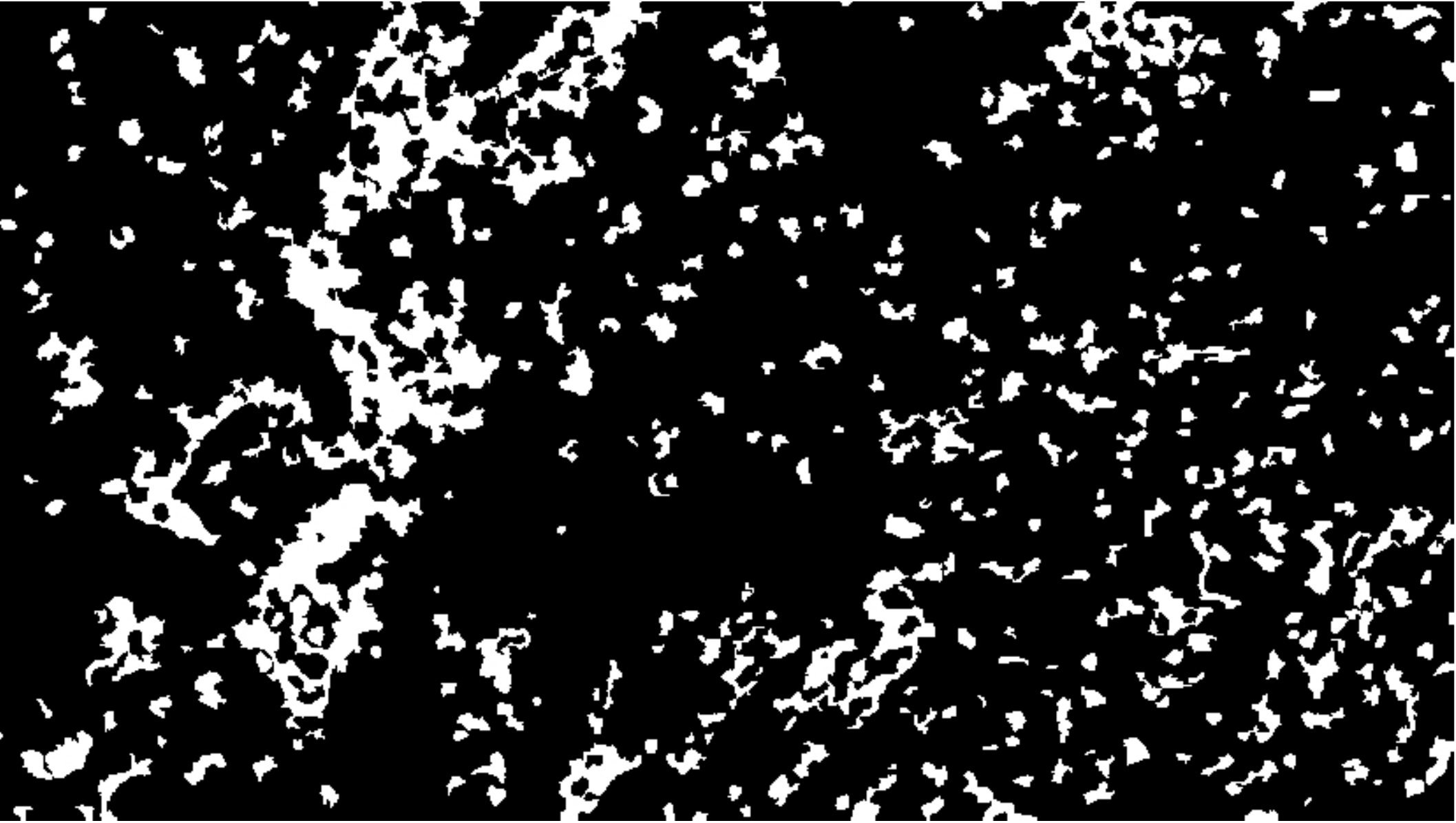}
					\caption{$\hat{\mathbf{m}}_{\mathrm{S}}$}
					\label{fig:SMAP_3}
			\end{subfigure}
			\begin{subfigure}{\subfwidth}
					\centering
					\includegraphics[width=\figsize]{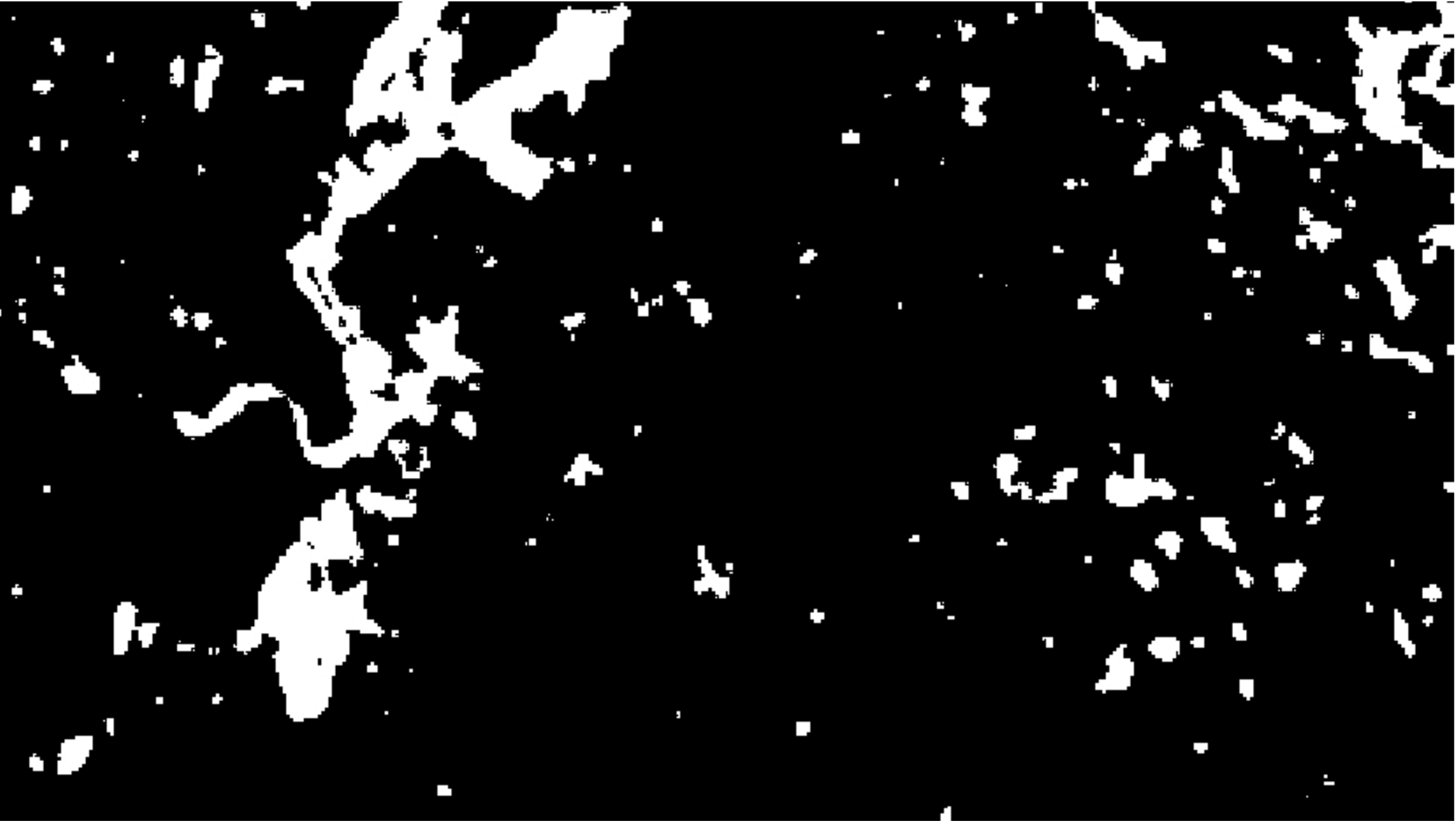}
					\caption{$\hat{\mathbf{m}}_{\mathrm{CDL}}$}
					\label{fig:CDLMAP_3}
			\end{subfigure}
\caption{Real images affected by real changes  with ground truth, Scenario 3: \protect\subref{fig:Yt1_3} observed MS optical image $\mathbf{Y}_{t_1}$ from Gloucester region acquired before the flooding by Google Earth , \protect\subref{fig:Yt2_3} observed radar image $\mathbf{Y}_{t_2}$ from Gloucester region acquired after the flooding by TerraSAR-X, \protect\subref{fig:mask_3} groud-truth mask $\mathbf{m}$ indicating changed areas constructed by photointerpretation,\protect\subref{fig:FMAP_3} change map $\hat{\mathbf{m}}_{\mathrm{F}}$ of the fuzzy method, \protect\subref{fig:RFMAP_3} change map $\hat{\mathbf{m}}_{\mathrm{RF}}$ of the robust fusion method, \protect\subref{fig:SMAP_2} change map $\hat{\mathbf{m}}_{\mathrm{S}}$ of the segmentation-based method, and \protect\subref{fig:CDLMAP_3} change map $\hat{\mathbf{m}}_{\mathrm{CDL}}$ of proposed method.}%
	\label{fig:real_3}%
\end{figure}

\noindent \textbf{Scenario 3: optical vs. SAR --} In order to test the performance of the different compared methods in a multi-modality situation, we consider two images acquired over the Gloucester region, UK, before and after a catastrophic flooding accident in 2007. The before-flooding image, presented in Figure \ref{fig:Yt1_3}, is a multispectral optical image with 3 channels acquired by Google Earth while the after-flooding image, depicted in Figure \ref{fig:Yt2_3}, is a radar image acquired by TerraSAR-X. These $2325\times 4133$-pixel images are both characterized by a $7.3$m spatial resolution. The ground-truth change mask is represented on Figure \ref{fig:mask_3}. Figure \ref{fig:real_3} depicts the observed images at each date, the ground-truth change mask and the change maps estimated by the four comparative methods.

Table \ref{table:ROCOPTOPT} (lines 5 and 6) reports the quantitative results for Scenario 3  and the corresponding ROC curves are displayed in Figure \ref{fig:rocSAROPT}. Similarly as in the previous scenarios, the analysis of these results shows that the proposed method outperforms the state-of-the-art methods even in this more challenging situation involving different image modalities with changes in rural and urban areas. As for Scenario 2, the TV regularization seems to be beneficial to smooth the fluctuations due to the nature of the noise in radar images, which may affect the other compared methods producing more false alarms.

Through this first illustration, we can observe that the segmentation-based method severly underperforms the other methods for scenarios 1 and 2 and underperfoms the proposed method under scenario 1. For brevity, in the following, we will compare only the F and RF methods to the proposed CDL one.

\subsubsection{Case of different resolutions without ground truth}\label{subsec:real_images_wogt}
The previous set of experiments considered pair of images characterized by the same spatial resolution. As a complementary analysis, this section reports experiments conducted on real images of different spatial resolutions with real changes. However, for these 3 pairs of images, corresponding to the three scenarios, no ground truth is available. We first consider a Sentinel-1 SAR image \citep{european_space_agency_sentinel-1_2017} acquired on October 28th 2016.  This image is a $540 \times 525$ interferometric wide swath high resolution ground range detected multi-looked SAR intensity image with a spatial resolution of $10$m according to 5 looks in the range direction. Moreover, we also consider two multispectral Landsat 8 \citep{united_states_geological_survey_landsat_2017} $180 \times 175$-pixel images with $30$m spatial resolution and composed of the RGB visible bands (Band 2 to 4), acquired over the same region on April 15th 2015 and on September 22th 2015, respectively. Unfortunately, no ground-truth information is available for the chosen dates, as experienced in numerous experimental situations \citep{bovolo_time_2015}. However, this region is characterized by interesting natural meteorological changes occurring along the seasons (e.g., drought of the Mud Lake, snow falls and vegetation growth), which helps to visually infer the major changes between observed images and to assess the relevance of the detected changes. All considered images have been manually geographically and geometrically aligned to fulfill the requirements imposed by the considered CD setup. Each scenario is individually studied considering the same denominations as in Section \ref{subsec:real_images} and the same compared methods as in Section \ref{subsec:compared}.\\

\begin{figure}
\centering
			\begin{subfigure}{\subfwidth}
					\centering	
					\includegraphics[width=\figsize]{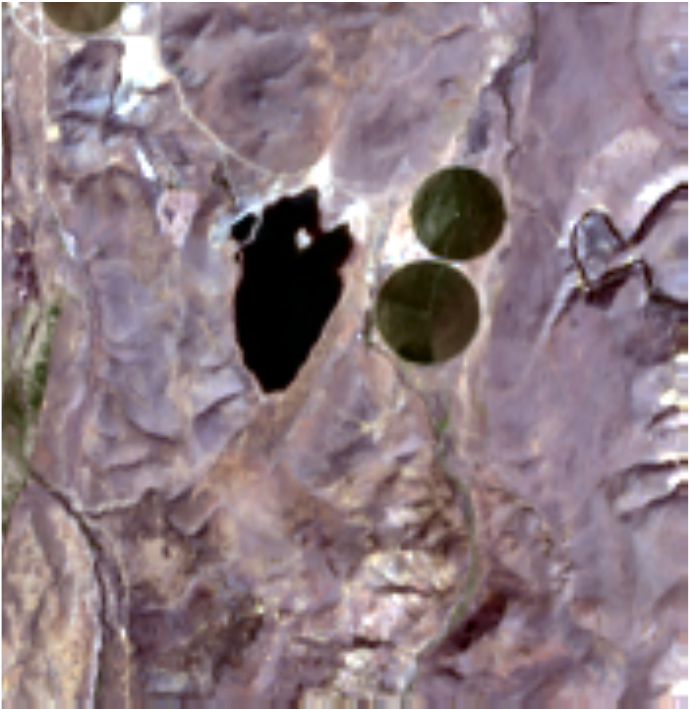}
					\caption{$\mathbf{Y}_{t_1}$}
					\label{fig:s2s2Yt1_2}
			\end{subfigure}
			\begin{subfigure}{\subfwidth}
					\centering	
					\includegraphics[width=\figsize]{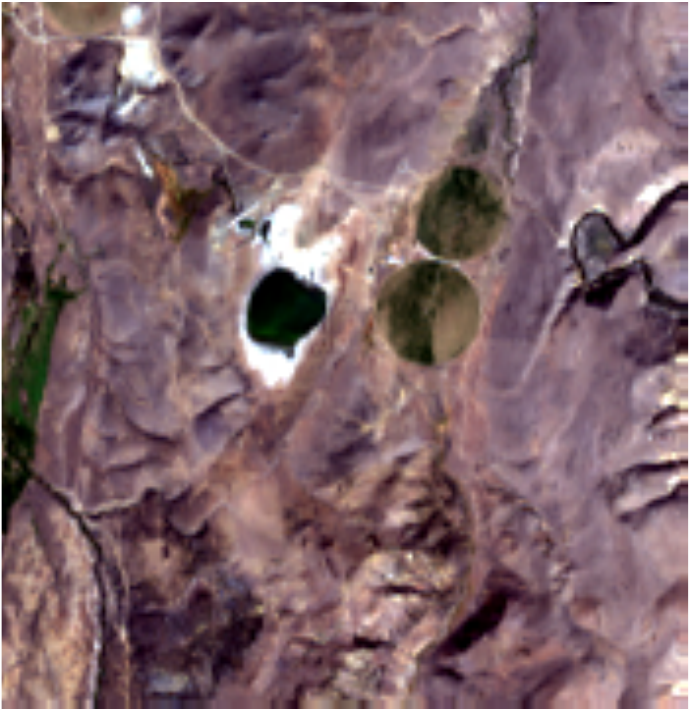}
					\caption{$\mathbf{Y}_{t_2}$}
					\label{fig:s2s2Yt2_2}
			\end{subfigure}
			\begin{subfigure}{\subfwidth}
					\centering
					\includegraphics[width=\figsize]{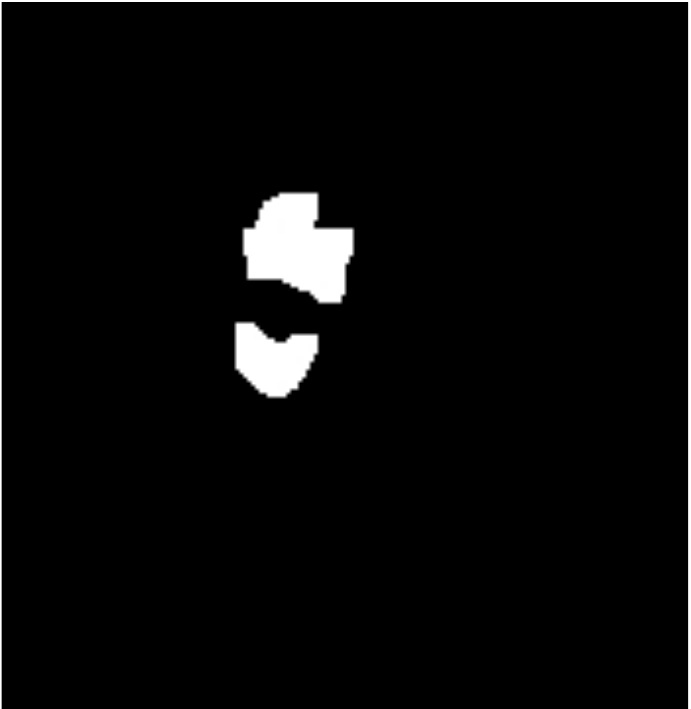}
					\caption{$\hat{\mathbf{m}}_{\mathrm{F}}$}
					\label{fig:s2s2FMAP_2}
			\end{subfigure}
            \begin{subfigure}{\subfwidth}
					\centering
					\includegraphics[width=\figsize]{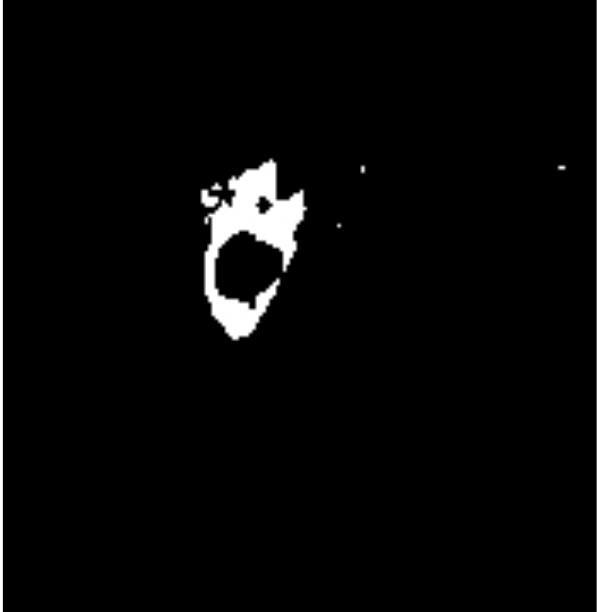}
					\caption{$\hat{\mathbf{m}}_{\mathrm{RF}}$}
					\label{fig:s2s2RFMAP_2}
			\end{subfigure}
            \begin{subfigure}{\subfwidth}
					\centering	
					\includegraphics[width=\figsize]{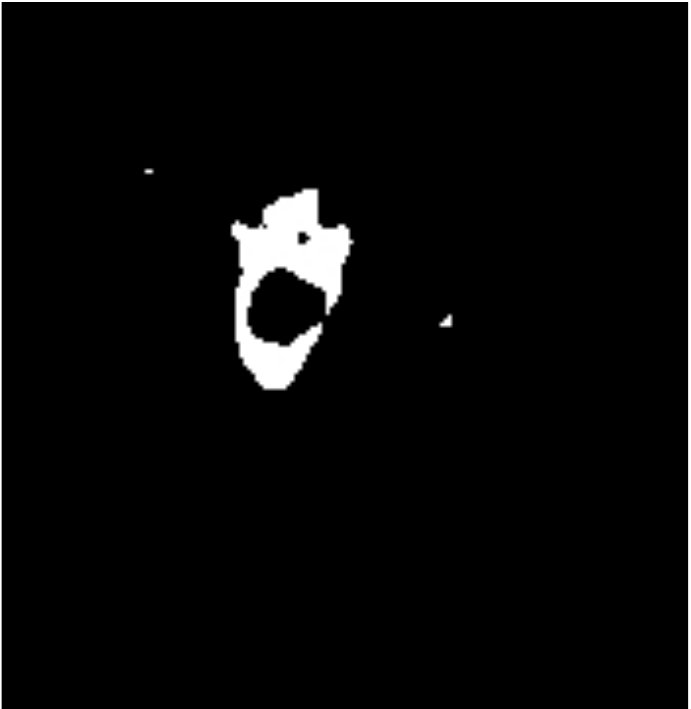}
					\caption{$\hat{\mathbf{m}}_{\mathrm{CDL}}$}
					\label{fig:s2s2DCMAP_2}
			\end{subfigure}
\caption{Real images affected by real changes  without ground truth, Scenario 1 (same spatial resolutions): \protect\subref{fig:s2s2Yt1_2}  observed  Landsat 8 MS image $\mathbf{Y}_{t_1}$ acquired on 04/15/2015, \protect\subref{fig:s2s2Yt2_2}  Landsat 8 MS image $\mathbf{Y}_{t_2}$  acquired on 09/22/2015, \protect\subref{fig:s2s2FMAP_2} change map $\hat{\mathbf{m}}_{\mathrm{F}}$ of the fuzzy method, \protect\subref{fig:s2s2RFMAP_2} change map $\hat{\mathbf{m}}_{\mathrm{RF}}$ of the robust fusion method
	 and \protect\subref{fig:s2s2DCMAP_2} change map $\hat{\mathbf{m}}_{\mathrm{CDL}}$ of the proposed method.}%
	\label{fig:realS2S2_2}%
\end{figure}

\noindent\textbf{Scenario 1: optical vs. optical --} In this scenario, two different situations are going to be explored, namely, observed images with the same or different resolutions. The first case considers both Landsat 8 images. Figure \ref{fig:realS2S2_2} depicts the two observed images and the change maps estimated by the three compared methods. These change maps have been generated according to \eqref{eq:CVArule} where the threshold has been adjusted such that each method reveals the most important changes, i.e., the drought of the Mud Lake. As expected, the robust fusion method presents better accuracy in detection since it was specifically designed to handle such a scenario. Nevertheless, the proposed method exhibits very similar results. It is worth noting that some of the observed differences are due to the patch decomposition required by the proposed method. The fuzzy method is able to localize the strongest changes, but low energy changes are not detected. The fuzzy method also suffers from resolution loss due to the size of the patches. Contrary to the proposed method, it does not take the patch overlapping into account, which contributes to decrease the detection accuracy.

\begin{figure}
\centering
			\begin{subfigure}{\subfwidth}
					\centering	
					\includegraphics[width=\figsize]{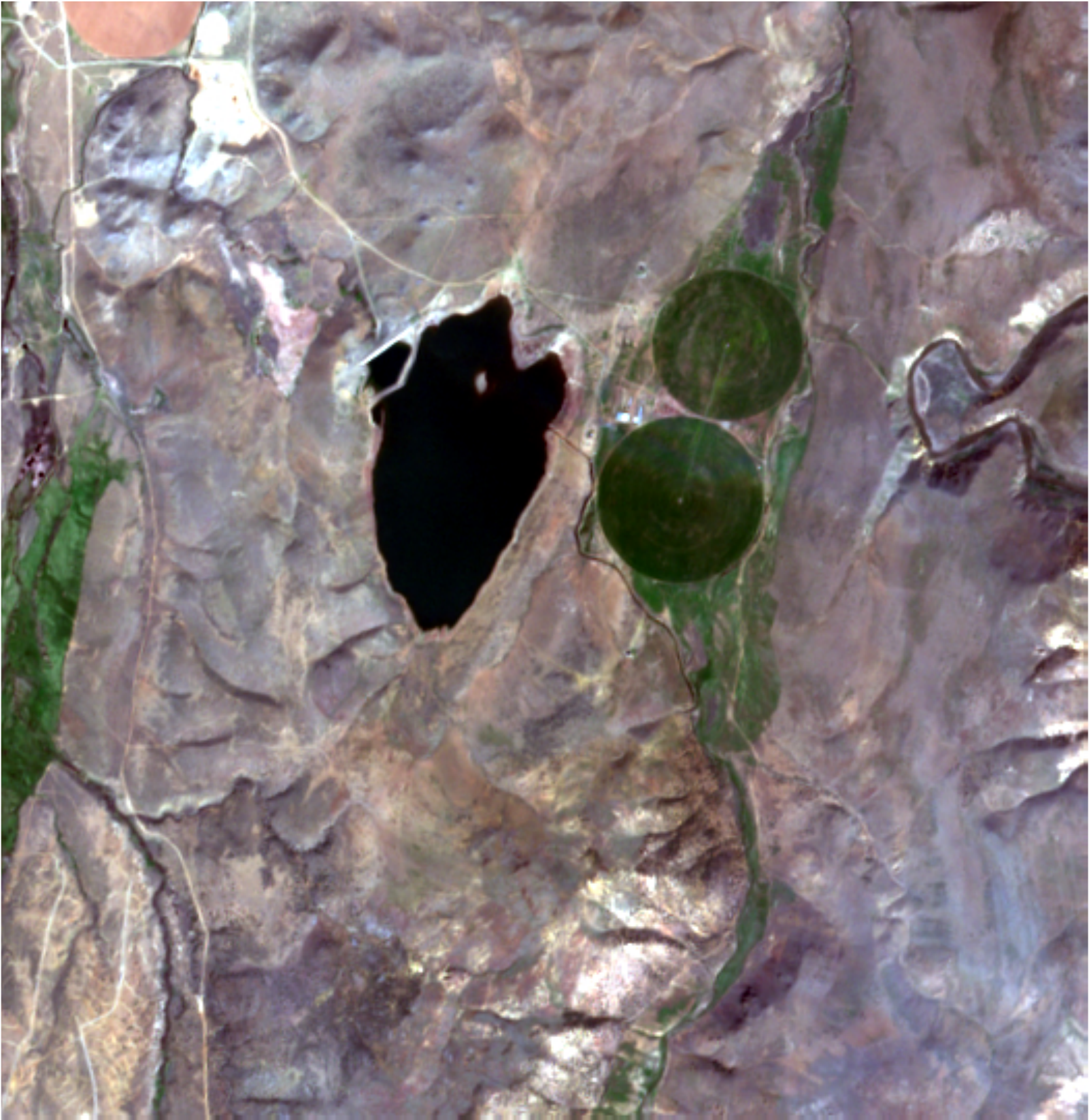}
					\caption{$\mathbf{Y}_{t_1}$}
					\label{fig:s2s2Yt1_1}
			\end{subfigure}
			\begin{subfigure}{\subfwidth}
					\centering	
					\includegraphics[width=\figsize]{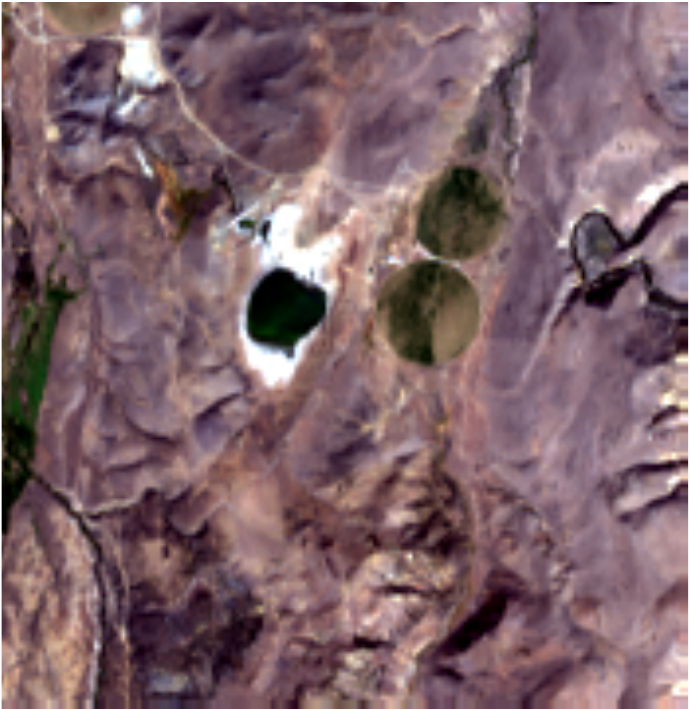}
					\caption{$\mathbf{Y}_{t_2}$}
					\label{fig:s2s2Yt2_1}
			\end{subfigure}
            \begin{subfigure}{\subfwidth}
					\centering
					\includegraphics[width=\figsize]{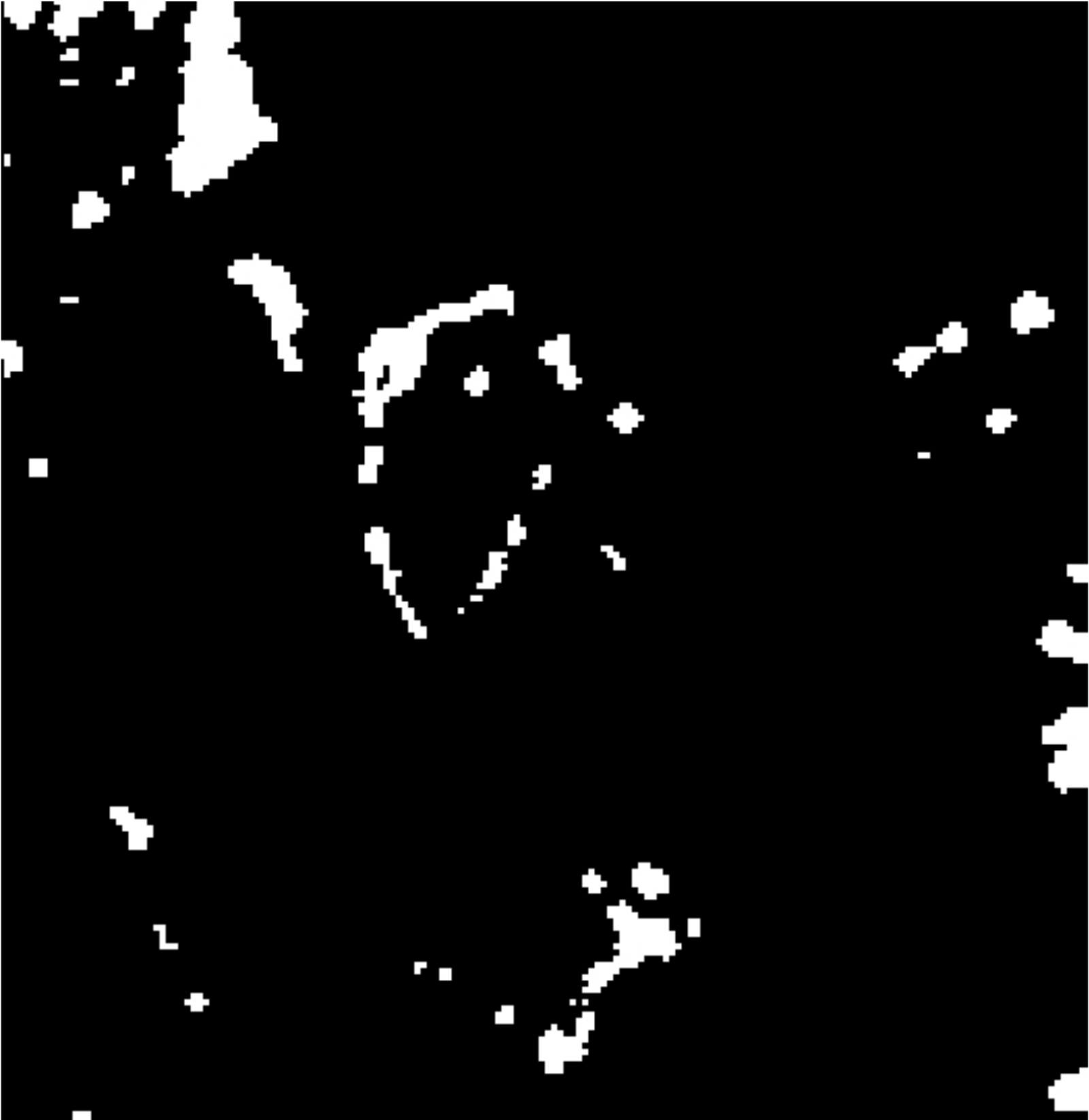}
					\caption{$\hat{\mathbf{m}}_{\mathrm{F}}$}
					\label{fig:s2s2FMAP_1}
			\end{subfigure}
            \begin{subfigure}{\subfwidth}
					\centering
					\includegraphics[width=\figsize]{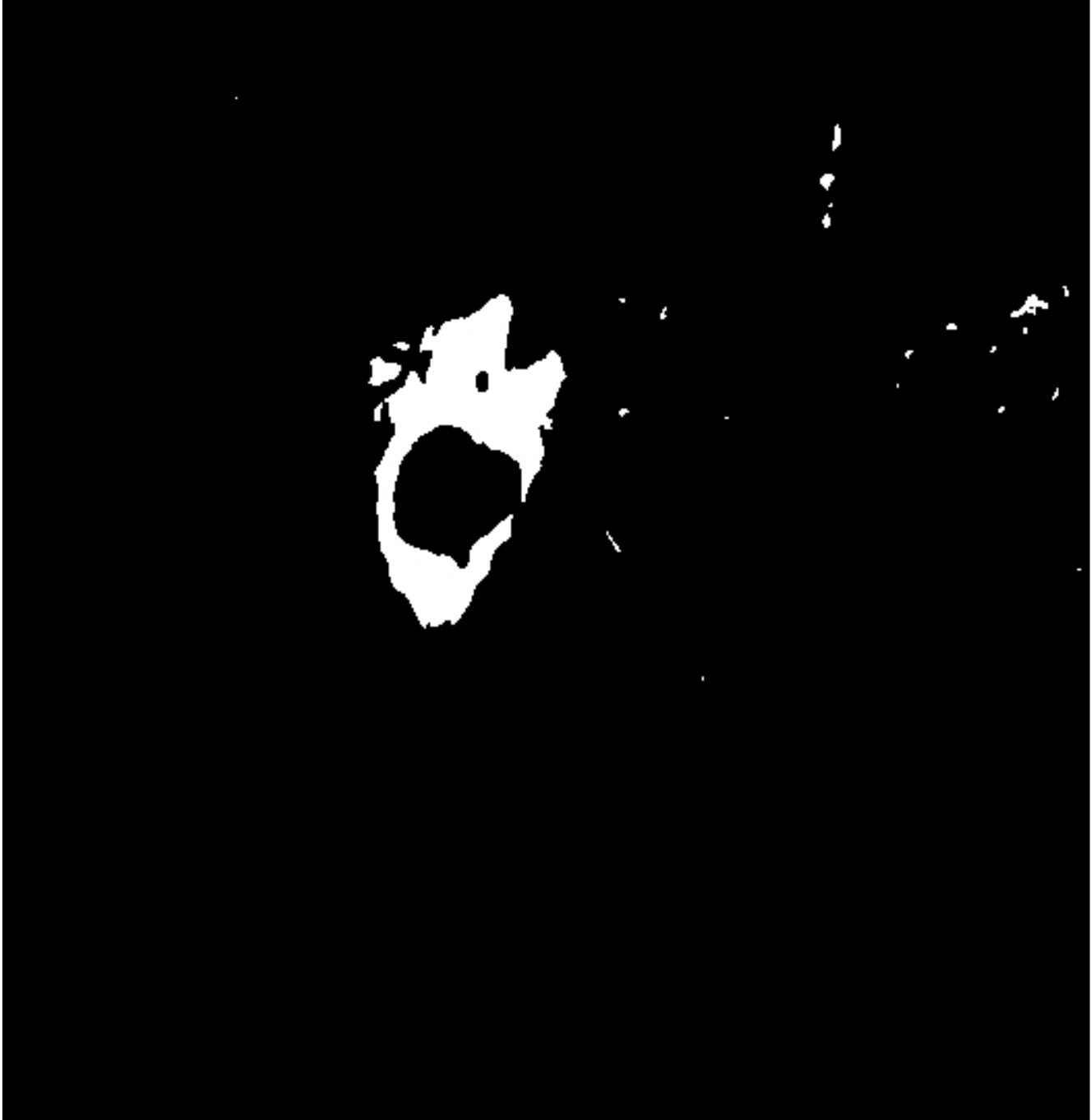}
					\caption{$\hat{\mathbf{m}}_{\mathrm{RF}}$}
					\label{fig:s2s2RFMAP_1}
			\end{subfigure}
			\begin{subfigure}{\subfwidth}
					\centering	
					\includegraphics[width=\figsize]{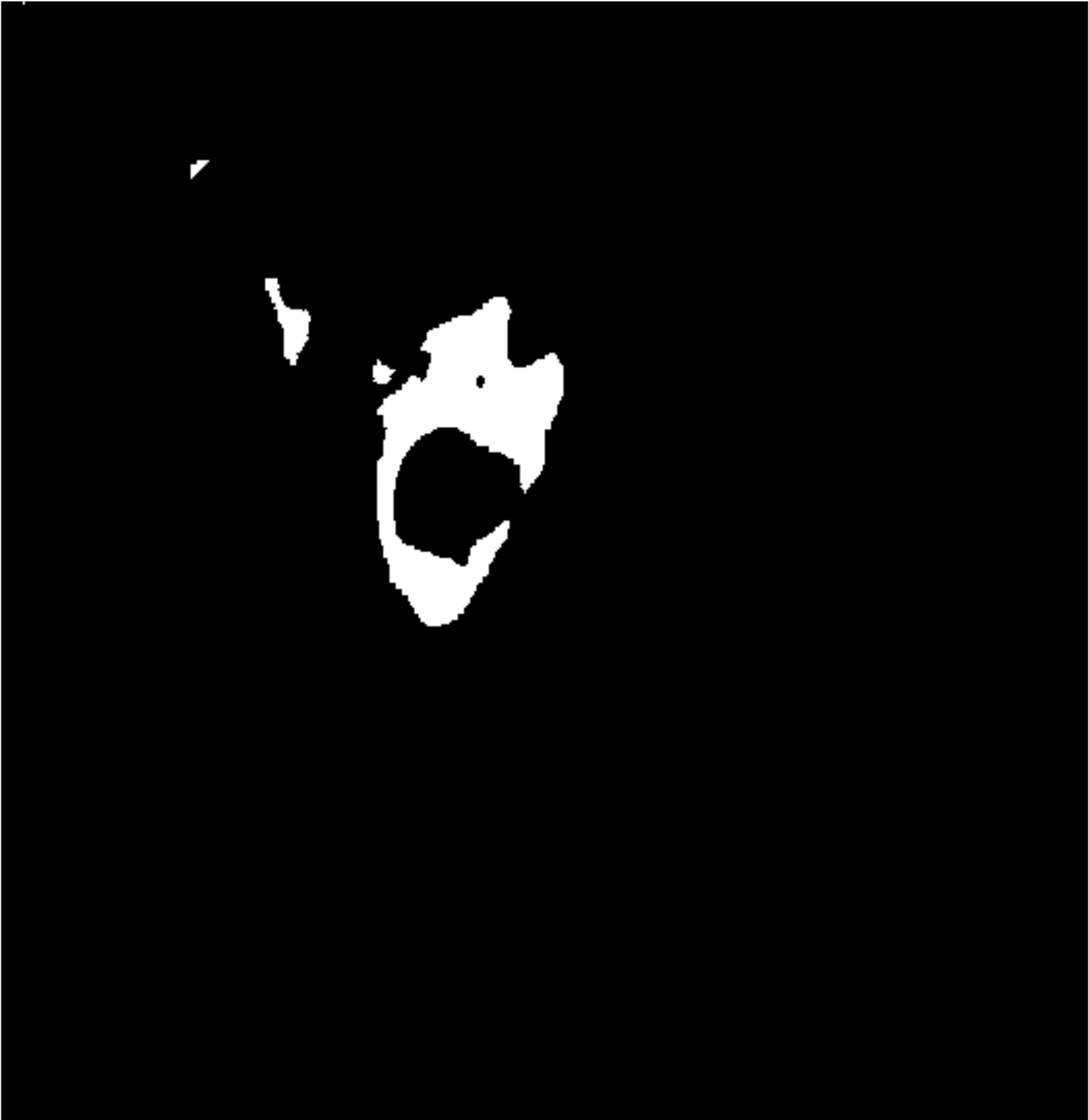}
					\caption{$\hat{\mathbf{m}}_{\mathrm{CDL}}$}
					\label{fig:s2s2DCMAP_1}
			\end{subfigure}
\caption{Real images affected by real changes  without ground truth, Scenario 1 (different  spatial resolutions): \protect\subref{fig:s2s2Yt1_1}  Sentinel-2 MS image $\mathbf{Y}_{t_1}$ acquired on 04/12/2016, \protect\subref{fig:s2s2Yt2_1}  Landsat 8 MS image $\mathbf{Y}_{t_2}$ acquired on 09/22/2015, \protect\subref{fig:s2s2FMAP_1} change map $\hat{\mathbf{m}}_{\mathrm{F}}$ of the fuzzy method, \protect\subref{fig:s2s2RFMAP_1} change map $\hat{\mathbf{m}}_{\mathrm{RF}}$ of the robust fusion method
	 and \protect\subref{fig:s2s2DCMAP_1} change map $\hat{\mathbf{m}}_{\mathrm{CDL}}$ of the proposed method.}%
	\label{fig:realS2S2_1}%
\end{figure}

Under the same scenario (i.e. optical vs. optical), an additional pair of observed images is used to better understand the algorithm behavior when facing to images of the same modality but with different spatial resolutions. The observed image pair is composed of the Sentinel-2 image acquired on April 12th 2016 and the Landsat 8 image acquired in September 22th 2015. Note that the two observed images have the same spectral resolution, but different spatial resolutions. Figure \ref{fig:realS2S2_1} depicts the observed images as well as the change maps estimated by the comparative methods. Once again, it is possible to state the similarity of the results provided by the robust fusion method and the proposed one. It also shows the very poor detection performance of the fuzzy method. This may be explained by the difficulty of coupling due to differences in resolutions.\\

\noindent \textbf{Scenario 2: SAR vs. SAR --} In this scenario, observed SAR images acquired by the same sensor (Sentinel-1) are used to assess the performance of the fuzzy method and the proposed one. The robust fusion method has not been considered due to the poor results obtained on the synthetic dataset (see Section \ref{subsec:synthetic_images} below). Figure \ref{fig:realS1S1} presents the observed images at each date and the change maps recovered by the two compared methods. The same strategy of threshold selection as for Scenario 1 has been adopted to reveal the most important changes. As expected, the proposed method presents a higher accuracy in detection than the fuzzy method. Possible reasons that may explain this difference are i) the fuzzy method is unable to handle overlapping patches and ii) the fuzzy method does not exploit appropriate data-fitting terms, in opposite to the proposed one. Besides, as SAR images present strong fluctuations due to their inherent image formation process, the additional TV regularization of the proposed method may contribute to smooth such fluctuations and better couple the dictionaries.\\

	\begin{figure}
		\centering
			\begin{subfigure}{\subfwidth}
					\centering	
					\includegraphics[width=\figsize]{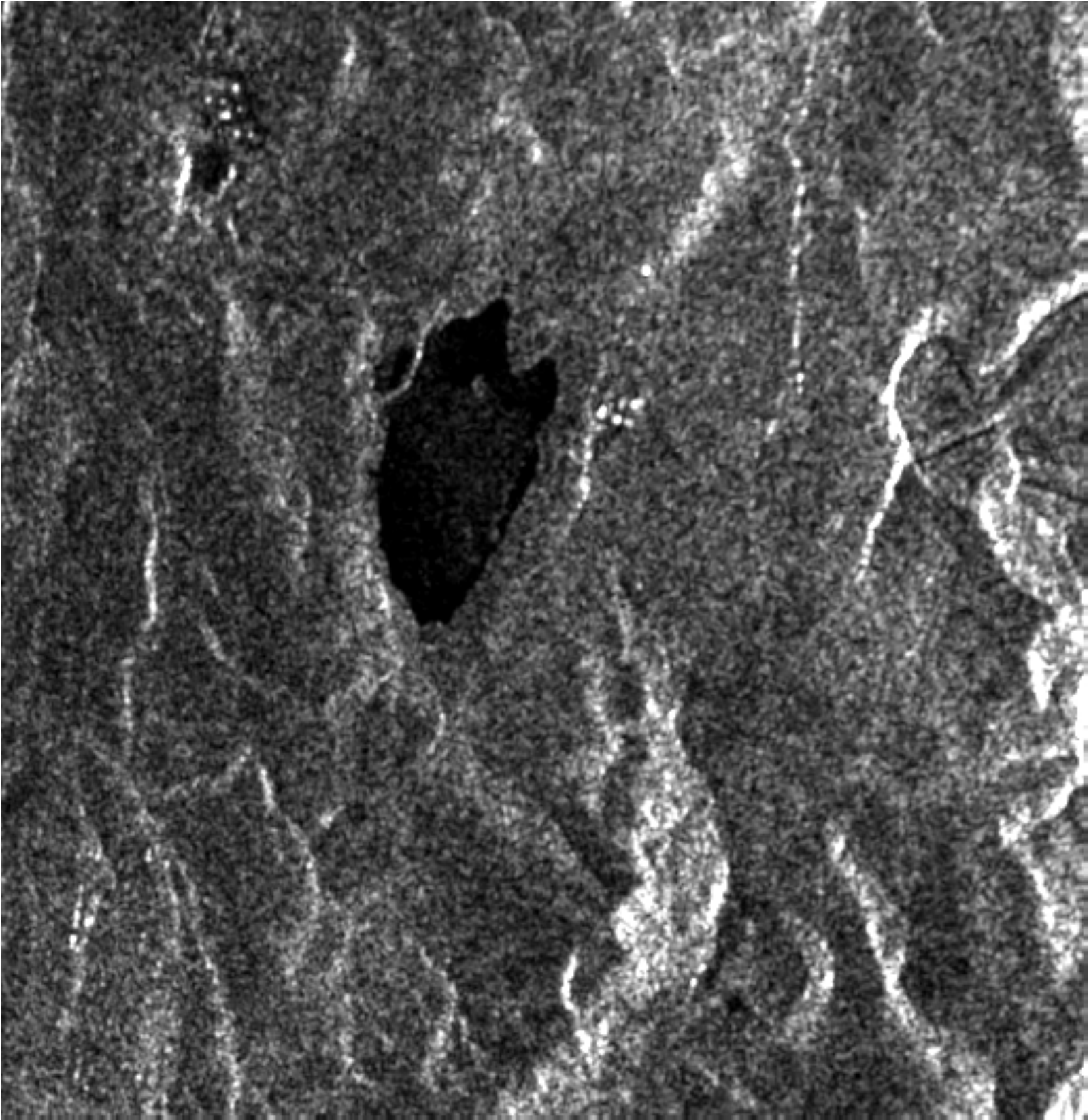}
					\caption{$\mathbf{Y}_{t_1}$}
					\label{fig:s1s1Yt1}
			\end{subfigure}
			\begin{subfigure}{\subfwidth}
					\centering	
					\includegraphics[width=\figsize]{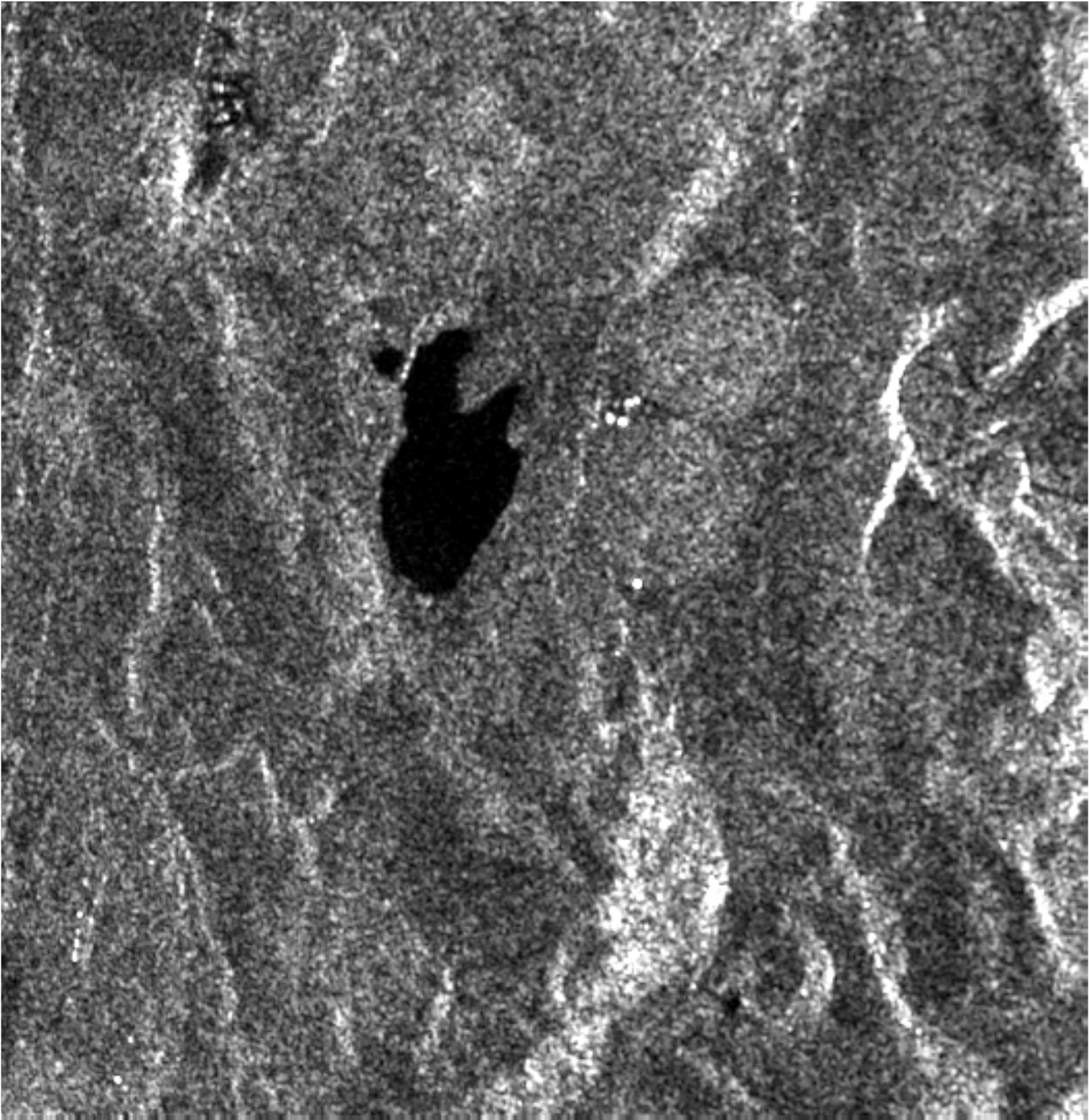}
					\caption{$\mathbf{Y}_{t_2}$}
					\label{fig:s1s1Yt2}
			\end{subfigure}
			\begin{subfigure}{\subfwidth}
					\centering	
					\includegraphics[width=\figsize]{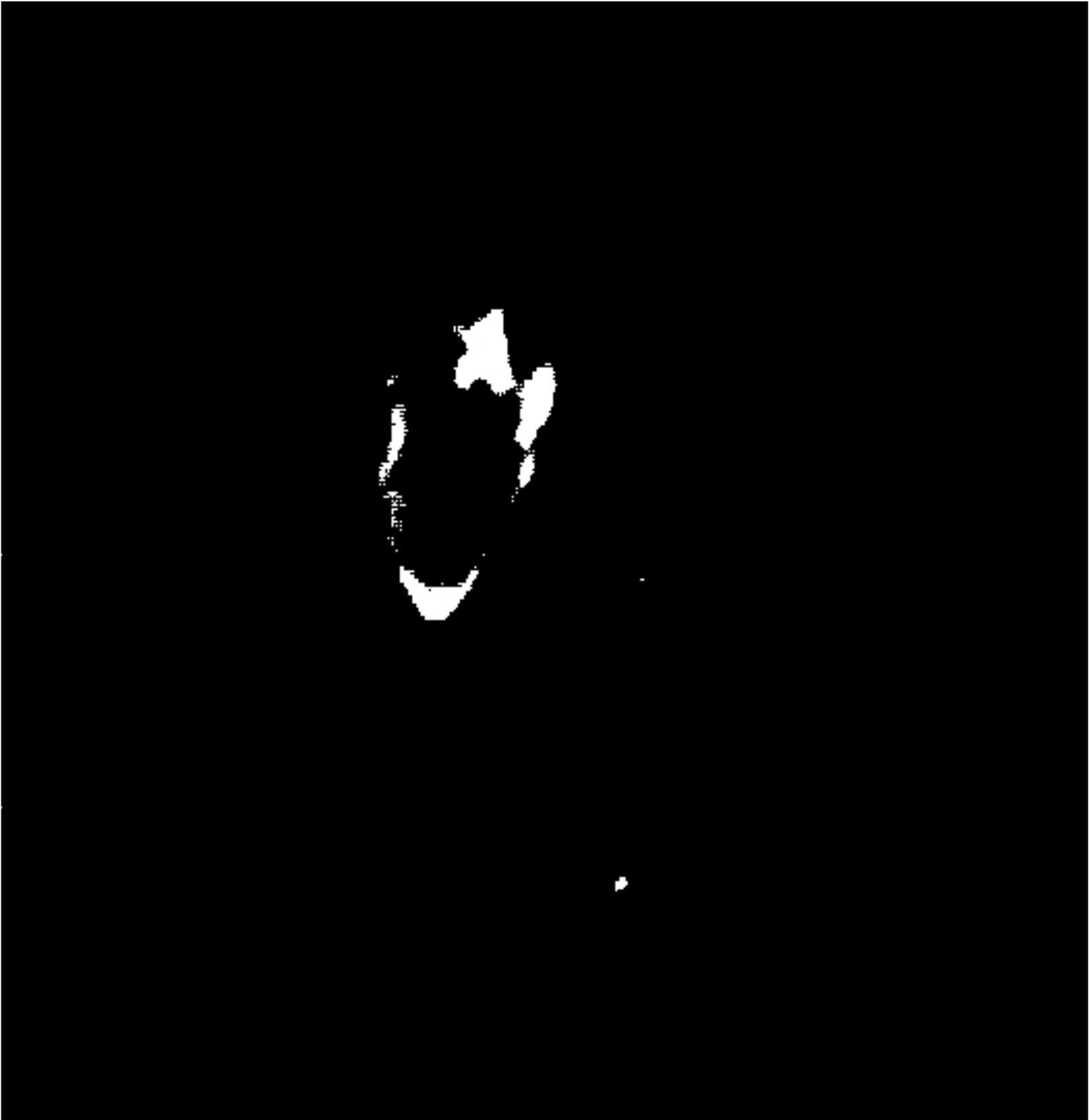}
					\caption{$\hat{\mathbf{m}}_{\mathrm{CDL}}$}
					\label{fig:s1s1DCMAP}
			\end{subfigure}
			\begin{subfigure}{\subfwidth}
					\centering
					\includegraphics[width=\figsize]{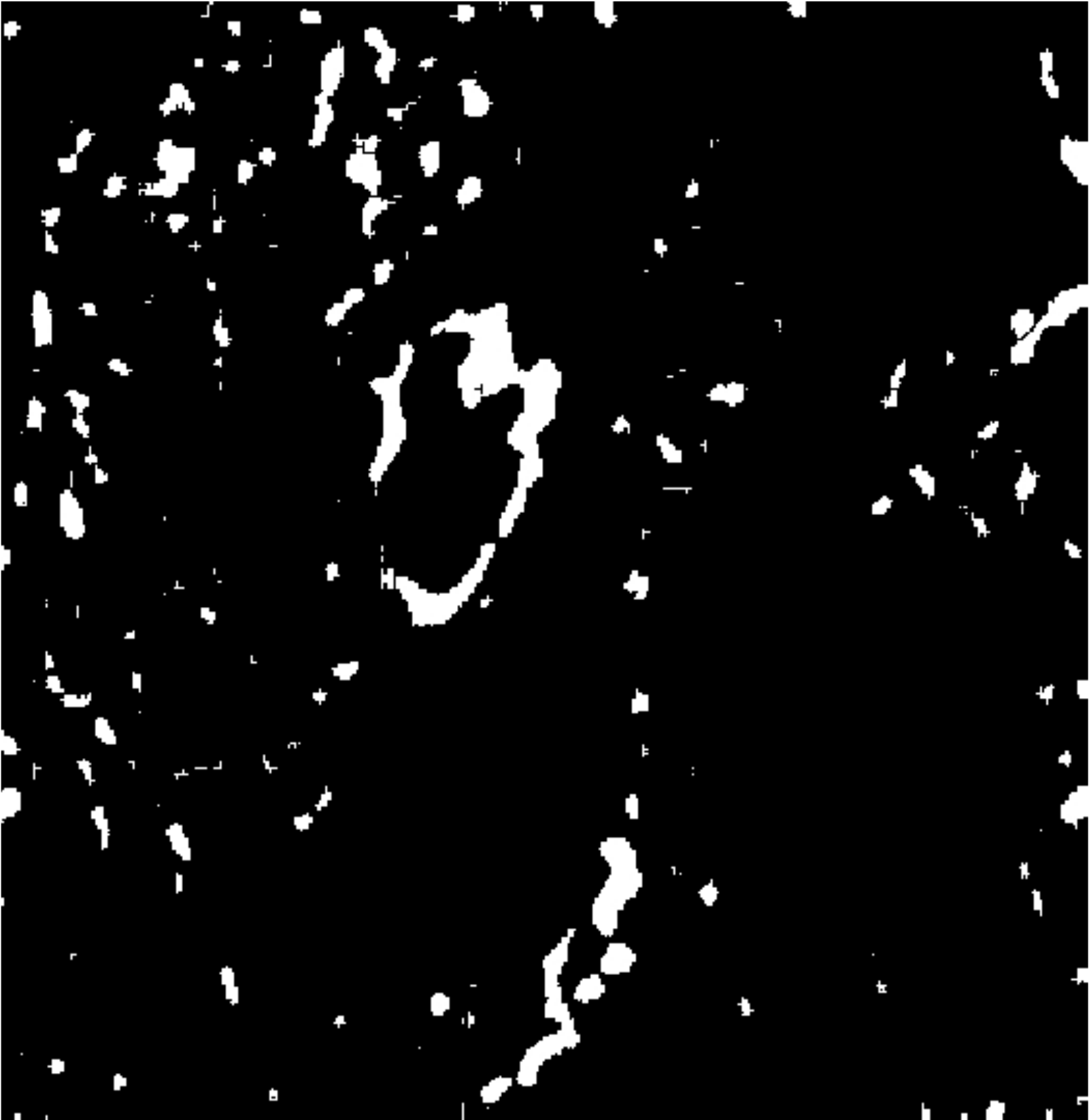}
					\caption{$\hat{\mathbf{m}}_{\mathrm{F}}$}
					\label{fig:s1s1FMAP}
			\end{subfigure}
\caption{Real images affected by real changes without ground truth, Scenario 2: \protect\subref{fig:s1s1Yt1}  Sentinel-1 SAR image  $\mathbf{Y}_{t_1}$ acquired on 04/12/2016, \protect\subref{fig:s1s1Yt2}  Sentinel-1 SAR image $\mathbf{Y}_{t_2}$ acquired on 10/28/2016, \protect\subref{fig:s1s1FMAP} change map $\hat{\mathbf{m}}_{\mathrm{F}}$ of the fuzzy method and \protect\subref{fig:s1s1DCMAP} change map $\hat{\mathbf{m}}_{\mathrm{CDL}}$ of the proposed method.}%
	\label{fig:realS1S1}%
\end{figure}

\begin{figure}
\centering
			\begin{subfigure}{\subfwidth}
					\centering	
					\includegraphics[width=\figsize]{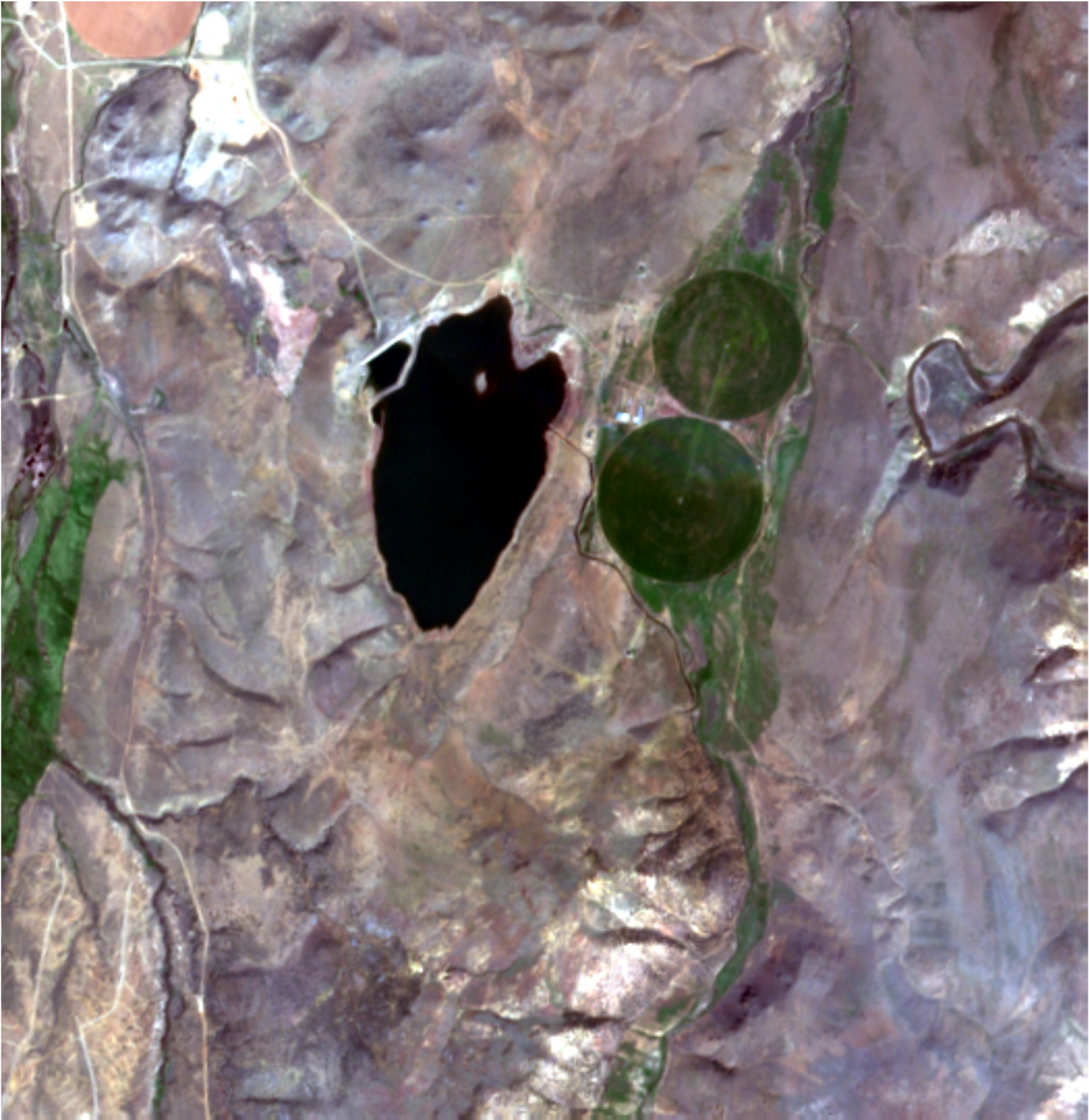}
					\caption{$\mathbf{Y}_{t_1}$}
					\label{fig:s1s2Yt1_2}
			\end{subfigure}
			\begin{subfigure}{\subfwidth}
					\centering	
					\includegraphics[width=\figsize]{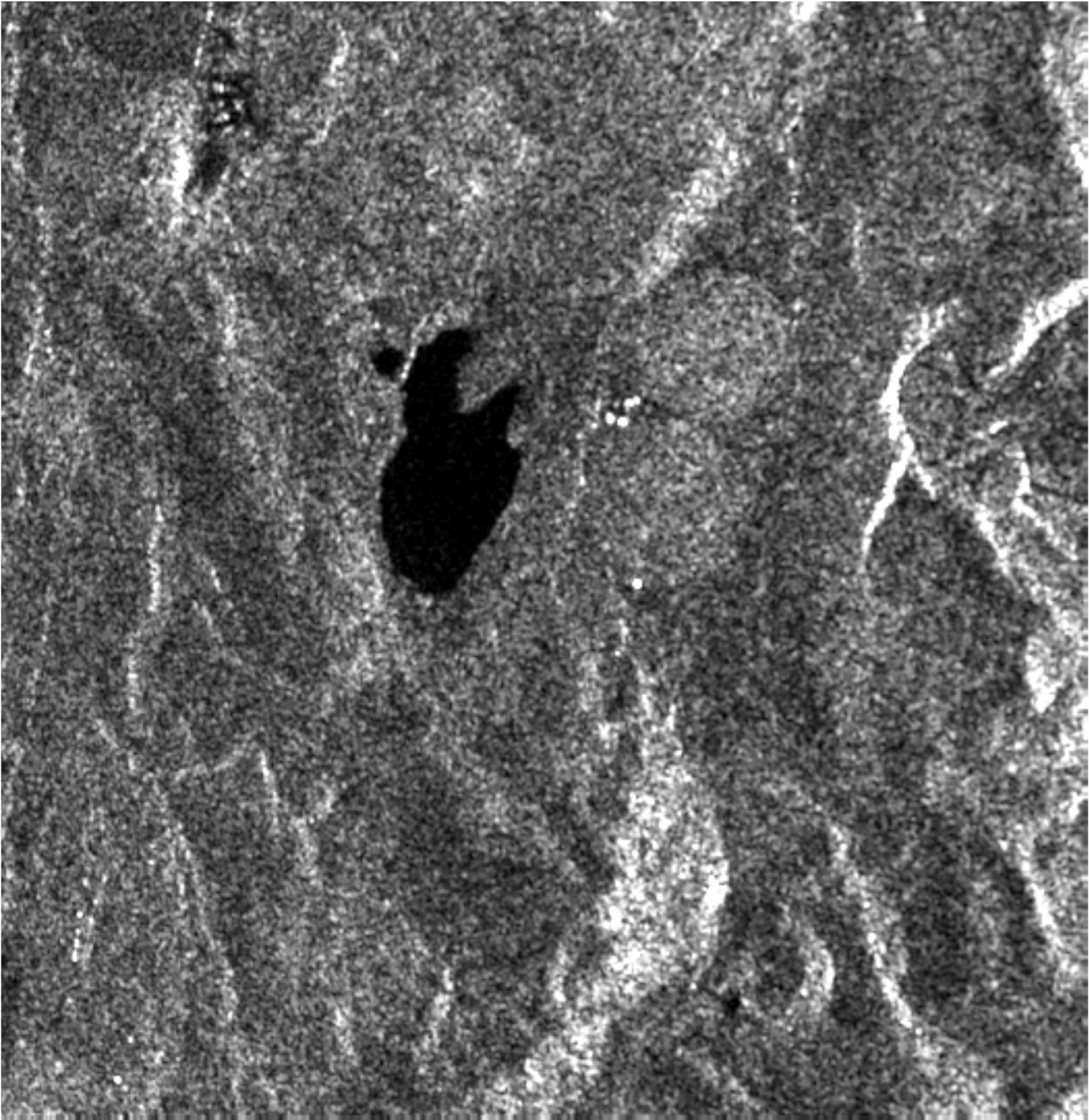}
					\caption{$\mathbf{Y}_{t_2}$}
					\label{fig:s1s2Yt2_2}
			\end{subfigure}
            \begin{subfigure}{\subfwidth}
					\centering
					\includegraphics[width=\figsize]{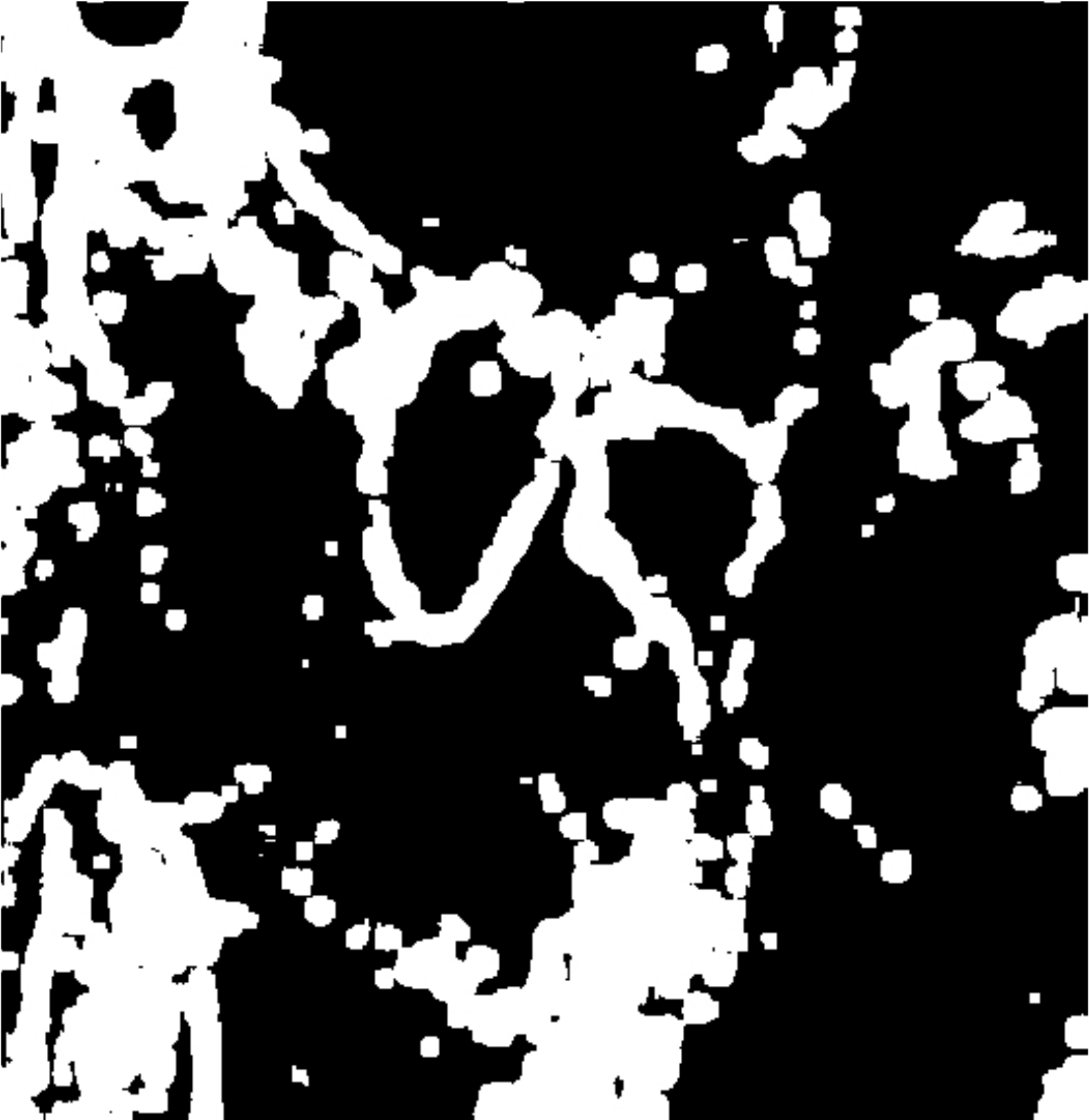}
					\caption{$\hat{\mathbf{m}}_{\mathrm{F}}$}
					\label{fig:s1s2FMAP_2}
			\end{subfigure}
			\begin{subfigure}{\subfwidth}
					\centering	
					\includegraphics[width=\figsize]{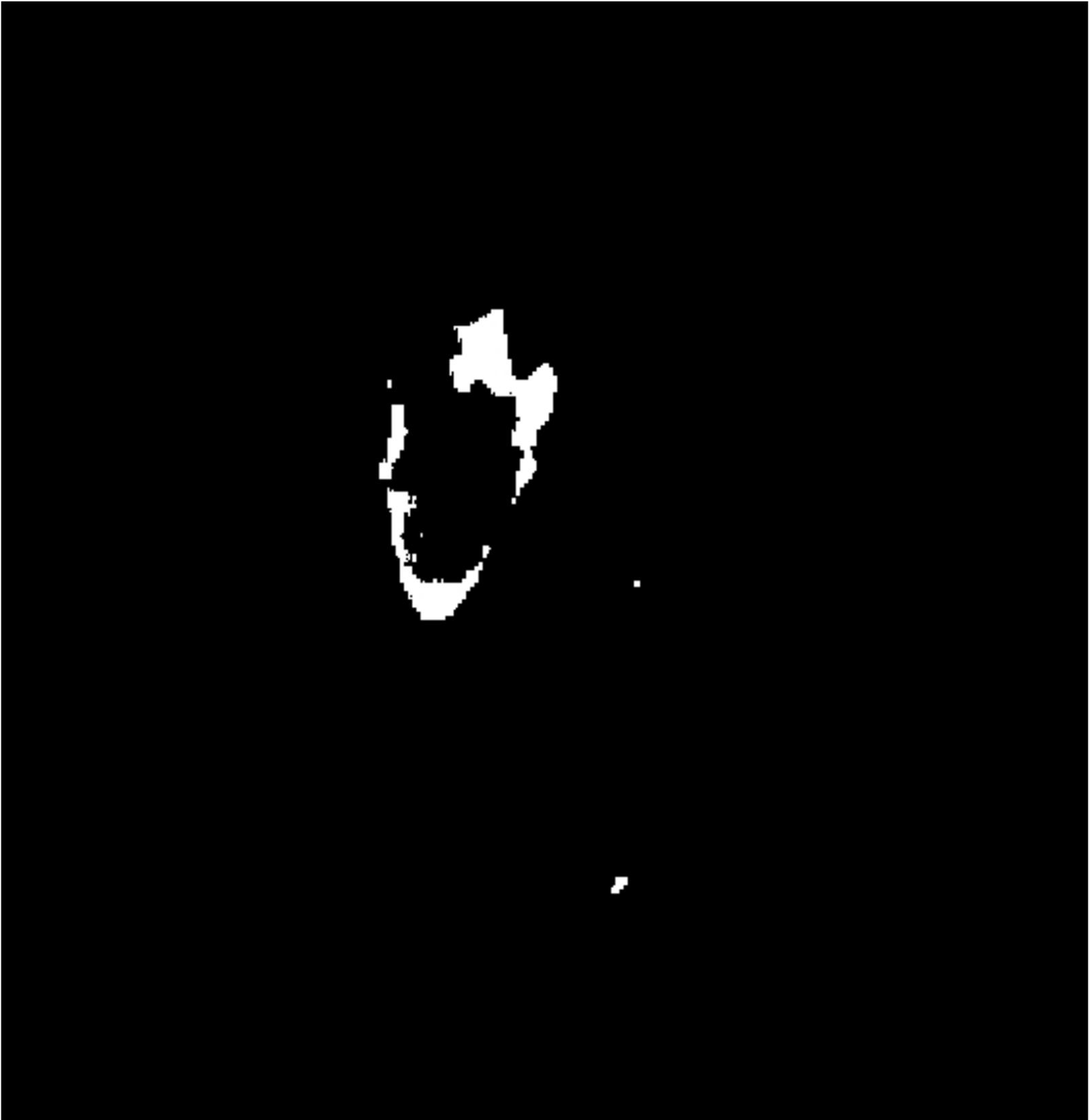}
					\caption{$\hat{\mathbf{m}}_{\mathrm{CDL}}$}
					\label{fig:s1s2DCMAP_2}
			\end{subfigure}
\caption{Real images affected by real changes  without ground truth, Scenario 3 (same spatial resolution): \protect\subref{fig:s1s2Yt1_2}  Sentinel-2 MS image $\mathbf{Y}_{t_1}$ acquired on 04/12/2016, \protect\subref{fig:s1s2Yt2_2}  Sentinel-1 SAR image $\mathbf{Y}_{t_2}$ acquired on 10/28/2016, \protect\subref{fig:s1s2FMAP_2} change map $\hat{\mathbf{m}}_{\mathrm{F}}$ of the fuzzy method and \protect\subref{fig:s1s2DCMAP_2} change map $\hat{\mathbf{m}}_{\mathrm{CDL}}$ of the proposed method.}%
	\label{fig:realS1S2_2}%
\end{figure}

\noindent \textbf{Scenario 3: optical vs. SAR --} For this scenario, once again, two different situations are addressed: images with the same or different spatial resolutions. The first one considers the Sentinel-2 MS image acquired on April 12th 2016 and the Sentinel-1 SAR image acquired in  October 28th 2016. Figure \ref{fig:realS1S2_2} presents the observed images and the change maps derived from the fuzzy and proposed methods. To derive the change maps, the thresholding strategy is the same as for all previous scenarios. Once again, the proposed method shows better detection accuracy performance than the fuzzy one. It is important to emphasize the similarity of the results achieved in Scenario 3 and Scenario 2 for images acquired at the same date. Note also that this similarity can be observed for the proposed method, which contributes to increase its reliability for CD between multimodal images.

\begin{figure}
\centering
			\begin{subfigure}{\subfwidth}
					\centering	
					\includegraphics[width=\figsize]{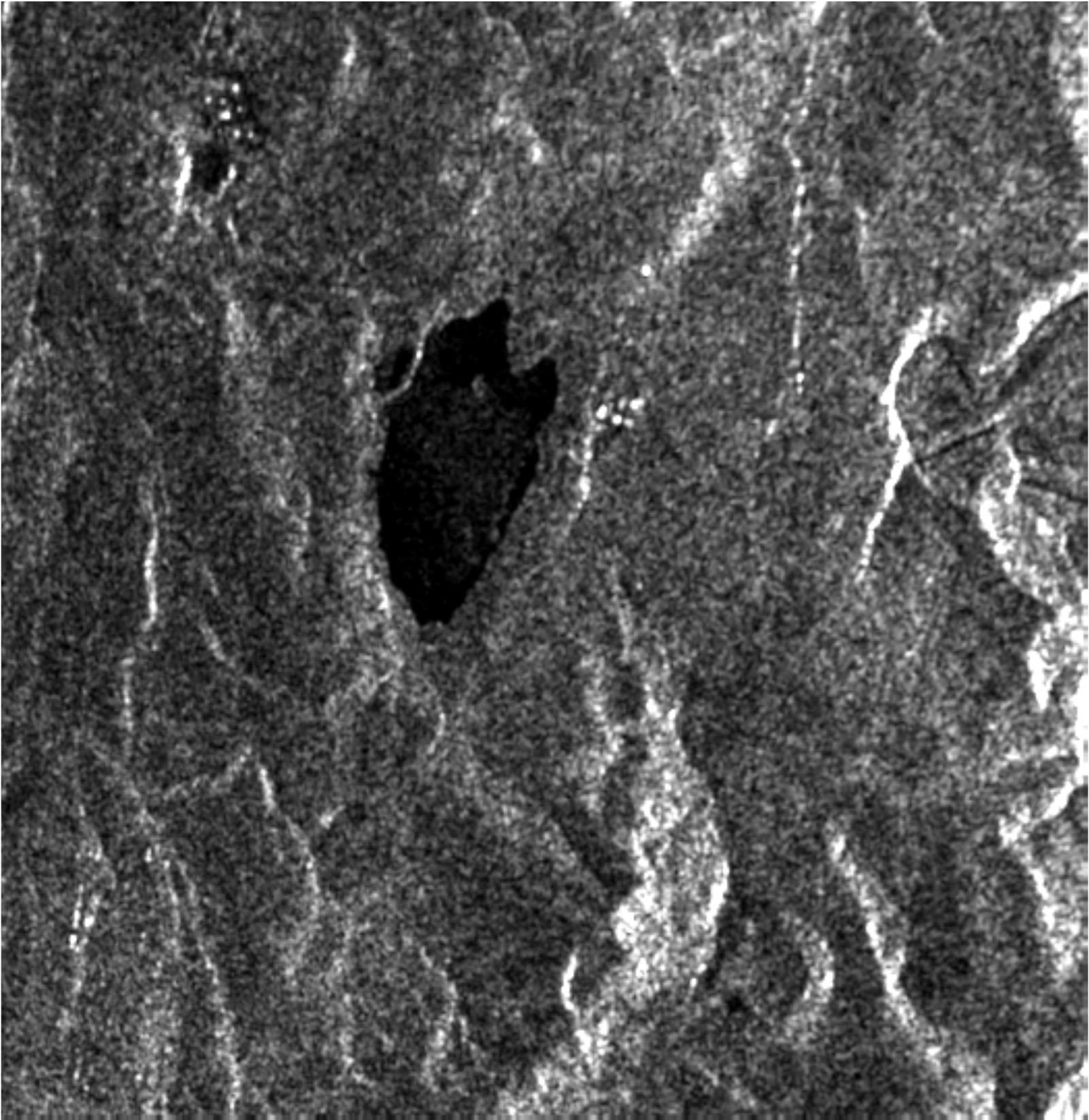}
					\caption{$\mathbf{Y}_{t_1}$}
					\label{fig:s1s2Yt1_1}
			\end{subfigure}
			\begin{subfigure}{\subfwidth}
					\centering	
					\includegraphics[width=\figsize]{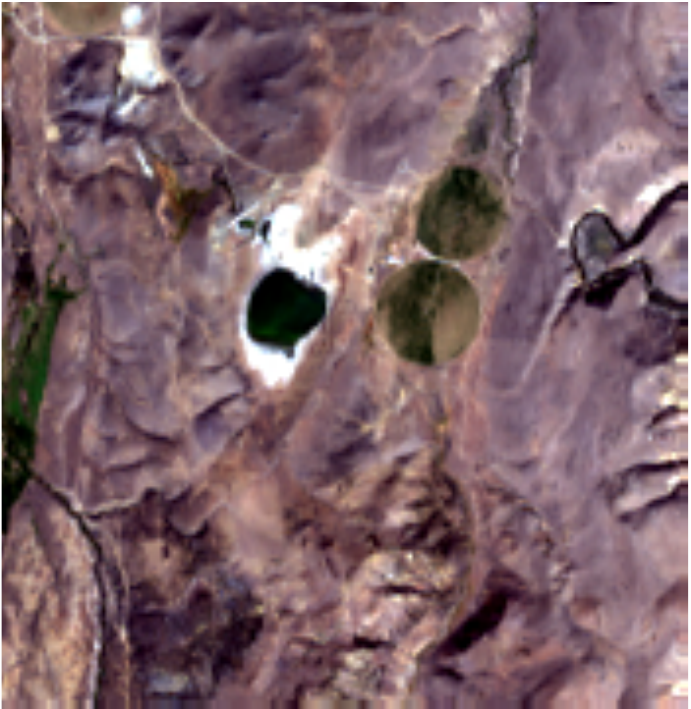}
					\caption{$\mathbf{Y}_{t_2}$}
					\label{fig:s1s2Yt2_1}
			\end{subfigure}
            \begin{subfigure}{\subfwidth}
					\centering
					\includegraphics[width=\figsize]{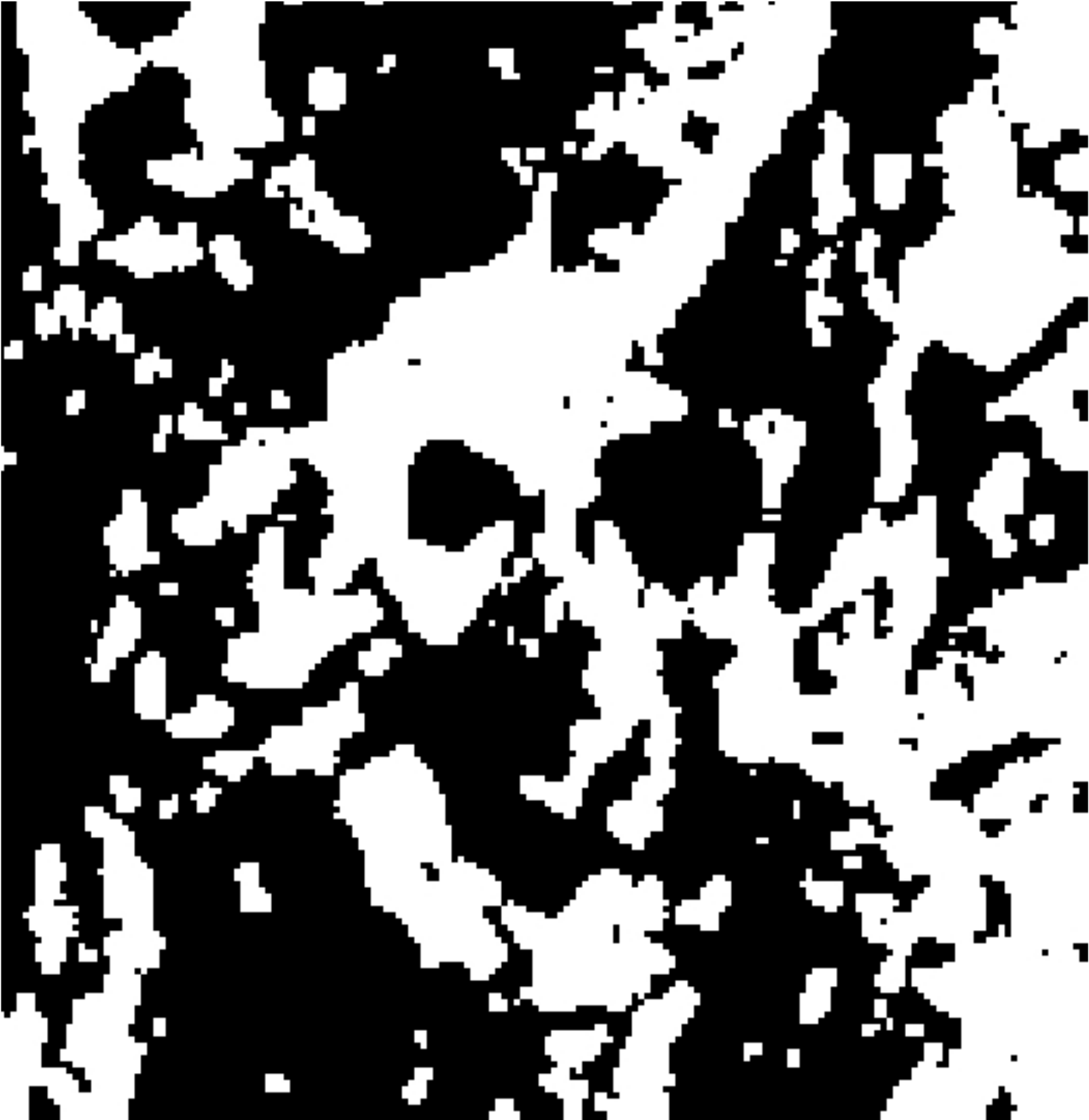}
					\caption{$\hat{\mathbf{m}}_{\mathrm{F}}$}
					\label{fig:s1s2FMAP_1}
			\end{subfigure}
			\begin{subfigure}{\subfwidth}
					\centering	
					\includegraphics[width=\figsize]{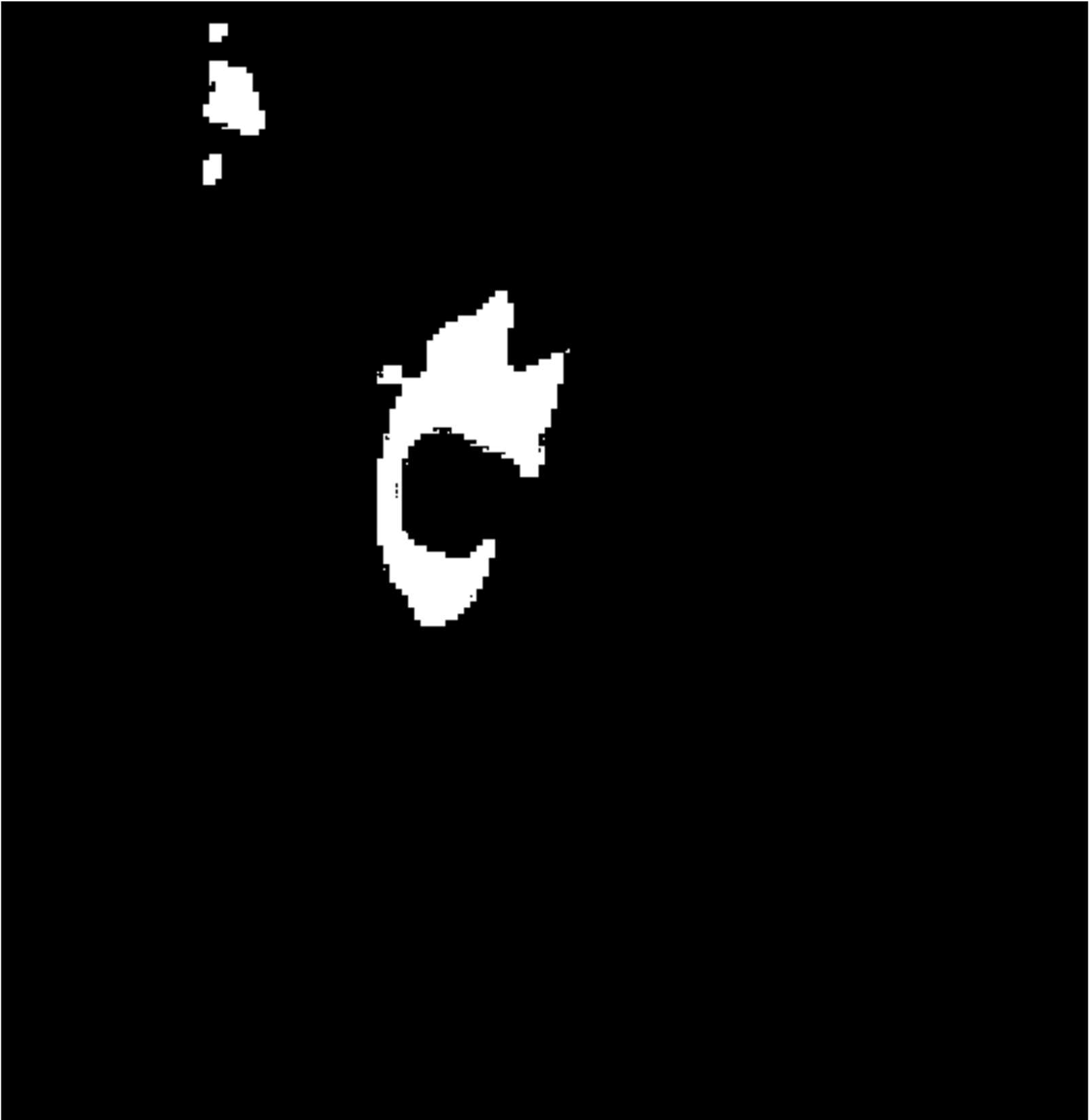}
					\caption{$\hat{\mathbf{m}}_{\mathrm{CDL}}$}
					\label{fig:s1s2DCMAP_1}
			\end{subfigure}
\caption{Real images affected by real changes without ground truth, Scenario 3 (different spatial resolutions): \protect\subref{fig:s1s2Yt1_1}  Sentinel-1 SAR image $\mathbf{Y}_{t_1}$ acquired on 04/12/2016, \protect\subref{fig:s1s2Yt2_1}  Landsat 8 MS image $\mathbf{Y}_{t_2}$ acquired on 09/22/2015, \protect\subref{fig:s1s2FMAP_1} change map $\hat{\mathbf{m}}_{\mathrm{F}}$ of the fuzzy method and \protect\subref{fig:s1s2DCMAP_1} change map $\hat{\mathbf{m}}_{\mathrm{CDL}}$ of the proposed method.}%
	\label{fig:realS1S2_1}%
\end{figure}

The second observed image pair consists in a Sentinel-1 SAR image acquired on April 12th 2016 and a Landsat 8 MS image acquired on September 22th 2015. This pair represents the most challenging situation among all presented images, namely differences in both modalities and resolutions. Figure \ref{fig:realS1S2_1} presents the observed images at each date and the recovered change maps. For this last experiment, the proposed method presents better accuracy in detection than the fuzzy one. All differences in all previous situations can be observed in this scenario, culminating in the poor detection performance of the fuzzy method and a reliable change map for the proposed one.

\subsection{Statistical performance assessment}\label{subsec:synthetic_images}
Finally, the last set of experiments aims at statistically evaluating the performance of compared algorithms thanks to simulations on real images affected by synthetic changes. More precisely, in the case of multi-band images, a dedicated CD evaluation protocol was proposed by \citet{ferraris_detecting_2017} based on a single high spatial resolution hyperspectral reference image. The experiments conducted in this work follow the same strategy. Two multimodal reference images acquired at the same date have been selected as change-free latent images. By conducting simple copy-paste of regions, as in \citet{ferraris_detecting_2017}, changes have been generated in both images as well as their corresponding ground-truth maps. This process allows synthetic yet realistic changes to be incorporated within one of these latent images, w.r.t. a pre-defined binary reference change mask locating the pixels affected by these changes and further used to assess the performance of the CD algorithms. This process is detailed in what follows.
  
\subsubsection{Simulation protocol} \label{subsubsec: simulation}

\noindent \textbf{Reference images --} The reference images $\mathbf{X}^{\mathrm{ref}}_{1}$ and $\mathbf{X}^{\mathrm{ref}}_{2}$ used in this experiment comes from two largely studied open access satellite sensors, namely Sentinel-1 \citep{european_space_agency_sentinel-1_2017} and Sentinel-2 \citep{european_space_agency_sentinel-2_2017} operated by the European Spatial Agency. These images have been acquired over the same geographical area, i.e., the Mud Lake region in Lake Tahoe, at the same date on April 12th 2016. To fulfil the requirements imposed by the considered CD setup, both have been manually geographically and geometrically aligned. The Sentinel-2 image used in this section is a $540 \times 525 \times 3$ image with $10$m spatial resolution and composed of $3$ spectral bands corresponding to visible RGB (Bands $2$ to $4$). On the other hand, Sentinel-1 reference image is a $540 \times 525$ interferometric wide swath high resolution ground range detected multi-looked SAR intensity image with a spatial resolution of $10$m according to 5 looks in the range direction.\\

\noindent  \textbf{Generating the changes --} Using a procedure similar to the one proposed by \citet{ferraris_detecting_2017}, given the reference images $\mathbf{X}^{\mathrm{ref}}_{\alpha}$ ($\alpha\in\left\{1,2\right\}$), and a previously generated change mask $\mathbf{m}\in \mathbb{R}^{N_\alpha}$, a change image $\mathbf{X}^{\mathrm{ch}}_{\alpha}$ can be generated as
\begin{equation}
\mathbf{X}^{\mathrm{ch}}_{\alpha} = \vartheta\left(\mathbf{X}^{\mathrm{ref}}_{\alpha},\mathbf{m}\right)
\end{equation}
where the change-inducing functions $\vartheta: \mathbb{R}^{\Ndim \times N_\alpha}\times \mathbb{R}^{N_\alpha} \rightarrow \mathbb{R}^{\Ndim \times N_\alpha}$ is defined to simulate realistic changes in some pixels of the reference images. A set of $10$ predefined change masks has been designed according to specific copy-paste change rules similar as the ones introduced by \citet{ferraris_detecting_2017}.\\

\noindent \textbf{Generating the observed images --} 
The observed images are generated under the previously defined $3$ distinct scenarios involving $3$ pairs of images, namely,
\begin{itemize}
  \item Scenario 1 considers two optical images,
  \item Scenario 2 considers two SAR images,
  \item Scenario 3 considers a SAR image and an optical image.
\end{itemize}
 Each test set pair $\left\{\mathbf{X}^{\mathrm{ref}}_{\alpha_1},\mathbf{X}^{\mathrm{ch}}_{\alpha_2}\right\}$ is formed by considering $\left(\alpha_1,\alpha_2\right) = \left(\alpha,\alpha\right)$ with $\alpha=1$ for Scenario 1 and $\alpha=2$ for Scenario 2. Conversely, for Scenario 3 handling multimodal images, two test pairs can be formed considering $\alpha_1\neq\alpha_2$, i.e., $\left(\alpha_1,\alpha_2\right) \in \left\{\left(1,2\right),\left(2,1\right)\right\}$.

	\begin{figure*}
    	\centering
        	\begin{subfigure}{\subfigwidthROC}
					\centering	
					\includegraphics[width=\figwidthROC]{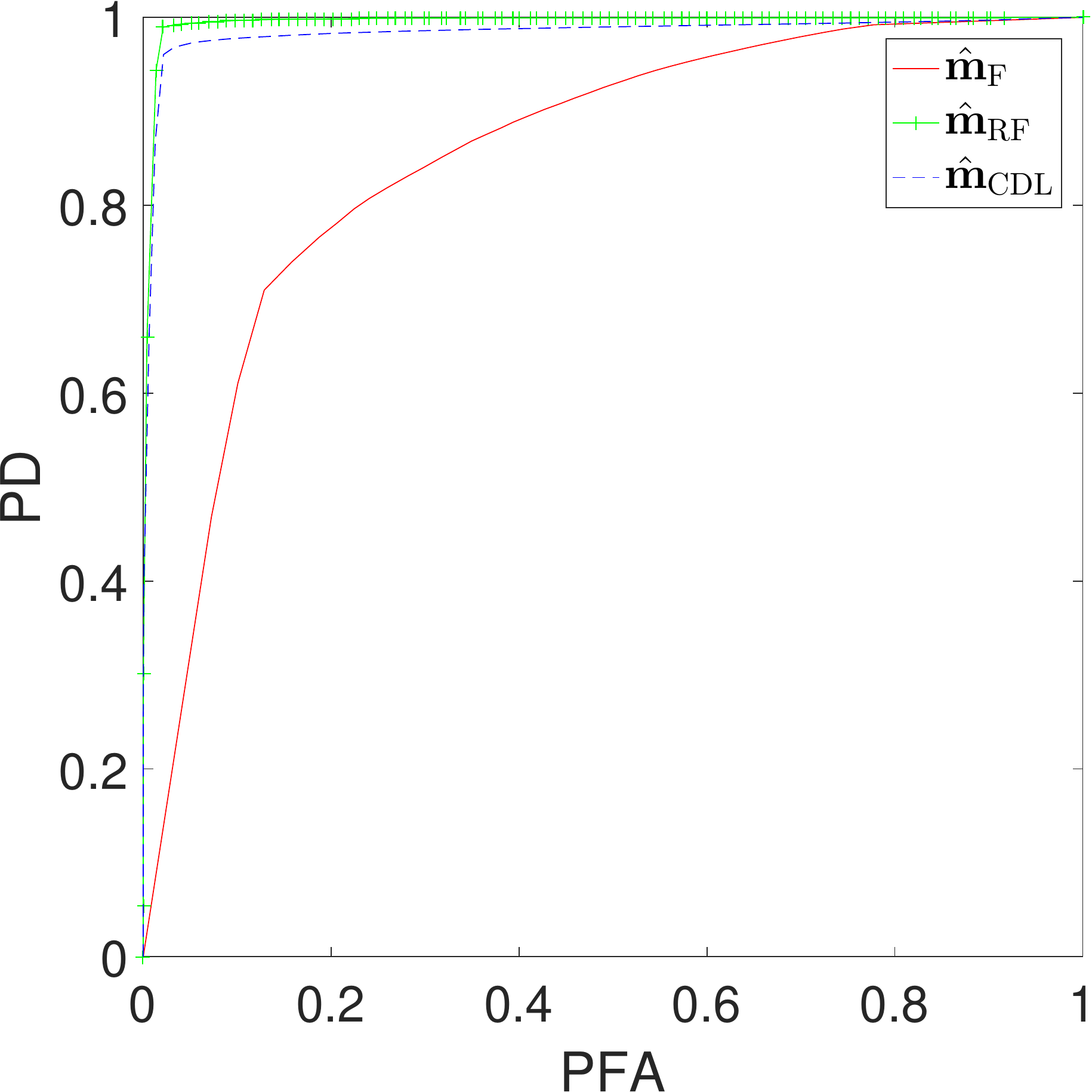}
					\caption{}
					\label{fig:rocS2S2}
			\end{subfigure}
			\begin{subfigure}{\subfigwidthROC}
					\centering	
				  	\includegraphics[width=\figwidthROC]{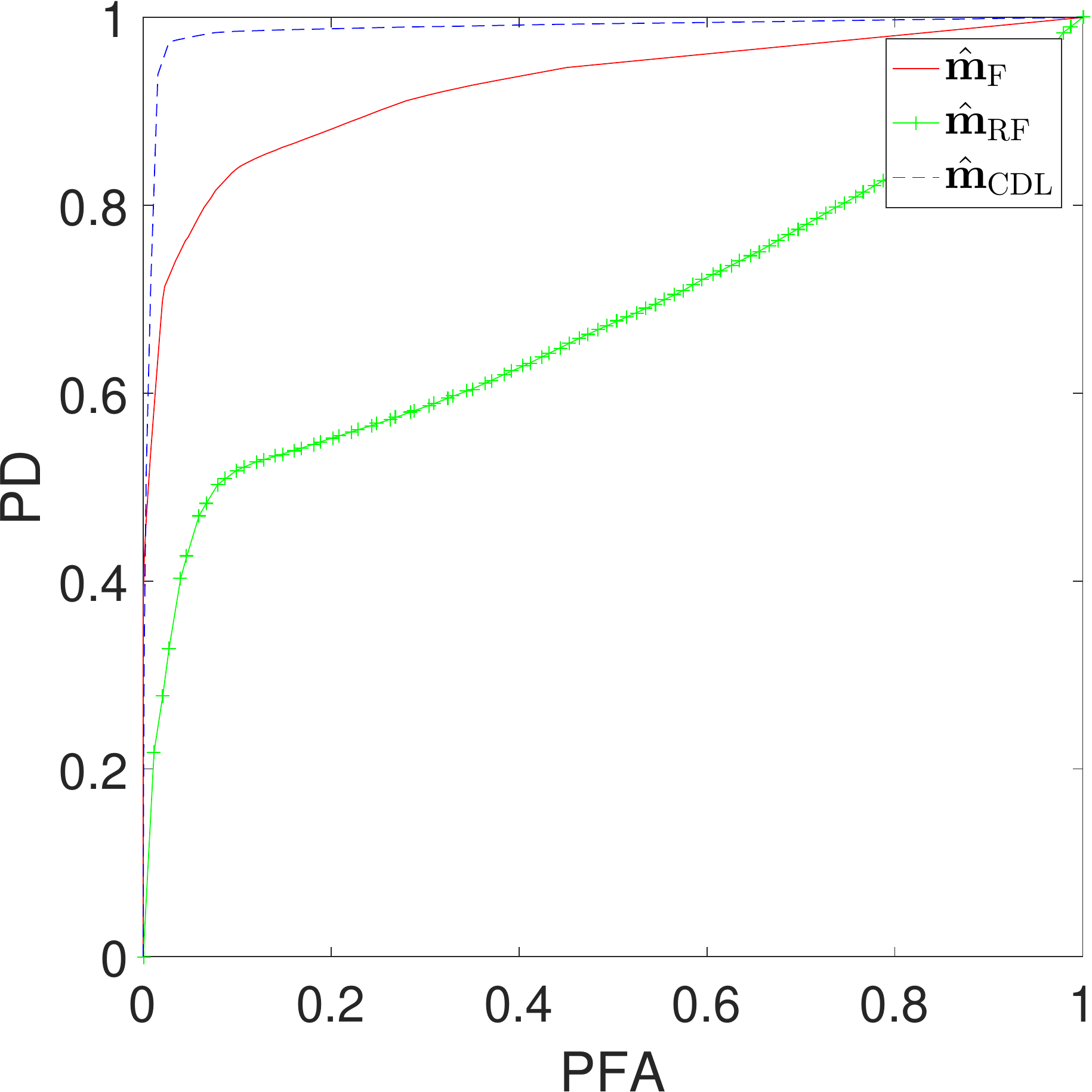}
					\caption{}
					\label{fig:rocS1S1}
			\end{subfigure}
			\begin{subfigure}{\subfigwidthROC}
					\centering
					\includegraphics[width=\figwidthROC]{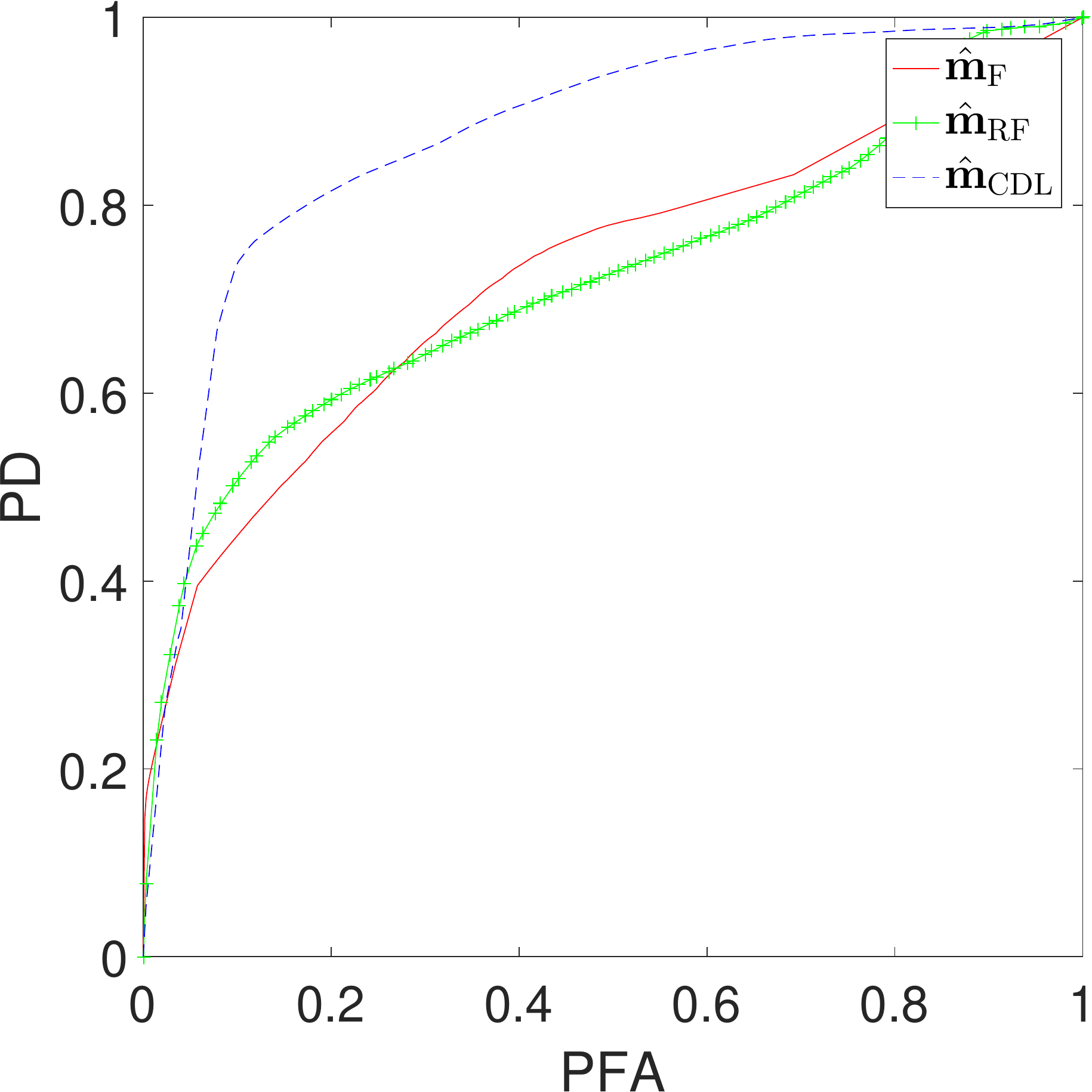}
					\caption{}
					\label{fig:rocS1S2}
			\end{subfigure}
			\caption{Real images affected by synthetic changes: ROC curves for  \protect\subref{fig:rocS2S2} Scenario 1,  \protect\subref{fig:rocS1S1} Scenario 2, \protect\subref{fig:rocS1S2} Scenario 3.}%
            \label{fig:ROC}%
	\end{figure*}

\subsubsection{Results}

The ROC curves displayed in Fig. \ref{fig:ROC} with corresponding metrics in Table \ref{table:ROCSEN} correspond to the CD results obtained for each specific scenario. These results are discussed below.\\

\noindent \textbf{Scenario 1: optical vs. optical --} The ROC curves displayed in Fig. \ref{fig:rocS2S2} with corresponding metrics in Table \ref{table:ROCSEN} (first two rows) correspond to the CD results obtained from a pair of optical observed images. These results show that the robust fusion achieves the best CD performance. This method has the benefit of exploring the joint model between the optical images, contrary to the fuzzy and proposed methods. Nevertheless, the proposed method achieves very similar performance. More importantly, they provide almost perfect detection even for very low PFA, i.e., for very low energy changes. On the other hand, the fuzzy method suffers from non detection and false alarm, even when applying the iterative strategy with a similar parameter selection approach as in \citet{gong_coupled_2016}. This happens mostly in low energy change regions. One possible explanation is that the iterative selection is not able to distinguish between low energy and unchanged pixels, which may bias the coupling of dictionaries. Also, the disjoint reconstruction cannot properly deal with low energy changes because coupling is not perfect. In addition, as the methods directly work with the observed images without estimating the latent image, noise could be interpreted as a change, thus increasing the false alarm rate.\\
	
\noindent \textbf{Scenario 2: SAR vs. SAR --} As in the previous case, this dual scenario considers homologous observed SAR images. In this case the ROC curves are displayed in Fig. \ref{fig:rocS1S1} with corresponding metrics in Table \ref{table:ROCSEN} ($3$rd and $4$th rows). Fig. \ref{fig:rocS1S1} shows that the proposed method offers the highest precision among the compared methods and keeps a high level of detection compared to the Scenario 1. The fuzzy method presents a better accuracy result compared to optical images. One of the reasons is that optical images are generally characterised by richer information, which makes the dictionary coupling more difficult than for two SAR images. At the end, the robust fusion CD method shows a very low detection accuracy as it is not suited to deal with SAR images.\\

\noindent \textbf{Scenario 3: optical vs. SAR --} This scenario corresponds to a more difficult problem than the previous one. The physical information extracted in each image cannot be directly related in the observational space, contrary to the previous scenarios. The ROC curve are displayed in Fig. \ref{fig:rocS1S2} with corresponding metrics in Table \ref{table:ROCSEN} (last two rows). As in Scenario 2, Fig. \ref{fig:rocS1S2} shows that the proposed method still offers the highest detection accuracy, while the other methods present a very poor performance. Regarding the fuzzy method, the dictionary and the subsequent sparse code estimations are severely affected by the differences in terms of dimensionality of measurements and dynamics. Even by tuning the algorithmic parameters to increase the weight of the image of lowest dynamics (or lowest resolution), the dictionaries are not properly coupled. Note that, to use the robust fusion method in this challenging scenario, a spectral degradation has been artificially applied to reach the same \emph{spectral} resolution for the two images. This has been achieved by considering a band-averaging to finally form a panchromatic image. Resulting detection performance is even poorer than the fuzzy method because it supposes the same physical information between images. Only strong related changes are detected in this case.

\begin{table}
    \caption{Real images affected by synthetic changes for Scenarios 1--3: quantitative detection performance (AUC and distance).}
    \centering
    \begin{tabular}{|c|c|c|c|c|c|}
    \cline{3-5}
    \multicolumn{2}{c|}{} & $\hat{\mathbf{m}}_{\mathrm{F}}$ & $\hat{\mathbf{m}}_{\mathrm{RF}}$ & $\hat{\mathbf{m}}_{\mathrm{CDL}}$ \\
    \hline
		\hline
		\multirow{2}{*}{\rotatebox{00}{Scenario 1}}     & AUC   & $     0.8520$& $\one{0.9946}$ & $\two{0.9838}$ \\
                                                     	& Dist. & $	0.7867$& $\one{0.9802}$ & $\two{0.9677}$ \\
		\hline
    \multirow{2}{*}{\rotatebox{00}{Scenario 2}}      	& AUC   & $\two{0.9251}$& $0.6819$ & $\one{0.9871}$ \\
                                                     	& Dist. & $\two{0.8587}$& $0.6185$ & $\one{0.9727}$ \\
    \hline
    \multirow{2}{*}{\rotatebox{00}{Scenario 3}}      	& AUC   & $\two{0.7277}$& $ 0.7227$ & $\one{0.8755}$ \\
                                                    	 & Dist.& $\two{0.6758}$& $0.6604$  & $\one{0.8097}$ \\
    \hline
    \end{tabular}
  \label{table:ROCSEN}
\end{table}


\subsection{Implementation details}

\vf{This paragraph briefly discusses some computational aspects of the proposed method. Concerning the algorithmic implementation, the code was implemented in Matlab and run on a Windows platform equipped with a Intel Core i7 8GB RAM CPU. Depending on the size of the images and on the number of iterations, analyzing a pair of images as those considered in the experiments described in this section may require one hour. The computational bottleneck is the memory required to store parameters for each image and possibly large temporary byproduct variables, in particular those relying on large matrix computation such as $\mathbf{A}_{\alpha}\mathbf{A}_{\alpha}^{T}$ or $\mathbf{D}_{\alpha}^T \mathbf{D}_{\alpha}$. The size of such matrices depends directly on the numbers $N_1$ and $N_2$ of pixels of the observed images and the chosen size $N_{\mathrm{d}}$ of the dictionaries.  Note however that some computations could have been conducted in parallel to speed up the procedure, e.g., optimizing independently w.r.t. the dictionaries $\mathbf{D}_{1}$ and $\mathbf{D}_{2}$ (see Section \ref{subsec:optim_dictionary}), optimizing independently w.r.t. the latent images $\mathbf{X}_1$ and $\mathbf{X}_2$ (see Section \ref{subsec:optim_latent}), and optimizing w.r.t. the scaling matrix $\mathbf{S}$ independently from the coding matrix $\mathbf{A}_2$ (see Section \ref{subsec:optim_scaling}). Moreover, we experimentally noticed that the solution reached after very few iterations is generally highly satisfactory. Nevertheless, after more iterations, the shapes of the dictionary atoms visually seem to be much more representative and the sparseness of the code is significantly enforced.}

\vf{Another important implementation aspect is the initialization of the optimization procedure, which is a critical issue because of the highly nonconvex nature of the problem. One may think of initializing the dictionary estimates by applying some techniques such as a (coupled) k-means clustering or the method proposed by \citet{gong_coupled_2016}. Nevertheless, as also noticed by \citet{seichepine_soft_2014}, it was empirically observed that a better strategy is to  randomly select coupled patches from the input data to form the initial dictionaries. This strategy may prevent the gradient to be initially stuck into a local minimum induced by weakly coupled dictionaries. As for the codes, since they are composed of a considerable number of block variables, one may pay attention to possibly exploding gradients due to high values. Initialization with zeros may induce the gradient to be stuck into local minima. Thus, one relevant strategy consists in initializing the codes with small random values.}

\section{Conclusion}
\label{sec:conclusion}

This paper proposed an unsupervised multimodal change detection technique to handle the most common remote sensing imagery modalities. The technique was based on the definition of a pair of latent images related to the observed images through a direct observation model. These latent images were modelled thanks to a coupled dictionary and sparse codes which provide a common representation of the homologous patches in the latent image pair. The differences between estimated codes were assumed to be spatially sparse, implicitly locating the changes. Inferring these representations, as well as the latent images, was formulated as an inverse problem which was solved by the proximal alternate minimization iterative algorithm dedicated to nonsmooth and nonconvex functions. Contrary to the methods already proposed in the literature, scaling problems due to differences in resolutions and/or dynamics were solved by introducing a scaling matrix relating coupled atoms. A simulation protocol allowed the performance of the proposed technique in terms of detection and precision to be assessed and compared with the performance of three algorithms. A real dataset collecting images from different multispectral and  SAR sensors at the same region was used to assess the reliability of the proposed method. Results showed that the method outperformed all state-of-the-art comparable methods in multimodal scenarios while presenting similar results as methods benefiting from prior knowledge of the scenario modelling. Future works include considering more complex image statistical models, such as non-Gaussian distribution for optical images \citep{zanetti_rayleigh-rice_2015}.

\begin{appendix}

\section{Data-fitting terms and corresponding proximal operators}
\label{ap:dft}

The data-fitting term $\mathcal{D}(\cdot|\cdot)$ is intimately related to the modality of the target image. This term defines the negative log-likelihood function relating the observed and latent images. Below, the most common data fitting terms and their associated proximal mappings are derived, defined as

\begin{equation}
  \mathrm{prox}^{\eta}_{\mathcal{D}(\mathbf{Y}|\cdot)}\left(\mathbf{U}\right) = \operatornamewithlimits{argmin}_{\mathbf{X}} \mathcal{D}(\mathbf{Y}|\mathbf{X})
  + \frac{\eta}{2} \left\|\mathbf{X}-\mathbf{U}\right\|_{\mathrm{F}}^2.
\end{equation}

\subsection{Multiband optical images}

Multiband optical images represent the most common modality of remotely sensed images. For this modality, the noise model may take into account several different noise sources \citep{deger_sensor_2015}. Nevertheless, it is commonly considered as additive Gaussian, up to some considerations in the acquisition, for instance sufficient number of the arriving photons. Therefore, the direct model $\mathit{T}_{\mathrm{MO}}[\cdot]$ in \eqref{eq:sensortransf} can be expressed as
\begin{equation}
			\mathbf{Y} = \mathbf{X} + \mathbf{N}
\end{equation}
where the noise matrix $\mathbf{N}$ is assumed to be distributed according to a matrix normal distribution (see, e.g., \citep{ferraris_robust_2017} for more details). Consequently, by assuming the noise components are independent and identically distributed\footnote{Pixelwise independence of the noise is a common assumption while spectral whiteness of the noise can be ensured by applying a whitening transform as pre-processing.} (i.i.d.), the data-fitting term associated with multiband optical images is  	
\begin{equation}
\mathcal{D}_{\mathrm{MO}}(\mathbf{Y}|\mathbf{X}) = \frac{1}{2}\left\|\mathbf{Y}-\mathbf{X}\right\|_{\mathrm{F}}^{2}.
\end{equation}
An explicit proximal operator associated with this function can be derived as
\begin{equation}
\mathrm{prox}^{\eta}_{\mathcal{D}_{\mathrm{MO}}(\mathbf{Y}|\cdot)}\left(\mathbf{U}\right) = \frac{\mathbf{Y} + \eta\mathbf{U}}{\eta+1}
\end{equation}
\subsection{Multi-look intensity synthetic aperture radar images}

SAR images correspond to the second most common modality of remote sensing images used in many applications. One of the main characteristics of such modality is that it allows to measure the scene in poor weather conditions and also during the night since SAR is an active sensor. Nevertheless, this configuration yields the speckle phenomenon, resulting from random fluctuations of the reflectivity of the backscattered signals. Many studies have been conducted to understand and mitigate the speckle phenomenon. A common approach that helps to decrease the speckle level while increasing the SNR consists in averaging samples of the same pixel acquired over independent observations. This procedure is usually referred to as multi-look processing. According to this strategy, the generative model is considered as a multiplicative perturbation by i.i.d random variables $\mathbf{N} = [n_i,\cdots,n_{N}]$ following a common gamma probability density function in intensity images with unit mean $\mathrm{E}[n_i] = 1$ and variance $\mathrm{var}[n_i] = \frac{1}{r}$ where $r$ is the number of looks. The direct model $\mathit{T}_{\mathrm{SAR}}[\cdot]$ can thus be written as
\begin{equation}
	\mathbf{Y} = \mathbf{X} \odot \mathbf{N}
\end{equation}
where $\odot$ denotes the termwise (i.e., Hadamard) product.

By assuming pixel independence, the data-fitting term for each pixel can be expressed as the sum of Itakura-Saito divergences
\begin{equation}
\mathcal{D}_{\mathrm{SAR}}(\mathbf{Y}|\mathbf{X}) =
 \sum_{i=1}^N \left(\frac{y_i}{x_i} - \log \frac{y_i}{x_i} - 1\right)
\end{equation}
This function has been widely considered for speckle removing \citep{aubert_variational_2008,woo_proximal_2013} and also music analysis \citep{fevotte_nonnegative_2009}. Nevertheless, it usually leads to a challenging non-convex problem which admits more than one global solution. In \citet{sun_alternating_2014}, the associated proximal operator is derived by computing the root of a $3$rd degree-polynomial equation. An alternative consists in considering an approximation by resorting to a log-transform of the data, e.g., leading to an I-divergence \citep{woo_proximal_2013,steidl_removing_2010}. Up to a constant, this divergence can be rewritten equivalently as a Kullback-Leibler divergence which is closely related to Poisson modeling \citep{figueiredo_restoration_2010}
\begin{equation}
\mathcal{D}_{\mathrm{SAR}}(\mathbf{Y}|\mathbf{X}) = \sum_{i=1}^N \left(x_i - y_i \log x_i\right).
\end{equation}
This data-fitting term leads to an explicit proximal operator for the $i$th component given by
\begin{equation}
\mathrm{prox}^{\eta}_{\mathcal{D}_{\mathrm{SAR}}(y_i|\cdot)}\left(u_i\right) = \frac{1}{2}\left(u_i - \frac{1}{\eta} + \sqrt{\left(u_i - \frac{1}{\eta}\right)^2 + \frac{4y_i}{\eta}}\right).
\end{equation}

\section{Usual proximal mappings involved in the parameter updates}
\label{ap:proj}

The projections and proximal operators involved on PALM algorithm \citep{bolte_proximal_2014} and described in Algorithm \ref{algo:PALM_SCDL_Diff} are properly defined as:
\begin{itemize}
\item The proximal map for $\mathbf{A}_{1}$ accounting for the sum $\lambda\left\|\cdot\right\|_1 + \iota_{\geq0}(\cdot)$ is explicitly given by:
\begin{equation}
\mathrm{prox}^{\eta}_{\lambda\left\|\cdot\right\|_1 + \geq0}\left(\mathbf{A}_{1}\right) =   \max\left(|a_{1,{(ji)}}|-\frac{\lambda}{\eta},0\right) \quad \forall (i,j)
\end{equation}
\item The proximal map for $\Delta\mathbf{A}$ accounting for the $\gamma\left\|\cdot\right\|_{2,1}$ is explicitly given by:
\begin{equation}
\hspace{-0.5cm}\mathrm{prox}^{\eta}_{\gamma\left\|\cdot\right\|_{2,1}}\left(\Delta\mathbf{A}\right) =  \begin{cases} \left(1 - \frac{\gamma}{\eta\left\|\Delta\mathbf{a}_i\right\|_{2}}\right)\Delta\mathbf{a}_i & \text{if} \left\|\Delta\mathbf{a}_i\right\|_{2}>\frac{\gamma}{\eta}\\
				0 & \text{otherwise.}
\end{cases}
\end{equation}
\item Projecting $\mathbf{D}$ onto set $\mathcal{S}$ can be computed explicitly based on \citet{thouvenin_modeling_2017,bolte_proximal_2014} which is given by:
\begin{equation}
\mathcal{P}_{\mathcal{S}}\left(\mathbf{D}\right) = \frac{\mathcal{P}_{+}(\mathbf{d}_{i})}{\left\|\mathcal{P}_{+}(\mathbf{d}_{i})\right\|^2_2} \quad \forall i = 1 \cdots \Natom
\end{equation}
with
\begin{equation}
\mathcal{P}_{\mathcal{+}}\left(\mathbf{d}_{i}\right) = \max\left(0,d_{(j,i)}\right) \quad \forall j = 1 \cdots \Ndim
\end{equation}
\item Projecting $\mathbf{S}$ onto set $\mathcal{C}$ is explicitly given by:
\begin{equation}
\mathcal{P}_{\mathcal{C}}\left(\mathbf{S}\right) =  \begin{cases} \max\left(0,s_{(j,i)}\right) & \forall i=j \\
0 & \text{otherwise}
\end{cases}
\end{equation}
\end{itemize}
\end{appendix}
\section*{Acknowledgments}
The authors would like to thank Prof. Jose M. Bioucas-Dias, Universidade de Lisboa, Portugal, for fruitful discussion regarding this work.

\bibliographystyle{elsarticle-harv}
\bibliography{strings_all_ref,publications_clean}
\end{document}